\documentclass{article}
\usepackage{eurosym}
\usepackage{amsfonts}
\usepackage{amssymb}
\usepackage{amsmath}
\usepackage{color}
\usepackage{graphicx}
\usepackage{pdfpages}
\usepackage[utf8]{inputenc}
\usepackage{booktabs}
\usepackage{float}
\usepackage{hyperref}

\newtheorem{theorem}{Theorem}

\newtheorem{assumption}{Assumption}

\newtheorem{corollary}{Corollary}

\newtheorem{definition}{Definition}

\newtheorem{lemma}{Lemma}

\newtheorem{proposition}{Proposition}

\input{tcilatex}
\begin{document}

\author{Marcia Schafgans\thanks{\textit{Corresponding author}: Economics
Department, London School of Economics, Houghton Street, London WC2A 2AE,
UK. E-mail address: m.schafgans@lse.ac.uk} \qquad \qquad Victoria Zinde-Walsh%
\thanks{%
Economics Department, McGill University, 855 Sherbrooke St. W., Montreal,
Quebec H3A 2T7, Canada. E-mail address: victoria.zinde-walsh@mcgill.ca.} \\
London School of Economics \quad McGill University and CIREQ }
\title{Multivariate kernel regression in vector and product metric spaces}
\date{January 2026}
\maketitle

\begin{abstract}
This paper derives limit properties of nonparametric kernel regression
estimators without requiring existence of density for regressors in $\mathbb{R}^{q}.$
In functional regression limit properties are established for multivariate
functional regression. The rate and asymptotic normality for the
Nadaraya-Watson (NW) estimator is established for distributions of
regressors in $\mathbb{R}^{q}$ that allow for mass points, factor structure,
multicollinearity and nonlinear dependence, as well as fractal distribution;
when bounded density exists we provide statistical guarantees for the
standard rate and the asymptotic normality without requiring smoothness. We
demonstrate faster convergence associated with dimension reducing types of
singularity, such as a fractal distribution or a factor structure in the
regressors. The paper extends asymptotic normality of kernel functional
regression to multivariate regression over a product of any number of metric
spaces. Finite sample evidence confirms rate improvement due to singularity
in regression over $\mathbb{R}^{q}.$ For functional regression the
simulations underline the importance of accounting for multiple functional
regressors. We demonstrate the applicability and advantages of the NW
estimator in our empirical study, which reexamines the job training program
evaluation based on the LaLonde data.\newline
\end{abstract}

\section{Introduction}

This paper extends nonparametric kernel regression to more general regressor
settings than those considered in the literature. The general regression
model 
\begin{equation}
Y=m(X)+u,\quad E(u|X)=0,  \label{model}
\end{equation}%
is free from the difficulty of choosing a parametric specification. We focus
on the Nadaraya-Watson (NW) estimator, introduced by Nadaraya (1965) and
Watson (1964), which recognizes that a continuous regression function can be
estimated pointwise by a weighted average that attaches higher weights to
close-by observations.

Here we emphasize the fact that the regressor $X$ can be a vector in $%
\mathbb{R}^{q}$, or alternatively $X$ may belong to a function space, a more
general metric space, or comprise of components from several such metric
spaces. In fact, the components of $X$ do not necessarily have to belong
to spaces of vectors or functions but could be intervals, graphs, or
networks, as long as a metric (or even a semi-metric) can be defined for
each space. $Y$ represents a scalar dependent variable, $u$ denotes an
unobserved error, and the conditional mean function $m$ satisfies some
smoothness assumptions.

The NW estimator has been used extensively with $X\in \mathbb{R}^{q}$ (see
e.g. the textbook Li and Racine, 2007, for discussion and examples) and has
recently been introduced to functional regression by Ferraty and Vieu
(2004). Well known limit distributional results for the NW estimator were
derived in $\mathbb{R}^{q}$ under restrictions requiring existence and
smoothness of the density. For functional regression (where there is no
density) the limit distributional results were derived in a univariate
context only.

In this paper we establish asymptotic normality of the NW estimator for
regression in the presence of a general regressor $X$ that could have a
multivariate {singular distribution} in $\mathbb{R}^{q}$ or is comprised of
any number of functional and vector regressors. A singular distribution does
not admit a density function that integrates to it.

In settings where data has both discrete and continuous components Racine
and Li (2007) obtained asymptotic normality of the NW estimator without
having to deal with the singularity by treating the discrete and
(absolutely) continuous components separately. However, sometimes the
distinction between discrete and continuous variables is not
straightforward; continuous variables could be discretized with different
levels of discretization. When data with both discrete and continuous
components is viewed as a vector in a Euclidean space, $X\in \mathbb{R}^{q},$
the distribution of $X$ is singular. In our simulations we demonstrate that
there may be no gain from avoiding the singularity by considering the
discrete regressors separately.

The presence of latent factors in the continuous regressors, common in
macroeconomic and finance models (e.g., Bai and Ng, 2006, for portfolio,
stock returns and macroeconomic data) could also imply a singular
distribution. For example, if the regressor is a \thinspace $q\times 1$
vector $X\sim N\left( 0,\Sigma \right) $ with $\Sigma $ a singular matrix of
rank $r<q$, the distribution is singular. Similarly, non-linear common
factors, such as in Hotelling's (1929) spatial model of horizontal
differentiation which assumes that each consumer has an `ideal' variety
identified by his location on the unit circle (see also Desmet et al.,
2010), imply a smaller effective dimension for the regressor space,
resulting in a singular distribution over $\mathbb{R}^{q}.$ This also is
true when there exists a functional relation between the regressors (e.g.
with exact collinearity that can arise in production functions, Ackerberg et
al., 2015).

Singularity also originates from a fractal structure in the data; examples
in economics include the daily prices in the cotton market (Mandelbrot,
1997), financial markets, and networks (see Takayasu et al., 2009). Fractals
are common to many geographic features, including coastlines, river networks
and landforms, and have been used in urban growth studies (e.g., Shen, 2002)
and spatial econometrics in general. Furthermore, singularities also result
when continuously distributed variables exhibit mass points (e.g.
Arulampalam et al., 2017, for neonatal mortality and Olson, 1998, for weekly
hours worked).

We demonstrate the benefit of extending the NW estimator to regression with
singular data by applying it to the data from a randomized experiment in a
job training program evaluation study by LaLonde (1986). Following the work
by Rosenbaum and Rubin (1983), Dehejia and Wahba (1999, 2002) applied
propensity score methods to the LaLonde data for estimation of causal
treatment effects in an attempt to generalize the experimental results to
nonexperimental data. The propensity score matching was used instead of a
multivariate nonparametric model with matching on individual characteristics
which was deemed impractical because of the high dimensionality of the
regressors. The benefits of kernel regression were analyzed in Heckman et
al. (1997, 1998) (without allowing for singularity). The LaLonde data and
methodologies were discussed by Angrist and Pischke (2009) and in a recent
review by Imbens and Xu (2024). As shown here the discreteness of most of
the regressors implies reduced dimension of the support of the joint
distribution; for continuous variables existence of continuous density is
not imposed and mass at zero in income is accounted for. Our asymptotic
results provide the validity of the NW estimator for this singular
distribution. The kernel estimators we employ give new insights into the
heterogeneous effects of the program, based on a variety of individual
characteristics and compare quite well with random forest estimates of the
conditional average treatment effect on the treated (CATT) (Wager and Athey,
2018).

Our results also extend to functional regression (see, e.g. Ramsey and
Silverman, 2005) where estimation and inference techniques have been
developed by Ferraty and Vieu (2004) and pointwise asymptotic normality was
established in regression for a Banach or metric space by Ferraty et al. (2007), Ferraty and Vieu (2006), and Geenens (2015) in the i.i.d. case. Masry
(2005) derived the limit distribution for a strongly mixing process.
Recently Kurisu et al. (2025) made a case for extending the univariate
set-up of functional regression by considering jointly a random vector and a
function to obtain an estimate for the propensity score used in evaluating
the average treatment effect. We establish asymptotic normality in
multivariate functional regression with regressors in any number of
heterogeneous metric spaces. This provides a basis for simultaneously
evaluating the impact of the different predictors rather than comparing
their performance in distinct models, as in Caldeira et al. (2020) and
Ferraty and Nagy (2022).\footnote{%
E.g., Caldeira et al. (2020) compares the model forecasting aggregate stock
market excess return on a function representing the history of returns with
regression models based on traditional predictors. Ferraty and Nagy (2022)
compare the performance of separate models for predicting adult height with
functional regressors (one being growth velocity profiles from ages 1-10 and
the other for 5-8).} Our simulations show that using multivariate rather
than univariate functional regression can improve the fit of the kernel
estimator.

We derive asymptotic normality results for a random regressor $X$ supported
on some domain in a vector space, $\mathbb{R}^{q},$ or metric, semi-metric
space, $\Xi ^{\left[ 1\right] },$ or a product of such spaces $\Xi ^{\lbrack
q]}\equiv \Xi _{1}^{[1]}\times \cdots \times \Xi _{q}^{[1]}$. The metrics on 
$\Xi _{l}^{[1]}$, $\left\Vert .\right\Vert _{l}$, may differ for each of the 
$q$ components of function spaces, thus as in Kurisu et al. (2025) one may
be the $\mathbb{R}^{1}$ space and the other one a function space. A key
ingredient in our technical derivations is small cube probability, which
characterizes local properties of $X$ in the general multivariate case in
place of the density. We introduce this concept here.

In the univariate metric space, $\Xi =\Xi ^{\lbrack 1]}$, the probability
measure is characterized by the small ball probability (e.g., see Ferraty
and Vieu, 2006): for the ball $B\left( x,h\right) =\left\{ X:\left\Vert
x-X\right\Vert \leq h\right\} $ {centered at $x$} in $\Xi ^{\left[ 1\right] }
$ {the probability measure} is denoted $P_{X}\left( B\left( x,h\right)
\right) $. Characterizing the measure locally via a ball is insufficient
when we wish to examine heterogeneous regressors in $\mathbb{R}^{q}$ {or, in
general, in product metric spaces $\Xi ^{\left[ q\right] }$.}

Kankanala and Zinde-Walsh (2024) introduced small cube probability for a
cuboid. A cuboid $C\left( x,h\right) $ centered around $x=\left(
x^{1},...,x^{q}\right) \in \mathbb{R}^{q}$ for a vector $h=\left(
h^{1},...h^{q}\right) ^{\prime }$ with positive finite components is defined
as the set $C\left( x,h\right) =$ $\left\{ X\in \mathbb{R}^{q}:\left\vert
X^{l}-x^{l}\right\vert \leq h^{l},l=1,\cdots ,q\right\} .$ With the
distribution function of $X$ given by $F_{X}$ the corresponding probability
measure is $P_{X}\left( C\left( x,h\right) \right) =\int_{C\left( x,h\right)
}dF_{X}.$ The small cube probability permits us to extend the regression on
univariate metric spaces to $\Xi ^{\lbrack q]}$ where the probability
measure $P_{X}$ is defined. For the cuboid 
\begin{equation}
C\left( x,h\right) =\left\{ X:\left\Vert X^{l}-x^{l}\right\Vert _{l}\leq
h^{l},l=1,\cdots ,q\right\} =\left\{ X:X^{l}\in B^{l}\left(
x^{l},h^{l}\right) ,l=1,\cdots ,q\right\} .  \label{cuboid}
\end{equation}%
the corresponding small cube probability is also denoted $P_{X}\left(
C\left( x,h\right) \right) .$\medskip 

One of our contributions is the derivation of auxiliary technical results
that express moments for the multivariate kernels and related functions in
terms of the small cube probabilities without appealing to differentiability
on which previous multivariate derivations relied. The moments and moment
bounds are derived for general multivariate local functions under arbitrary
distributions over $\mathbb{R}^{q}$ or probability measures over $\Xi ^{%
\left[ q\right] }$. Bounds on a moment functional expressed via power of
small cube probability pinpoint the rate of growth of the functional. These
results generalize the derivations for the univariate kernel used in
functional regression to the multivariate setting. The full details of these
auxiliary technical results are presented in the supplemental material
(Appendix A). The moment expressions could find use in other contexts, for
instance for local linear and local polynomial estimation in $\mathbb{R}^{q}$
or in products of suitable metric spaces, $\Xi ^{\left[ q\right] },$ kernel
estimation of distribution functions and conditional distributions in $%
\mathbb{R}^{q}$ as well as to kernel regression of objects in metric spaces
on objects in products of spaces.

Implementation of the NW estimator relies on a tuning bandwidth parameter.
We show that in $\mathbb{R}^{q}$ a popular cross-validation method of
choosing a bandwidth with properties that were worked out for the absolutely
continuous (a.c.) case, has similar properties in some empirically relevant
classes of singular distributions with dimension-reducing singularity. We
also examine adaptive bandwidth selection for regressor distributions that
are represented by a mixture of a continuous distribution with some mass
points.

We provide simulation evidence on some important features of the behavior of
the NW estimator under possible singularity of the distribution of
regressors, $F_X$, in $\mathbb{R}^{q}$, in particular on the pointwise rate
of convergence and specific impact of mass points. We examine the behavior
of the NW estimator for models with dependence on both a functional object
in $\Xi ^{\left[ 1\right] }$ and a random variable.

The structure of the paper is as follows. Section 2 provides the set-up
suitable for the multivariate vector and functional regression highlighting
the probability measure for the regressor. Section 3 gives the asymptotic
normality results under the most general distributional assumptions. Section
4 discusses implementation, in particular, bandwidth selection. Section 5
provides a sketch of the simulation results and Section 6 is devoted to the
empirical study. The supplementary material collects various auxiliary
results and the proofs as well as the details of the Monte Carlo simulations
and the empirical study.

\section{The set-up and assumptions}

This section provides the formula for the Nadaraya-Watson (NW) kernel
estimator over $\Xi ^{\left[ q\right] }$, introduces some useful notation
and gives formal assumptions. The distributional assumptions are very
general in that they do not restrict the distribution over $\mathbb{R}^{q}$
to have absolutely continuous components, and apply to the probability
measure over the multivariate metric space $\Xi ^{\left[ q\right] }$ for an
arbitrary $q.$

\begin{assumption}
\label{A.measure on prod} [Probability Measure] Given the metric measure
spaces $\Xi _{l}^{\left[ 1\right] },$ $l=1,...,q$ with corresponding
sigma-algebras and probability measures $P_{X^{l}}$ assume that the
sigma-algebra for $\Xi ^{\left[ q\right] }=\prod_{l=1}^{q}\Xi _{l}^{\left[ 1%
\right] }$ is generated by the products of sets from sigma algebras for $\Xi
_{l}^{\left[ 1\right] }$ and a probability measure $P_{X}$ is defined on
this sigma algebra$;$ the mapping of $X=\left( X^{1},...,X^{q}\right) $ into
each of the components $X^{l}\in \Xi _{l}^{\left[ 1\right] }$ is measurable $%
\left( P_{X^{l}}\right) $ with respect to the joint measure.
\end{assumption}

In the product space $\Xi ^{\left[ q\right] }$ we define a vector $w$ as $%
\left( w^{1},...,w^{q}\right) ^{T}$ where each component is in the
corresponding space, thus for $\Xi =\mathbb{R}^{q},$ $w$ is a $q$%
-dimensional vector of reals, in $\Xi =\Xi ^{\lbrack q]}$ each $w^{l}\in \Xi
_{l}^{[1]},\ l=1,...,q.$ The bandwidth vector is $h=\left(
h^{1},...,h^{q}\right) \in \mathbb{R}^{q}$ with $0<\underline{h}=\min
\left\{ h^{1},...,h^{q}\right\} >0$ and $\bar{h}=\max \left\{
h^{1},...,h^{q}\right\} $. We use the same notation $\left\Vert \cdot
\right\Vert $ for the absolute value of a scalar in $\mathbb{R}^{1}$, the
Euclidean norm for a vector in $\mathbb{R}^{q}$ or norm for a function in $%
\Xi =\Xi ^{\lbrack 1]},$ with $\Xi ^{\lbrack 1]}$ a Banach space, or metric
(semi-metric) in a metric space $\Xi ^{\lbrack 1]}$; where the meaning is
not clear from the context we shall specify.

\subsection{The Nadaraya-Watson (NW) estimator}

The NW estimator, $\widehat{m}(x)$, for a sample $\left\{ \left(
Y_{i},X_{i}\right) \right\} _{i=1}^{n}$ generated by (\ref{model}) is
defined below. Generically the argument of the kernel function is 
\begin{align}
& W_{X}\left(x\right )  \notag \\
=& \left\{ 
\begin{array}{lll}
h^{-1}(x-X) & =\left( \left( h^{1}\right) ^{-1}(x^{1}-X^{1}),\cdots ,\left(
h^{q}\right) ^{-1}(x^{q}-X^{q})\right) & \text{on }\Xi =\mathbb{R}^{q} \\ 
h^{-1}\left\Vert x-X\right\Vert & =\left( \left( h^{1}\right)
^{-1}\left\Vert x^{1}-X^{1}\right\Vert _{1},\cdots ,\left( h^{q}\right)
^{-1}\left\Vert x^{q}-X^{q}\right\Vert _{q}\right) & \text{on }\Xi =\Xi ^{%
\left[ q\right] }.%
\end{array}%
\right.  \label{W(x)}
\end{align}

The NW estimator is given by 
\begin{eqnarray}
\widehat{m}\left( x\right) &=&B_{n}^{-1}\left( x\right) A_{n}\left( x\right)
,\text{ with}  \label{NW} \\
B_{n}\left( x\right) &=&\frac{1}{n}\sum_{i=1}^{n}K\left( W_{i}(x)\right) ;%
\text{ }A_{n}\left( x\right) =\frac{1}{n}\sum_{i=1}^{n}K\left(
W_{i}(x)\right) Y_{i}.\;\;\;\;\;\;\;  \label{B, A}
\end{eqnarray}%
where $K(W_{i}(x))=K(W_{X_{i}}\left(x\right) )$ is a multivariate
(non-negative) kernel function and $h$ usually depends on $n$; $x$ such that
at least for some $i$ we have that $K\left( W_{i}(x)\right) >0$. The kernel
function $K$ and bandwidth vector $h$ determine the properties for the NW
estimator. In the metric space $\Xi =\Xi ^{\lbrack 1]}$ the kernel function $%
K$ is defined for a univariate non-negative argument; in the case $\Xi =\Xi
^{\lbrack q]}$ with $q>1$ different bandwidths could appear for the
different components $W_{X}^{l}\left( x\right) ,l=1,..,q.$ With a symmetric
kernel on $\mathbb{R}^{q}$ we can just write $W_{X}\left( x\right)
=h^{-1}\left\Vert x-X\right\Vert $ for any $\Xi .$

\subsection{The kernel}

We restrict the multivariate kernel functions on $\mathbb{R}^{q}$ to have
bounded support and be suitably differentiable in the interior.

Let $I_{\xi }$ denote any subset of the set $\left\{ 1,...,q\right\} $ $\ $%
of consecutive non-negative integers; there are $2^{q}$ such subsets
including the empty set $\varnothing ;$ denote by $q\left( \xi \right) $ the
cardinality of the set $I_{\xi }=\left\{ j_{1},...,j_{q({\xi })}\right\} $
with $j_{1}<...<j_{q({\xi })}.$ We use $\prod_{j\in I_{\xi }}^{{}}\left(
\partial _{j}\right) $ to denote an operator that, when applied to a
differentiable function $g\left( z\right) =g\left( z^{1},...,z^{q}\right) $
at $z$, maps it to its partial derivative for $j_{1}<...<j_{q({\xi })}$,
that is 
\begin{equation*}
\left( \prod_{j\in I_{\xi }}^{{}}\left( \partial _{j}\right) \right) g\left(
z\right) =\frac{\partial ^{q({\xi })}}{\partial _{j_{1}}...\partial _{j_{q({%
\xi })}}}g\left( z\right) .
\end{equation*}%
We call a function $g(z)$ \textquotedblleft sufficiently
differentiable\textquotedblright\ if for any set $I_{\xi }$ the derivative $%
\left( \prod_{j\in I_{\xi }}^{{}}\left( \partial _{j}\right) \right) g\left(
z\right) $ exists and is continuous at any point on the interior of its
support.\medskip

The following assumption is made on the kernel function.

\begin{assumption}
\label{A.kernel} [Kernel]

\begin{itemize}
\item[(a)] The kernel function $K\left( w\right) =K\left(
w^{1},...,w^{q}\right) $ is a sufficiently differentiable density function.

\item[(b)] $K\left( w\right) $ is non-negative; $K\left( w\right) $ is
non-increasing for $w:w^{j}\geq 0, \ j=1,...,q$.

\item[(c)] $K\left( w\right) $ is either symmetric (with respect to zero)
with support on $\left[ -1,1\right] ^{q}$ or $K\left( w\right) $ is
supported on $\left[ 0,1\right] ^{q}$.

\item[(d)] $K\left( w\right) $ satisfies $K\left(\iota\right ) >0$ where $%
\iota=(1,...,1)^{\prime }$.
\end{itemize}
\end{assumption}

Assumptions \ref{A.kernel}(a--c) are satisfied by the commonly employed
product kernels of Epanechnikov or quartic kernels. Assumption \ref{A.kernel}%
(d) is not usual for kernel regression on $\mathbb{R}^{q};$ in the context
of univariate functional regression\ it is satisfied by a Type I kernel
defined in Ferraty and Vieu (2006) as \thinspace $K:C_{1}I_{\left[ 0,1\right]
}\leq K\leq C_{2}I_{\left[ 0,1\right] }$ with some $0<C_{1}\leq C_{2}<\infty
.$ Condition (d) in conjunction with (a-c) provides the same type of
univariate kernel. Extended to a multivariate setting it can be said that a
kernel that satisfies Assumption \ref{A.kernel} (a-d) is a type I kernel.
The uniform kernel is an example. The functional regression literature
demonstrates that with kernels of type I asymptotic normality can be
established in more general univariate settings. As commonly used in $%
\mathbb{R}^{q}$ kernels are not of type I, the asymptotic normality results
are given separately to apply under Assumption \ref{A.kernel}(a,b,c) and
under the full Assumption \ref{A.kernel}.

\subsection{Additional Assumptions}

Consider the process $\left\{ \left( X_{i},Y_{i}\right) \right\} _{i\in 
\mathbb{N}}.$ An i.i.d sequence would provide the simplest characterization,
but strong mixing makes it possible to extend the results to time series
data. Denote by $\mathcal{F}_{a}^{b}$ the sigma algebra generated by $%
\left\{ \left( X_{i},Y_{i}\right) \right\} _{i=a}^{b}.$ Define 
\begin{equation*}
\alpha \left( l\right) =\underset{t}{\sup }\underset{A\in \mathcal{F}%
_{-\infty }^{t};B\in \mathcal{F}_{t+l}^{\infty }}{\sup }\left\vert P\left(
AB\right) -P\left( A\right) P\left( B\right) \right\vert .
\end{equation*}%
Recall that the process is strong mixing if $\alpha \left( l\right)
\rightarrow 0$ as $l\rightarrow \infty .$

\begin{assumption}
\label{A.mom} [Data Generating Process and Moments]

\begin{itemize}
\item[(a)] The sequence $\left\lbrace(Y_{i},X_{i})\right\rbrace$ for $%
i=1,\cdots,n$ with $Y_{i}\in \mathbb{R};X_{i}\in \Xi ^{\left[ q\right] }$ is
stationary and strong mixing with $\alpha \left( l\right) $ that satisfies
for some $\zeta >0$%
\begin{equation*}
\alpha \left( l\right) <C l^{-\kappa };\ \text{ }\kappa >\frac{2\left(
2+\zeta \right) }{\zeta}.
\end{equation*}

\item[(b)] $E(u|X=x)=0;$ $\mu _{2}\left( x\right) =E\left( u^{2}|X=x\right) $
satisfies $0<L_{\mu _{2}}<\mu _{2}\left( x\right) <M_{\mu _{2}}<\infty ,$ $%
\mu _{2}\left( x\right) $ is continuous in the neighborhood of $x.$

\item[(c)] $E\left\vert Y_{i}\right\vert ^{2+\zeta }<\infty $ and $%
E(\left\vert u \right\vert^{2+\zeta}|X=x)<\infty.$

\item[(d)] For $x\in \Xi ^{\left[ q\right] }$ and $i\neq j$ the bivariate
function 
\begin{equation*}
\mu \left( x_{1},x_{2}\right) =E\left( \left\vert u_{i}u_{j}\right\vert
|X_{i}=x_{1},X_{j}=x_{2}\right)
\end{equation*}
is continuous in a neighborhood of the point $\left( x,x\right) \in \Xi ^{%
\left[ q\right] }\times \Xi ^{\left[ q\right] }.$

\item[(e)] The conditional expectation $E\left( \left\vert
Y_{i}Y_{j}\right\vert |X_{i},X_{j}\right) \leq C<\infty $ for all $i,j.$
\end{itemize}
\end{assumption}

\noindent The assumption requires a polynomial bound on the rate of decline
of the mixing coefficient with a link to the moment of $Y;$ it is similar to
those in Masry (2005) and Hong and Linton (2020).

\begin{assumption}
\label{A.mx} [Conditional mean] The function $m\left( x\right) $ on the
space $\Xi ^{\left[ q\right] }$ is such that%
\begin{equation*}
\left\vert m\left( x\right) -m\left( z\right) \right\vert \leq M_{\Delta m}%
\underset{l}{\max }\left\Vert x^{l}-z^{l}\right\Vert _{l}^{\delta };\text{ }%
\delta >0.
\end{equation*}
\end{assumption}

Assumption \ref{A.mx} requires Holder continuity of $m\left( x\right) ;$ it
would follow from differentiability or Lipschitz continuity in $\mathbb{R}%
^{q}$ with $\delta =1$. In the above assumptions, and below, $L$ and $M$
denote lower and upper bounds of functions where the subscript typically
denotes the function whose bounds are provided. The bounds could depend on
the point $x.$

\subsection{The probability measures}

For the probability measure $P_{X}$ on a generic space $\Xi ,$ that could
coincide with $\mathbb{R}^{q},$ $\Xi ^{\left[ 1\right] },$ or $\Xi ^{\left[ q%
\right] },$ any point $x\in \Xi $ is a point of support if for $\underline{h}%
>0$ {the measure $P_{X}\left( C\left( x,\underline{h}\right) \right) >0.$}

\subsubsection{Measures on $\mathbb{R}^{q}$}

By the Lebesgue decomposition, the distribution $F_{X}$ on $\mathbb{R}^{q}$
can be represented as a mixture of an absolutely continuous distribution, $%
F^{a.c.},$ a singular distribution (the distribution function is continuous
but there is no function that integrates to it), $F^{s},$ and a discrete
distribution, $F^{d}:$ 
\begin{equation*}
F_{X}\left( x\right) =\alpha _{1}F^{a.c.}\left( x\right) +\alpha
_{2}F^{s}\left( x\right) +\alpha _{3}F^{d}\left( x\right) ;\alpha _{l}\geq
0,\ {l=1,2,3};\ \sum\nolimits_{l=1}^{3}\alpha _{l}=1.
\end{equation*}

In a multivariate setting as soon as at least one variable is continuously
distributed, mass points do not arise and the joint distribution is a
continuous function, but with some discrete components or mass points in
some of the continuous components the distribution can no longer be
absolutely continuous and is singular. In many applications at least one of
the variables is assumed continuous and in a semiparametric regression often
an index model is assumed (single index in Ichimura, 1993; multiple index in
Donkers and Schafgans, 2008) to avoid singularity as well as to reduce
dimensionality of the model. In a general multivariate distribution the
presence of singularity achieves reduction of dimension (see, e.g. examples
2-4 in Kankanala and Zinde-Walsh, 2024) that will have a similar beneficial
effect on the convergence of the kernel estimator.

\subsubsection{Measures on metric spaces and products}

The discussion in this section applies to the space $\mathbb{R}^{q}$ as a
special case. Particular classes of probability measures considered in
univariate functional regression (e.g. Ferraty et al., 2007, Ferraty and Vieu, 2006) have a small ball probability centered at a point $x$ of support
either with a polynomial (fractal) rate of decline $P_{X}(B(x,h))\sim
C\left( x\right) h^{\tau }>0$ ($\tau >0),$ or with an exponential type rate
of decline $P_{X}(B(x,h))\sim C\left( x\right) \exp (-h^{-\tau _{1}}\log
h^{-\tau _{2}})$ ($\tau _{1}>0,\tau _{2}>0)$ as $h\rightarrow 0.$ This
characterization can be applied to the multivariate setting by replacing $%
B(x,h)$ with the cuboid and the univariate bandwidth in the rate with $\bar{h%
}.$ The exponential rate of decay of the small ball probability requires a
type I kernel and leads to slow convergence for the estimators (curse of
dimensionality). There are ways to mitigate the curse of dimensionality
arising from such exponential decay. It is common to apply finite
dimensional approximation of these functionals as suggested in Gasser et al.
(1998). Indeed, the case where functional data can be accurately
approximated in a finite dimensional space is not rare (corresponds to
observation of smooth curves with common shape) as noted by Ferraty and Nagy
(2022).\vspace{0.1in}

Kernels of type I play an important role in establishing pointwise
asymptotic normality in the absence of any restrictions on the decline of
the small cube measure. For kernels that may not be of type I sufficient
conditions on the shrinkage of the probability measure as $h\rightarrow 0$
were proposed in Assumption $H_{3}$ in Ferraty et al. (2007), Ferraty and
Vieu (2006) and were referred to in various subsequent papers on functional
regression, e.g. Hong and Linton (2020). The assumption below generalizes these
conditions to apply to $C\left( x,h\right) $ on $\Xi ^{\left[ q\right] };$
the assumption is both necessary for the conditions to hold (see the
supplementary material, Appendix B) and at the same time sufficient for the convergence
results.

\begin{assumption}
\label{A.dist} [Small ball probability measure] {Given any point $x\in \Xi ^{%
\left[ q\right] }$ in the support of the probability measure $P_{X}$} for
all $h$ with $\underline{h}>0$ and for some $0<\varepsilon <1,$ there is a
constant $1<C_{\varepsilon }\,<\infty $ such that 
\begin{equation}
\text{ }\frac{P_{X}(C(x,h))}{P_{X}(C(x,\varepsilon h))}<C_{\varepsilon
}<\infty .  \label{general band}
\end{equation}
\end{assumption}

\begin{definition}
$\mathcal{D}$ is the class of probability measures that satisfies (\ref%
{general band}).\footnote{%
Condition (\ref{general band}) is equivalent to the doubling property (e.g.
Vol'berg, Konyagin, 1988) that states that (\ref{general band}) applies with 
$\varepsilon =1/2.$ Indeed for any $\varepsilon $ there are positive
integers $\kappa _{1},\kappa _{2}:$ $\varepsilon \geq 2^{-\kappa _{1}}$ and $%
2^{-1}\geq \varepsilon ^{\kappa _{2}}.$ If the measure is doubling for
constant $C_{1/2}$, then (\ref{general band}) holds for $C_{\varepsilon
}=C_{1/2}^{\kappa _{1}};$ if ( \ref{general band}) holds, then the constant
for doubling is $C_{1/2}=C_{\varepsilon }^{\kappa _{2}}.$ We introduce the
form (\ref{general band}) in case there is a preference for some $%
\varepsilon .$\medskip}
\end{definition}

A polynomial decay condition places a measure into class $\mathcal{D}.$
Indeed if the small cube probability satisfies 
\begin{equation}
0<L_{P}\left( x\right) (2\underline{h})^{s\left( x\right) q}\leq
P_{X}(C(x,h))\leq M_{P}\left( x\right) \left( 2\bar{h}\right) ^{s\left(
x\right) q}<\infty  \label{LM F bounds}
\end{equation}%
{where for some $c$, $H(x),$ $1\leq c<\infty ,$ $0<H\left(x\right) <\infty $
and $\bar{h}=c\underline{h}<H\left( x\right) ,$ $0\leq s\left( x\right) \leq
1$ and $M_{F}\left( x\right) /L_{F}\left( x\right)<B<\infty $ at all points
of support $x$, (\ref{general band}) holds with $C_{\varepsilon }=B\left(
c/\varepsilon \right) ^{q}.$}

Condition (\ref{LM F bounds}) applies quite widely and holds for many
distributions of regressors used in econometric models. In $\mathbb{R}^{q}$
it is satisfied by any absolutely continuous distribution with\ a positive
bounded density function $f_X\left( x\right) $ where $M_{P}\left( x\right)
\geq \underset{\tilde{x}\in C\left( x,H\right) }{\sup }f_X\left( \tilde{x}%
\right) ;$ $L_{F}\left( x\right) =\underset{\tilde{x}\in C\left(
x,H/c\right) }{\inf }f_X\left( \tilde{x}\right) $ and $s\left( x\right) =1.$
If $x$ is an isolated mass point then (\ref{LM F bounds}) applies with $s=0.$
If $X$ has a linear structure with $r$ common factors, the probability
measure is singular and satisfies (\ref{LM F bounds}) with $s\left( x\right)
=s=\frac{q}{r}.$ For a fractal distribution that is singular with constant $%
s,$ $0<s<1$, the bounds also apply.

Condition (\ref{LM F bounds}) is satisfied by the general class of Ahlfors
(1966) regular (A-r) distributions common in statistics, where for this
class $s\left( x\right) =s,$ and $L_{P}\left( x\right) =L$ and $M_{P}\left(
x\right) =M$ are constants, as well as by a finite mixture of such
distributions (as proved in the supplementary material, Appendix B). Thus an absolutely
continuous distribution ($s=1)$ or, more generally, a measure given by a
continuous possibly singular distribution function that satisfies (\ref{LM F
bounds}) contaminated with some mass points ($s=0$) is in $\mathcal{D}$%
\textbf{$;$ }this applies to the empirical example examined here, ensuring
the pointwise asymptotic normality of the NW estimator with standard kernels.

\subsubsection{Joint measure}

Consider the product space $\Xi ^{\left[ 2q\right] }=\Xi ^{\left[ q\right]
}\times \Xi ^{\left[ q\right] };$ the measure on this product space has
marginals $P_{X}$ on each $\Xi ^{\left[ q\right] }$ (see, e.g., Pollard,
2001). {The joint measure $P_{s,t}\left( C\left( x,h\right) \times C\left(
x,h\right) \right) ,$ defined as $\Pr \left( X_{t}\in C\left( x,h\right)
,X_{s}\in C\left( x,h\right) \right) ,$ is a product of the measures of the
cuboid in the case of independency.} With dependence an additional
assumption is made on how the joint measure relates to the small cuboid
measure. {We provide the same assumption as in e.g. Masry (2005) and Hong
and Linton (2020) for the small cube probability.}

\begin{assumption}
\label{A.Jointmeasure} [Joint Measure] The joint measure $P_{s,t}\left(
C\left( x,h\right) \times C\left( x,h\right) \right) $ is such that for some 
$0<M_{FF}<\infty $%
\begin{equation}
\underset{t\neq s}{\sup }P_{s,t}\left( C\left( x,h\right) \times C\left(
x,h\right) \right) \leq M_{FF} \left( \overset{\textcolor{white}.}{P}%
_{X}\left( C\left( x,h\right) \right) \right) ^{2}.  \label{general product}
\end{equation}
\end{assumption}

\section{Asymptotic normality of the NW estimator}

Consider the NW estimator as given by (\ref{NW}), (\ref{B, A}). As the
sample size increases the bandwidths {are assumed} go to zero. For $\Xi =%
\mathbb{R}^{q}$ the denominator, $B_{n}(x)$ is proportional to the usual
kernel density estimator, given by $h^{-q}B_{n}(x),$ at point $x.$ {When the
density, $f_{X}\left( x\right) $, exists and is continuous}, the estimator $%
h^{-q}B_{n}(x)$ consistently estimates $f_{X}\left( x\right) ,$ but if the
density does not exist, $h^{-q}B_{n}\left( x\right) $ diverges to infinity.
Consistency of the NW estimator $\widehat{m}\left( x\right) $ over a
univariate metric space was established in Gy\"{o}rfi et al. (2002), the
limit distribution in Ferraty et al. (2007), Masry (2005) and Geenens (2015).

The key to the asymptotic normality result is the derivation of the moments
for multivariate functions of the form $g(X)K^{m}\left( h^{-1}\left\Vert
x-X\right\Vert \right) $ for general probability measures and establishing
lower and upper bounds (derivations in the supplementary material, Appendix B). The
bounds provide expressions in terms of the small cube probability:%
\begin{equation}
L_{EgK^{m}}\left( x\right) {P_{X}(C(x,h))}\leq \left\vert E\left[
g(X)K^{m}\left( h^{-1}\left\Vert x-X\right\Vert \right) \right] \right\vert
\leq M_{EgK^{m}}\left( x\right) {P_{X}(C(x,h))}  \label{EK^mg}
\end{equation}%
with constants $L_{EgK^{m}}\left( x\right) $ and $M_{EgK^{m}}\left( x\right) 
$ at $x.$ Most important, (\ref{EK^mg}) provides a lower bound on $%
EB_{n}\left( x\right) =EK\left( K^{m}\left( h^{-1}\left\Vert x-X\right\Vert
\right) \right) $, given by $L_{EK}P_{X}\left( C\left( x,h\right) \right) $,
with appropriate conditions for $L_{EK}$ to be strictly positive to ensure
that the denominator of the NW estimator is such that it exists and the
limit does not blow up. Type I kernel automatically entails that $L_{EK}>0$,
but for kernels such as Epanechnikov the bound requires Assumption {\ref%
{A.dist}. }With $g\left( X\right) $ that is continuous at $x$%
\begin{equation*}
E\left[ g(X)K^{m}\left( h^{-1}\left\Vert x-X\right\Vert \right) \right]
=g\left( x\right) E\left[ K^{m}\left( h^{-1}\left\Vert x-X\right\Vert
\right) \right] \left( 1+o\left( 1\right) \right) .
\end{equation*}%
These moment expressions for distributions over $\mathbb{R}^{q}$ hold under
the standard assumptions of existence and continuity of (bounded) density $%
f_X,$ and the function $g$, {where} 
\begin{equation}
E\left[ g(X)K^{m}\left( h^{-1}\left\Vert x-X\right\Vert \right) \right]
=\prod\limits_{i=1}^{q}\left( -h^{i}\right) g\left( x\right) f_X\left(
x\right) \int K^{m}\left( v\right) dv\left( 1+o\left( 1\right) \right) ,
\label{EgfK}
\end{equation}%
with more details about the $o\left( 1\right) $ term under smoothness of $f_X
$ (see, e.g. derivations in Li and Racine, 2007). Once the moments and the
bounds are derived, the proofs of asymptotic normality proceed along similar
lines to those in Masry (2005).

The point-wise limit normality is provided in the next theorem under two
alternative types of conditions: (i) with type I kernel without imposing
further constraints on $F_X$, and (ii) not imposing the type I kernel but
with the distributional Assumption \ref{A.dist}. Denote the bias of the
estimator given $x$, $E\left( \widehat{m}\left( x\right) \right) -m\left(
x\right) ,$ by $bias\left( \widehat{m}\left( x\right) \right) .$ The
difference $\widehat{m}\left( x\right) -m\left( x\right) $ is delivered by $%
\frac{A_{n}^{c}(x)}{B_{n}(x)}$ with the \textquotedblleft
centered\textquotedblright\ $A_{n}^{c}(x)=A_{n}(x)-m\left( x\right)
B_{n}(x). $

\setcounter{theorem}{0}

\begin{theorem}
\label{T.1} {Under either of the following sets of assumptions (i)
Assumptions \ref{A.measure on prod}-\ref{A.mx} and \ref{A.Jointmeasure} or
(ii) Assumptions \ref{A.measure on prod}, \ref{A.kernel}(a-c), \ref{A.mom}-%
\ref{A.Jointmeasure} } for $h\rightarrow 0$ as $n\rightarrow \infty $ such
that $nP_{X}\left( C\left( x,h\right) \right) \rightarrow \infty $

\begin{itemize}
\item[(a)] 
\begin{equation*}
\frac{\sqrt{n}E\left[ K\left( h^{-1}\left\Vert x-X\right\Vert \right)\right] 
}{\sqrt{\mu _{2}\left( x\right) E\left[ K^{2}\left( h^{-1}\left\Vert
x-X\right\Vert\right) \right] }}\left( \widehat{m}\left( x\right) -m\left(
x\right)-bias(\widehat{m}(x)\right) \rightarrow _{d}Z\sim N\left( 0,1\right)
;
\end{equation*}

\item[(b)] the rates are\ 
\begin{eqnarray*}
&& bias(\widehat{m}(x) =O(\bar{h}^{\delta })+O\left( nP_{X}\left(
C(x,h)\right) \right) ^{-1}; \\
&& \frac{\sqrt{n}E \left[ K\left( h^{-1}\left\Vert x-X\right\Vert \right)%
\right] }{ \sqrt{\mu_{2}\left( x\right) E\left[ K^{2}\left( h^{-1}\left\Vert
x-X\right\Vert\right)\right] }} \simeq O\left( (nP_{X}\left( C\left(
x,h\right)\right) ^{1/2}\right) .
\end{eqnarray*}

\item[(c)] for $h$ such that $\bar{h}^{2\delta }\left( nP_{X}\left(
C(x,h)\right) \right) \rightarrow 0$ 
\begin{equation*}
\frac{\sqrt{n}E\left[K\left( h^{-1}\left\Vert x-X\right\Vert \right)\right] 
}{\sqrt{\mu _{2}\left( x\right) E\left[ K^{2}\left( h^{-1}\left\Vert
x-X\right\Vert \right)\right] }}\left( \widehat{m}\left( x\right) -m\left(
x\right) \right) \rightarrow _{d}Z\sim N\left( 0,1\right) .
\end{equation*}
\end{itemize}
\end{theorem}

\noindent \textbf{Remarks.}

\begin{enumerate}
\item A sequence of bandwidths at $x$ that satisfy the conditions of the
theorem always exists. Indeed, whatever the rate of monotonic decline in $%
P_{X}(C(x,h))$ as $h\rightarrow 0$ for $n\rightarrow \infty $ a sequence of $%
h$ that depends on $n$ such that $nP_{X}\left( C\left( x,h\right) \right)
\rightarrow \infty $ always exists. T{he rate for the bias of $\widehat{m}%
\left( x\right) $ in $\Xi ^{\left[ q\right] }$ is established in the theorem
as $O\left( \bar{h}^{\delta }\right) +O\left( \left( nP_{X}\left(
C(x,h)\right) \right) ^{-1}\right) .$ For the bias (squared) to disappear in
the limit }$\bar{h}^{2\delta }P_{X}\left( C(x,h)\right) n$ needs to go to
zero. If $P_{X}\left( C\left( x,h\right) \right) \rightarrow 0$ a bandwidth
sequence that simultaneously satisfies $nP_{X}\left( C\left( x,h\right)
\right) \rightarrow \infty $ and $\bar{h}^{2\delta }P_{X}\left(
C(x,h)\right) n\rightarrow 0$ can always be found; when $x$ is a mass point $%
P_{X}\left( C\left( x,h\right) \right) $ will be bounded from below, but
selecting $h=o\left( n^{-1/2\delta }\right) $ for such a point makes the
bias term go to zero.

\item The assumptions of Theorem \ref{T.1} and the moment computations in
the supplementary material (Appendix B) imply that $E\left[ K\left( h^{-1}\left\Vert
x-X\right\Vert \right)\right]$ has the same rate as $P_{X}(C(x,h))$ while $%
varA_{n}^{c}(x)$ declines at the rate ${P_{X}\left( C\left( x,h\right)
\right) }/{n}.$ The rate for the asymptotic variance for $\widehat{m}\left(
x\right) $ equals $\left( nP_{X}\left( C\left( x,h\right) \right) \right)
^{-1}$ (this goes to zero).

\item The limit result shows that when density exists for a distribution on $%
\mathbb{R}^{q},$ the standard convergence rate $n^{1/2}h^{q/2}$ applies
since then $P_{X}\left( C\left( x,h\right) \right) =O\left( h^{q}\right) .$
This rate holds even when the density is discontinuous. Without the usual
smoothness assumptions made in the literature, statistical guarantees for
the rate\textbf{\ }and for asymptotic normality are thus shown to hold.

\item If there is singularity at the point $x$ that satisfies (\ref{LM F
bounds}) with $s<1,$ then the rate is $n^{1/2}h^{sq/2}$, which is faster
than in the absolutely continuous case $\left(
n^{1/2}h^{sq/2}>n^{1/2}h^{q/2}\right) $, mitigating somewhat the
\textquotedblleft curse of dimensionality\textquotedblright . When $x$ is an
isolated mass point then at that point the parametric rate $n^{1/2}$ holds.

\item Under continuous differentiability the rate of the bias can be reduced
by employing a local linear estimator (see, e.g. the standard derivations in
Li and Racine, 2007, and for univariate functional regression in Ferraty and
Nagy, 2022). Establishing the distributional properties of the local linear
estimator with arbitrary probability distributions in $\mathbb{R}^{q}$ and
multivariate probability measures in a metric space can proceed similarly,
but requires stronger assumptions.
\end{enumerate}

The convergence rate in (c) of Theorem \ref{T.1} is $O\left( (nP_{X}\left(
C\left( x,h\right) \right) ^{-1/2}\right).$\footnote{%
This convergence rate obtains under $h\rightarrow 0.$ In the presence of an
irrelevant regressor, say $x^{(2)}$, such that $m\left( x\right) =m\left(
x^{\left( 1\right) }\right) $ for all $x=\left( x^{\left( 1\right)
},x^{\left( 2\right) }\right) ,$ this requirement can be restricted to the
function $m\left( x^{\left( 1\right) }\right) $ with the irrelevant $%
x^{\left( 2\right) }$ eliminated. For the estimator this elimination can be
achieved by setting the bandwidth on components of $x^{\left( 2\right) }$ to
be larger than the range of those variables, possibly infinite.} Existence
of a limit variance $\sigma _{\widehat{m}\left( x\right) }^{2}$ requires
that $\left( nP_{X}\left( C\left( x,h\right) \right) \right)\frac{\left[%
EK\left( h^{-1}\left\Vert x-X\right\Vert \right) \right]^2}{{\mu _{2}\left(
x\right) E\left[ K^{2}\left( h^{-1}\left\Vert x-X\right\Vert\right)\right] }}
$ converges. Without additional assumptions it is possible that the ratio
does not converge; see example in the supplementary material (Appendix B) that provides a
case when convergence does not hold; this happens when the small cube
probability declines very rapidly and the kernel is not uniform. Suitable
additional assumptions on the distribution, such as $H_{3}$ in Ferraty et
al. (2007) and Condition 3(i) in Masry (2005) and similar ones in subsequent
papers provide restrictions on the probability measure on $\Xi ^{\left[ 1%
\right] }$ that are sufficient for the convergence. Generally, one needs to
ensure that the limits given below on the expectation of the kernel function
and its square hold.\footnote{%
This implies that the extra condition is also required for the Corollary 1
of Hong and Linton (2020).}

\begin{assumption}
\label{A.lim} As $n\rightarrow \infty ,$ $h\rightarrow 0$ 
\begin{equation*}
\left( P_{X}(C(x,h))\right) ^{-1}E\left[K^{s}\left( h^{-1}\left\Vert
x-X\right\Vert \right) \right] \rightarrow \bar{B}_{s}\left( x\right) ;s=1,2.
\end{equation*}
\end{assumption}

This assumption holds quite widely. From the moment expressions it can
easily be shown that it holds for the uniform kernel without any additional
distributional assumptions. In the case of continuous density it holds by
virtue of (\ref{EgfK}) with 
\begin{equation}
\bar{B}_{1}\left( x\right) =f_X\left( x\right) \int K\left( v\right)
dv,\quad \bar{B}_{2}\left( x\right) =f_X\left( x\right) \int K^{2}\left(
v\right) dv.  \label{lim for a a c}
\end{equation}%
{Suppose that singularity arises, because of combining discrete and
continuous variables in $\mathbb{R}^{q}$ or functional dependence between
the regressors, that restrict the support of the distribution to be in some
subspace of dimension $r<q$, $V\left(r\right) \subset $ $\mathbb{R}^{q}.$}
If the distribution on $V\left( r\right) $ is absolutely continuous with a
continuous density, then derivations provide similar limits to (\ref{lim for
a a c}) with integration over $V\left( r\right) $ and density restricted to $%
V\left( r\right) .$

Define now 
\begin{equation*}
\alpha \left( n,h\right) ={nP_{X}\left( C\left( x,h\right) \right) };\quad
\sigma _{\widehat{m}\left( x\right) }^{2}=\mu _{2}(x)\bar{B}_{2}(x)/(\bar{B}%
_{1}(x))^{2}\ .
\end{equation*}

\begin{theorem}
\label{T.2} Under the conditions of Theorem \ref{T.1} and Assumption \ref%
{A.lim} with $\alpha \left( n,h\right) \rightarrow \infty $ and for $h$ such
that $\alpha \left( n,h\right) \bar{h}^{2\delta}\rightarrow 0$

\begin{equation*}
\sqrt{\alpha \left( n,h\right) }\left( \widehat{m}\left( x\right)
-m\left(x\right) \right) \rightarrow _{d}N\left( 0,\sigma _{\widehat{m}%
\left( x\right) }^{2}\right) .
\end{equation*}
\end{theorem}

This limit extends the results that were obtained in the literature on
kernel estimation in $\mathbb{R}^{q}$ under smoothness assumptions on the
distribution $F_{X}.$ {For functional regression our assumptions are
comparable to those of Ferraty et al. (2007), Masry (2005), and subsequent
papers while they make the extension to multivariate functional
regression possible. }

\section{\protect\bigskip \textbf{Implementation and bandwidth selection}}

Estimation of $m\left( x\right) $ requires a selection of the kernel, $K,$
and bandwidth, $h.$ As may be clear from the results here and the
literature, type I kernel (such as the uniform) is preferred but other
kernels can also deliver asymptotic rates provided the small cube
probability does not decline exponentially fast. Aside from the estimator of
the conditional mean, estimators of variance and mean squared error are
needed to evaluate the performance of the estimator. While in the literature
on kernel regression on $\mathbb{R}^{q},$ the leading term of the limit
variance is expressed via the density function, often in the actual
implementation the corresponding estimators do not make use of plug-in
expressions, instead estimating the variance directly from the data and
possibly with bootstrap (see Hall and Horowitz, 2013).

Cross-validation procedures in popular statistical packages (such as R)
provide a single bandwidth (vector) that was shown to be consistent for the
\textquotedblleft optimal\textquotedblright\ bandwidth: minimizer of
weighted integrated mean squared error, WIMSE, (e.g. Li and Racine, 2007). The
proofs of consistency relied on absolute continuity of the regressors. The
consistency results extend to some classes of singular distributions.

WIMSE is defined for an absolutely continuous distribution with density
function $f_X\left( x\right) $ as%
\begin{equation*}
\int E\left( \widehat{m}\left( x\right) -m\left( x\right) \right)
^{2}M\left( x\right) f_X\left( x\right) dx
\end{equation*}%
with some weighting function $M\left( x\right) $ chosen to mitigate boundary
effects. The expression can be written with $dF_X$ replacing $f_X\left(
x\right) dx$ (valid in the case of singularity): 
\begin{equation}
\int E\left( \widehat{m}\left( x\right) -m\left( x\right) \right)
^{2}M\left( x\right) dF_X=\int \left[ var\left( \widehat{m}\left(
x\right)\right) +bias^{2}\left( \widehat{m}\left( x\right) \right) \right]
M\left( x\right) dF_X.  \label{WIMSE}
\end{equation}%
This function depends on the bandwidth vector $h$ used in the estimator (see
the review of bandwidth selection methods, including cross-validation and
plug-in in K\"{o}hler et al., 2014). The \textquotedblleft
optimal\textquotedblright\ bandwidth vector $h^{0}$ is a minimizer of the
WIMSE criterion function based on a trade-off between the variance and bias
of the NW estimator.

In the cross-validation procedure the finite sample analogue of WIMSE
replaces the expectation by 
\begin{equation*}
CV=n^{-1}\sum_{i=1}^{n}\left( Y_{i}-\widehat{m}_{-i}\left( X_{i}\right)
\right) ^{2}M\left( X_{i}\right)
\end{equation*}%
employing the leave-one-out kernel estimator, $\widehat{m}_{-i},$ and
provides the bandwidth vector $h_{cv}$ by minimizing the CV criterion.

Hall et al. (2007) gave a general result about consistency of the
cross-validated bandwidth for regression over $\mathbb{R}^{q}$ with discrete
and continuous regressors, with some of the regressors possibly being
irrelevant. Their general result in Theorem 2.1 was obtained under a set of
assumptions that required independent identically distributed observations,
restrictions on the support of the probability measure, two continuous
derivatives for density, the regression function, and the conditional
variance of the error; in addition, for the $d$ continuous relevant
regressors $h^{o}=n^{-\frac{1}{4+rd}}a^{o}$ holds with the vector $a^{o}$
having unique, positive and finite components. This result was extended to
weakly dependent data by Li et al. (2009) under assumptions that replaced
the i.i.d. assumption by requiring strict stationarity and $\beta -$mixing
in the process for $\left\{ x,y\right\} $ and martingale difference error,
with suitable restrictions on the mixing parameters.

The result on the cross-validated bandwidth applies more widely. For instance, consider a singular distribution of $X\in \mathbb{R}^{q}\,$\ where there is a functional dependence among the continuous variables in the presence of possibly some discrete covariates such that the support of the distribution is restricted to a subspace $V\left( r\right) \subset \mathbb{R}^{q}$ of dimension $r<q$ represented by a union of affine subspaces. If, restricted to $V\left(
r\right) ,$ the distribution function is such that the conditions of Theorem
2.1 of Hall et al. (2007) or Theorem 1 of Li et al. (2009) are satisfied
(Assumption CV) then the conclusions of those theorems are valid and the
consistency of the bandwidth and automatic dimension reduction by smoothing
out irrelevant regressors hold for this singular distribution. More details
are provided in the supplementary material (Appendix B).

Importantly, no knowledge of $V\left( r\right) $ or $r$ is required. This
implies that for functionally dependent continuous regressors the knowledge
of the number of factors is not required for the consistency of the
cross-validated bandwidth or the automatic dimension reduction. We
conjecture that in many other cases with possible singularity the
cross-validation procedure will facilitate dimension reduction by smoothing
out irrelevant variables.

Bandwidth selection could benefit from adaptation to different types of
singularity. The treatment of adaptive bandwidth selection in the literature
(Fan and Gijbels, 1996, Sain, 1994, Demir et al., 2010) typically focuses on
adjusting the smoothing parameter to accommodate the varying data density,
but not dealing with singularity or mass points. Adaptive bandwidths can
provide a better fit of the criterion function by increasing the number of
observations used to estimate the function at a point of sparsity.\footnote{%
Given some initial bandwidth $\tilde{h}$ and density estimate at this
bandwidth, $\hat{f}_{X},$ an adaptive bandwidth is defined for each point as 
$h\left( X_{i}\right) =\tilde{h}\left( \frac{\hat{f}_{X}\left( X_{i}\right) 
}{G}\right) ^{-\alpha }$ where $G=\left( \prod \hat{f}_{X}\left(
X_{j}\right) \right) ^{1/n}$ is the geometric mean of the densities and $%
\alpha $ is typically selected to be $1/2.$ {One could construct $\hat{f}%
_{X}(x)$ $\ $\ with a uniform kernel in which case it is identical to an
estimate of $P(C(x,\tilde{h}))$ by the proportion of observations in the $%
\tilde{h}$ cuboid around $x.$}} Such bandwidths can similarly be constructed
for cases of singular distributions. But these adaptation procedures still
need to be investigated in the case of general mixtures of singular
distributions. However, singularity adaptation simplifies considerably for
the empirically important case of a mixture of an absolutely continuous
distribution with mass points, where the two levels of singularity can be
separated. The approach is detailed in the supplementary material (Appendix B).

\section{Simulations}

This section provides the highlights of various simulations that show
features of the finite sample performance of the NW estimator under
singularity. Additional details and features are in the supplementary
material (Appendix C).

\subsection{Univariate (Point mass example)}

In this example we consider the regression distribution with mass points.
Alongside we examine the trinormal mixture considered in Kotlyarova et al.
(2016), an a.c. distribution which represents features (high density
derivatives) that makes it comparable to a singular distribution.

The distribution with mass points, following Jun and Song (2019), is given
by 
\begin{equation*}
F_{X}(x)=pF^{d}(x)+(1-p)\Phi (x)\quad \text{with }p=0.2,
\end{equation*}%
where $F^{d}$ is the discrete uniform distribution function with $%
D=\{-1,0,1\}$ the set of mass points; $\Phi $ is the standard Gaussian
distribution function.

We simulated 500 random samples $\{(Y_{i},X_{i})\}_{i=1}^{n}$ using the
model 
\begin{equation*}
Y_i=\sin (2.5X_i)+\sigma \varepsilon_i ,
\end{equation*}%
for different sample sizes. The error $\{\varepsilon _{i}\}_{i=1}^{n}$ is
drawn independently of the regressor and has a standard Gaussian
distribution; $\sigma $ is selected to yield a given signal to noise ratio, $%
snr$, here selected to equal one. We use the Epanechnikov kernel $K(u)=\frac{%
3}{4}(1-{u^{2}})1(u^{2}\leq 1)$ and obtain the leave-one-out cross-validated
bandwidth.

We analyze the pointwise RMSE at a coarse grid of points across samples of
size $n$ equal to 50, 100, 200, 400, 800, 1600, 3200 based on 500
replications from the above DGP. To obtain empirical rates of convergence we
regress $\log (RMSE)$ on $\log (n)$ and a constant. The coefficient on $\log
(n)$ is the \textquotedblleft realized\textquotedblright\ rate of
convergence; for example if $RMSE\propto n^{-2/5}$ (univariate kernel
regression with smooth density and second order kernel) then $\log
(RMSE)=\alpha _{0}+\alpha _{1}\log (n)$ and $\alpha _{1}$ should be close to
-0.4.\footnote{%
The authors thank Jeff Racine for suggesting this insightful exercise. See
also Hall and Racine (2015).}

In Table 1, illustrative results are provided for the regressor distribution
with mass points and the trinormal distribution on a set of support points. 
\begin{table}[t]
\caption{Empirical rate of convergence (i.e., $-\protect\alpha_1$ for $O(n^{-%
\protect\alpha_1})$) in the mass point and high derivative setting.}%
\begin{tabular}{p{.5cm}p{1.25cm}p{1.25cm}p{1.25cm}p{1.25cm}}
\multicolumn{5}{c}{$F_X(x)=0.2F^d(x)+0.8\Phi(x)$} \\ 
\cmidrule(lr){1-5} & X & NW & NW$_a$ &  \\ 
& 0.00 & -0.455 & -0.515 &  \\ 
& 0.10 & -0.186 & -0.443 &  \\ 
& 0.20 & -0.411 & -0.438 &  \\ 
& 0.30 & -0.465 & -0.431 &  \\ 
& 0.40 & -0.461 & -0.425 &  \\ 
&  &  &  & 
\end{tabular}
\begin{tabular}{lp{1.25cm}p{1.25cm}}
\multicolumn{3}{c}{$F_X(x)=$ trinormal $(x)$} \\ 
\cmidrule(lr){1-3} & X & NW \\ 
& 0.00 & -0.449 \\ 
& 0.50 & -0.381 \\ 
& 0.75 & -0.416 \\ 
& 1.00 & -0.413 \\ 
\  &  &  \\ 
&  & 
\end{tabular}
\newline
\begin{minipage}{1.0\textwidth}{Note: The column labeled NW$_a$ contains the results implementing the adaptive bandwidth selection procedure in the presence of masspoints.}
\end{minipage}
\end{table}

For the distribution with mass points, the NW estimator with cross-validated
bandwidth performs remarkably well at points sufficiently far from our mass
points (faster than the expected rate of -0.4). The empirical rate at mass
points is close to -0.5 when the bandwidth is set equal to zero. The
empirical convergence rate is slow for points close to the mass points
(within the small ball probability measure under cross validated bandwidth)
due to the boundary weight associated with mass in the neighborhood. {%
Bandwidth adaptive to masspoints improves the rate.} The convergence rates
for the trinormal distribution, are reflective of usual smooth nonparametric
regression although are somewhat faster at points with high derivatives.

\subsection{Bivariate (with effective dimension 1)}

We consider a model where $m(X)=\log (X_{1})+\log (X_{2})$ with regressors $%
X_{1}$ and $X_{2}$ satisfying $X_{1}+X_{2}=d(k),$ with fixed $d(k)$
corresponding to $k=1,2,3$.\footnote{%
An example could be where $X_{1}$ and $X_{2}$ represent earnings of the
husband and wife and, for tax purposes, their combined income is set at some 
$d(k)$.} This is equivalent to a model with one continuous and one discrete
regressor $m(X)=\log (X_{1})+\log (D-X_{1}),$ with $D=d(k)$.

We simulated 500 random samples $\{(Y_{i},X_{1i},X_{2i}\}_{i=1}^{n}$ using
the model 
\begin{equation*}
Y_i=\log (X_{1i})+\log (X_{2i})+\sigma \varepsilon_i
\end{equation*}%
for different sample sizes with the additive error chosen as in the previous
simulation. The probability of an observation belonging to a sub-population
with $k=1,2,3$ is set equal to $0.5$, $0.3$, and $0.2$ respectively and $%
d(1)=4,d(2)=6,d(3)=7$; $X_{1}$ is drawn from the uniform distribution: $%
U[1,3]$.

We implement the NW estimator first using $X_{1}$ and $X_{2}$ as regressors
(NW.c) and second using $X_{1}$ and $D$ as regressors (NW.d) and obtain the leave-one-out cross-validated bandwidths. For the
discrete regressor $D$ we use special discrete kernel weights proposed by
Wang and van Ryzin (1981) in accordance with Racine and Li (2004).\ 

In Table 2 we provide illustrative results comparing the empirical rate of
convergence of the NW.c and NW.d at a grid of points. 
\begin{table}[t]
\caption{Empirical Rates of the NW.c and NW.d estimators.}%
\begin{tabular}{llllp{1.25cm}p{1.25cm}p{1.25cm}}
\  &  &  &  &  &  &  \\ 
& $X_1$ & $X_2$ & $d(k)$ & \multicolumn{1}{c}{NW.c} & \multicolumn{2}{c}{NW.d
} \\ 
&  &  &  & $(X_1,X_2)$ & \multicolumn{2}{c}{$(X_1,d(k))$} \\ 
\cmidrule(lr){6-7} &  &  &  &  & ordered & unordered \\ 
& 1.5 & 2.5 & 4 & -0.451 & -0.422 & -0.419 \\ 
& 2.0 & 2.0 & 4 & -0.436 & -0.429 & -0.426 \\ 
& 2.5 & 1.5 & 4 & -0.429 & -0.406 & -0.404 \\ 
& 1.5 & 4.5 & 6 & -0.457 & -0.410 & -0.422 \\ 
& 1.5 & 5.5 & 7 & -0.445 & -0.392 & -0.410 \\ 
\  &  &  &  &  &  &  \\ 
&  &  &  &  &  & 
\end{tabular}%
\newline
\begin{minipage}{1.0\textwidth}{Note: The column labeled ``ordered'' contains the NW.d estimator where the discrete kernel is used for the discrete regressor; the column labeled ``unordered'' uses the Epanechnikov kernel.}
\end{minipage}
\end{table}
The reduced dimensionality is reflected in the estimates of the pointwise
rate of convergence which are around $-0.40$ rather than the slower rate of $%
-0.33$ the presence of two continuous regressors would suggest ($q=2$). The
estimate of the empirical rate for NW.c is slightly faster than NW.d,
moreover, indicating that there is no gain from separate treatment of
discrete regressors. With the reduced dimension structure here therefore one
gets the rate corresponding to the Hausdorf dimension of the regressor space
automatically without the need to recognize that it is possible to transform
the regressors to one discrete, and one continuous variable.

\subsection{Bivariate (in the presence of a functional regressor)}

Here we examine a functional regressor in a multivariate setting. Consider a
bivariate conditional mean function $m(X)=m(X_{1},X_{2})$, where $X_{1}$ is
a functional regressor and $X_{2}\in \mathbb{R}$ \ may be correlated with
some $m_{1}(X_{1})$. Let 
\begin{equation*}
Y_i=m_{1}(X_{1i})+X_{2i}+\sigma \varepsilon_i.
\end{equation*}%
Following Ferraty et al. (2007), the functional regressor is defined as 
\begin{equation*}
X_{1i}(t)=\sin (w_it)+(a_i+2\pi )t+b_i,\quad t\in (-1,1)
\end{equation*}%
with $a_i$ and $b_i$ drawn from $U(-1,1)$, $w_i$ drawn from $U(-\pi ,\pi )$
and 
\begin{equation*}
m_{1}(X_{1i})=\int_{-1}^{1}|X_{1i}^{\prime }(t)|(1-\cos (\pi t))dt.
\end{equation*}

For $X_{2}$ we consider two possibilities: (a) a N(0,1) random variable
independent of $X_{1}$; (b) $X_{2}=m_{1}(Z)$ where $Z(t)$ is a functional
regressor similar to $X_{1}(t)$ with $(a_i,b_i,w_i)$ replaced by $%
(a_i^{\prime },b_i^{\prime },w_i^{\prime })$ where the correlation between $%
(a_i^{\prime },b_i^{\prime },w_i^{\prime })$ and $(a_i,b_i,w_i)$ is given by 
$\rho $ (and set equal to either $0$ or $0.8$).

For the functional regressor $X_{1}$ we use the same metric as in Ferraty et
al. (2007), that is $\left\Vert x_{1}-X_{1}\right\Vert _{1}=\sqrt{%
\int_{-1}^{1}\left( x_{1}^{\prime }(t)-X_{1}^{\prime }(t)\right) ^{2}dt}.$
We use a product kernel with kernel $K(u)=1-u^{2}$ defined on $[0,1]$ for
the functional regressor and the Epanechnikov kernel defined on $[-1,1]$ for 
$X_{2}$.

Table 3 shows RMSE of the NW estimator at the cross-validated bandwidths as
well as RMSE where either the functional or scalar regressor is dropped. The
loss from misspecifying the functional regression as univariate can be
substantial. 
\begin{table}[t]
\caption{RMSE of the NW estimator in the presence of functional regressor $%
X_1$ at cross validated bandwidth, $n=250$.}%
\begin{tabular}{lccc}
\  &  &  &  \\ 
& \multicolumn{1}{c}{$X_2 = N(0,1)$} & \multicolumn{2}{c}{$X_2 = m_1(Z)$} \\ 
\cmidrule(lr){3-4} &  & $\rho =0.0$ & $\rho=0.8$ \\ 
\textbf{In-sample} &  &  &  \\ 
\  &  &  &  \\ 
RMSE & 0.746 & 0.915 & 1.058 \\ 
\  &  &  &  \\ 
\quad \textbf{Misspecification:} &  &  &  \\ 
\quad RMSE$_1$ & 1.140 & 1.918 & 2.099 \\ 
\quad RMSE$_2$ & 1.833 & 1.854 & 1.746 \\ 
\  &  &  &  \\ 
\textbf{Out-of-sample} &  &  &  \\ 
\  &  &  &  \\ 
RMSE & 0.915(4) & 1.026(19) & 1.210(18) \\ 
\  &  &  &  \\ 
&  &  & 
\end{tabular}
\begin{minipage}{1.0\textwidth}{Note: RMSE$_1$ stands for the RMSE where the $X_2$ regressor is excluded and RMSE$_2$ stands for the RMSE when ignoring the functional regressor. The number in brackets indicates the number of simulations (out of 500) where at the cross-validation bandwidth no neighbor to the out-of-sample observation exists.}
\end{minipage}
\end{table}

\section{Empirical study}

The causal inference literature has made extensive use of the LaLonde (1986)
data on the National Supported Work Demonstration (NSW) program following
the release of that data by Dehejia and Wahba (1999, 2002). Their finding,
that propensity score-based methods provide a way to generalize the
experimental results on the impact of training to nonexperimental data, was
influential and led to significant methodological advances and practical
changes as discussed in the review by Imbens and Xu (2024). Here, we
consider the experimental sample to analyze potential heterogeneous
treatment effects using multivariate kernel estimation. Kernel-based
matching on individual characteristics, advocated in Heckman et al. (1997,
1998), was not considered due to the claimed high dimensionality of the
regressors. We show that it is both feasible and insightful for this data
due to the dimension reduction implied by the presence of several discrete,
discretized and categorical regressors. The mass point of the regressor on
pre-treatment earnings at zero further contributes to the regressor
singularity and we do not require continuity or indeed existence of density
over positive values, thus kinks or mass at positive values are not
excluded. The kernel estimator is applicable to such singular distributions.

We focus here on the full LaLonde NSW male sample which contains 297 treated
individuals and 425 controls where the pre-intervention variables are
well-matched.\footnote{%
In the sub-sample with 1974 earnings data in Dehejia and Wahba (1999) the
distribution of 1975 earnings exhibits a significantly different mass at
zero between the treated (68\%) and untreated (60\%).}

With $Y$ denoting the post-treatment outcome, $T$ the treatment and $X$ the
individual pre-treatment characteristic(s), we use the nonparametric
regression model 
\begin{equation*}
m(x,j)=E(Y|X=x,T=j)\text{ for }j=0,1 
\end{equation*}%
to evaluate the heterogeneous effects of the treatment as 
\begin{eqnarray*}
\tau (x)&=& m(x,1)-m(x,0).
\end{eqnarray*}
The heterogeneous effect of treatment on the treated, also known as the
conditional average treatment effect, CATT, is given by 
\begin{eqnarray*}
\tau_T (x)&=& E(m(x,1)-m(x,0) |T=1)
\end{eqnarray*}
Focusing on the latter, we use the NW estimates to evaluate 
\begin{equation*}
\hat{\tau}_T(x_i)= \widehat{m}(x_i,1) - \widehat{m}(x_i,0) \quad
i=1,\cdots,n_T 
\end{equation*}
for all treated individuals $n_T$ (i.e., we use both the actual and the
counterfactual treatment for our estimates).

First, we consider a bivariate kernel regression model where we only use the
pre-treatment earnings (re75) as regressor $X$. Following that, we estimate
the multivariate model with the full set of variables $X$, where in addition
to the pre-treatment earnings we include years of education, high school
``no degree'' status, race, age, marital status, and pre-treatment
unemployment status, u75. It is not unreasonable to attempt nonparametric
estimation for this problem where the only truly continuous regressor is
earnings (and possibly age and education) as singularity provides dimension
reduction.

As with our simulations, we use the np package in R for the nonparametric
estimation where we consider the Epanechnikov (e), Uniform (u) and discrete
(d) kernel.\footnote{%
We use the discrete kernel proposed by Aitchison and Aitkin, 1976, where $%
K((d-d_{i})/h)=1-h$ if $d=d_{i}$, else $h$ where $h\in \lbrack 0,1/2]$.}
Bandwidth selection is based on cross validation and we consider the
adaptive bandwidth selection approach that accounts for the masspoint. As
was shown in our simulations the rate improvement associated with
singularities does not require special attention to discrete variables to
benefit from it.

Estimation results are reported in detail in the supplementary material
(Appendix D). Below the main findings are summarized.

For the bivariate regression model, the cross validated bandwidths confirm
that we should not smooth across treated and untreated observations and that
local heterogeneous treatment effects as related to pre-treatment earnings
are present. Figure 1, displays estimates of the conditional expectation
using the Epanechnikov kernel by treatment status and pre-treatment earnings
together with the bootstrapped confidence bounds. It suggests that treatment
for individuals at low levels of pre-treatment earnings, in particular, is
beneficial. 
\begin{figure}[H]
\caption{Nonparametric fit of the conditional expectation by pre-treatment
earnings and treatment status (cross validated bandwidth, Epanechnikov
kernel)}\includegraphics[scale=.7]{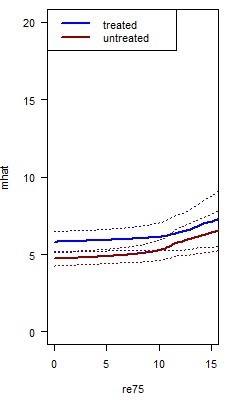}\newline
\begin{minipage}{1\textwidth}{Note: All graphs related to the empirical application are rescaled with all numbers denoted in '000\$s. }
\end{minipage}
\end{figure}

The adaptive bandwidth results in a slightly better in-sample correlation
between the post-treatment outcome, $re78$, and its fit (increasing from
0.2098 to 0.2116 (for OLS the correlation is 0.1697)); bandwidths obtained
using non-masspoint-observations only are quite similar to those obtained
when including the masspoints in this case. The NW estimates with the
adaptive bandwidth provide values of CATT that on average equal $\$920$
(76), \$920 (76), and \$906 (80) (standard error in brackets) for the (e,e), (d,e), (d,u) kernels on $(T,X)$,
respectively.\footnote{%
As discussed in the supplemental material (Appendix D), we denote the kernel with two
arguments: the first argument denotes the kernel applied to all binary
regressors (treat, u75, nodegree, black, hispanic, and married) and the
second argument denotes the kernel applied to the other regressors (re75,
educ, and age).} For comparison, the local linear kernel based estimates on
average equal \$822 (49) with the (e,e) kernel, while the average of the
CATT estimates based on random forest (RF) equal \$848 (52). The CATT
results of the kernel regression based approach are more variable than those
obtained using the random forest approach. For observations at mass points,
CATT estimates using adaptive bandwidth are closer to those obtained using
the random forest based approach.

For the multivariate model the cross-validated bandwidths provide important
insights. Firstly, even though pre-treatment earnings is still relevant, the
bandwidth is much larger than in the baseline model for all kernels,
suggesting a reduction of the heterogeneous impact with individual's
pre-treatment earnings. The bandwidths selected for nodegree, hispanic and
married are large, signaling that these variables are not relevant (these
regressors are automatically smoothed out from the regression function). At
the same time, the bandwidths for education and age imply a heterogeneous
impact associated with those characteristics, although the size of the
bandwidth for age is fairly large.

The inclusion of additional controls yields an improvement in the in-sample
correlation between the post-treatment outcome and its fit. For the (e,e)
kernel we see an increase in correlation from 0.210 in the bivariate model
to 0.338 (for comparison, for OLS the correlation equals 0.209 when age
squared is included as well); the results for the (d,e) and (d,u) kernel are
comparable. 

To highlight the heterogeneity of the treatment effect of education and its
interplay with race, we display in Figure 2 estimates of the conditional
expectation by treatment status, years of education, and race for an
individual with median age and pre-treatment earnings together with the
bootstrapped confidence bounds. 
\begin{figure}[H]
\caption{Nonparametric fit of the conditional expectation by years of
education, race, and treatment status with median pre-treatment earnings and
age) (cross validated bandwidth, Epanechnikov kernel)}
\medskip \text{\hspace{.61in} Black =0 (N=144) \hspace{.9in} Black=1 (N=579)}%
\newline
\includegraphics[scale=.9]{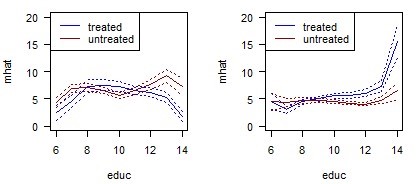}\newline
\begin{minipage}{1\textwidth}{Note: The median pre-treatment earnings equals $\$936$ and the median age is $23$. The estimates are rescaled and are denoted in '000\$. }
\end{minipage}
\end{figure}
The graph reflects a heterogeneity of the impact of treatment whereby the
more educated individuals identified as black appear to benefit more from
treatment than their nonblack counterparts. Gains of treatment arise where
the confidence band around the estimated nonparametric fit $\widehat{m}(x,1)$
lies above that of $\widehat{m}(x,0)$; for non-black individuals this is at
the middle range of education, for black individuals this starts around 10
years of education and is rising over that range. These results are further
supported when evaluating the average CATT for black individuals across
different levels of education (see supplemental material, Appendix D).

Box-plots of the CATT estimates for the multivariate model using the NW
regression estimate and the RF estimates are presented in Figure 3. The
limit distributional results of Wager and Athey (2018) do not apply here as
many components of $X$ are not continuously distributed. 
\begin{figure}[]
\caption{Box-plots of the CATT estimates (NW and RF)}
\medskip \includegraphics[scale=.7]{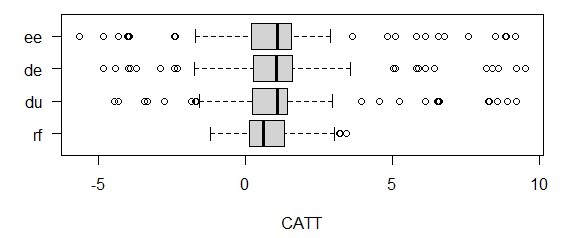}\newline
\begin{minipage}{1\textwidth}{Note: The estimates are rescaled and are denoted in '000\$.
}
\end{minipage}
\end{figure}
The kernel based regression CATT results remain more variable than those
provided by the random forest approach, but their interquartile range is
comparable. The NW kernel based estimates of the CATT on average exceed the
RF based estimates: $\$1,045$ (107), \$1,018 (108), and \$1,019 (104) for
the (e,e), (d,e) and (d,u) kernel on $(T,X)$ against \$794 (54) based on the
random forest. \medskip

The NW based results are stable across kernel, give interpretable insights
and with cross-validation make it possible to detect irrelevant regressors.

\newpage

\noindent \textbf{Acknowledgements:} The authors thank the participants at
the Econometric Study Group conference in Bristol, the Canadian Econometric
Study Group conference, and Saraswata Chaudhuri for their comments. We thank
Jeffrey Racine for his discussion and valuable suggestions at the CESG 2023
and Sid Kankanala for insightful comments on earlier versions of the paper.
We thank the Associate Editor and three anonymous referees for their careful
reading of the paper and very helpful comments and suggestions. \vspace{.2in}

\noindent \textbf{Funding:} Victoria Zinde-Walsh gratefully acknowledges
financial support from the Natural Sciences and Engineering Research Council
of Canada (NSERC) grant 253139.

\pagebreak
\appendix
\begin{center}
{\Large {{Multivariate kernel regression in vector and product metric spaces}} } \\ \ \\
\large {Marcia Schafgans and Victoria Zinde-Walsh} \\ \ \\
\large{January 2026}\vspace{1in}
\end{center}

\begin{center}
{\Large {{Supplemental Material}} }
\end{center}

Appendix A gives the {derivations for the moments and moment bounds}.
Appendix B provides the proofs and a remark on bandwidth selection with mass
points. Appendix C gives details on the Monte Carlo simulations. Appendix D
provides the details on the empirical study.\vspace{3in}
\pagebreak

\renewcommand{\theequation}{A.\arabic{equation}} \setcounter{equation}{0}%
\renewcommand{\thelemma}{A.\arabic{lemma}} \setcounter{lemma}{0}%
\renewcommand{\thecorollary}{A.\arabic{corollary}} \setcounter{corollary}{0}

\section{The general technical results}

In this appendix, we provide derivations for moments and bounds on the
moments that are used in establishing the limit properties of the estimators.

We provide the expectation of multivariate functions $\tilde{\psi}\left(
W\right) $ for $W=\left( W^{1},\cdots,\right.$ $\left. W^{\tilde{q}}\right) $%
, defined locally relative to the measure $P_{\tilde{X}}$ . We distinguish
two cases depending on whether $\tilde{X}\in \Xi ^{\lbrack q]}$ or $\tilde{X}%
\in \Xi ^{\lbrack 2q]}=\Xi ^{\lbrack q]}\times \Xi ^{\lbrack q]}$ (product
space).\medskip \noindent

\noindent \textbf{Case 1.} If $\tilde{X}\in \Xi ^{\lbrack q]}$ then $\tilde{q%
},\Xi ^{\left[ \tilde{q}\right] },\tilde{X}$, $\tilde{x},\tilde{h}$ coincide
with $q,\Xi ^{\left[ q\right] },X,x,h$. \newline
The function $\tilde{\psi}\left( W\right) $ is 
\begin{equation}
\tilde{\psi}\left( W\right) =\psi \left( \frac{x-X}{h}\right) =\psi \left( 
\frac{x^{1}-X^{1}}{h^{1}},\cdots ,\frac{x^{q}-X^{q}}{h^{q}}\right) \qquad
\qquad \qquad  \label{general fn}
\end{equation}%
when $\Xi ^{\left[ q\right] }=\mathbb{R}^{q}$; correspondingly, on general $%
\Xi ^{\left[ q\right] }$: 
\begin{equation}
\tilde{\psi}\left( W\right) =\psi _{+}\left( \frac{\left\Vert x-X\right\Vert 
}{h}\right) =\psi _{+}\left( \frac{\left\Vert x^{1}-X^{1}\right\Vert _{1}}{%
h^{1}},\cdots ,\frac{\left\Vert x^{q}-X^{q}\right\Vert _{q}}{h^{q}}\right) ,
\label{product function}
\end{equation}
{where $\psi _{+}$ denotes the multivariate function as defined on $\mathbb{R%
}^{q+}$.}

\noindent \textbf{Case 2.} If $\tilde{X}\in \Xi ^{\lbrack 2q]}$ then $\tilde{%
q},\Xi ^{\left[ \tilde{q}\right] },\tilde{X}$, $\tilde{x},\tilde{h}$
coincide with $2q$, $\Xi ^{\left[ q\right] }\times \Xi ^{\left[ q\right] }$, 
$(X_{t},X_{s})=\left( X_{t}^{1},\cdots
,X_{t}^{q},X_{s}^{1},\cdots,X_{s}^{q}\right), (x,x),$ and $(h,h).$\newline
The function $\tilde{\psi}\left( W\right) $ for the corresponding product
spaces $\Xi ^{\left[ \tilde{q}\right] }=\Xi ^{\left[ q\right] }\times \Xi ^{%
\left[ q\right] }$ is 
\begin{eqnarray*}
\tilde{\psi}\left( W \right) &=&\psi \left( \frac{x-X_{t}}{h}\right) \psi
\left( \frac{x-X_{s}}{h}\right), \text{ or} \\
\tilde{\psi}\left(W \right) &=& \psi _{+}\left( \frac{\left\Vert
x-X_{t}\right\Vert }{h}\right) \psi _{+}\left( \frac{\left\Vert
x-X_{s}\right\Vert }{h}\right) .
\end{eqnarray*}

\noindent {Note that Case 1 is the general case. Case 2 is specific to
evaluation of covariances arising from any Case 1.}\medskip

We consider the set of {consecutive} indices $\left\{ 1,2,\cdots ,\tilde{q}%
\right\} ;$ there are $2^{\tilde{q}}$ subsets of this set including the
empty set, $\varnothing .$ Denote each subset by $I_{\xi };$ $\xi
=0,1,\cdots ,2^{\tilde{q}}-1$ with $I_{0}=\varnothing $. The indices $\xi $
are ordered such that the indices are non-decreasing in the cardinality of
the set and are ordered lexicographically for each cardinality. ${q}({\xi })$
denotes the cardinality of the subset $I_{\xi }$. The complement of the
subset $I_{\xi }$ is denoted by $I_{\xi }^{c}$. Thus, for instance, for $%
\tilde{q}=q=3,$ there are 8 subsets: $I_{0}=\varnothing ;I_{1}=\left\{
1\right\} ,I_{2}=\left\{ 2\right\} ,I_{3}=\left\{ 3\right\} ,I_{4}=\left\{
1,2\right\} ,I_{5}=\left\{ 1,3\right\} ;I_{6}=\left\{ 2,3\right\}
,I_{7}=\left\{ 1,2,3\right\} .$ 
\medskip

We use $\prod_{j\in I_{\xi }}^{{}}\left( -\partial _{j}\right) $ to denote
an operator that, when applied to a sufficiently differentiable function $%
g\left( z\right) =g\left( z^{1},\cdots,z^{\tilde{q}}\right) $ at $z$, maps
it to its partial derivative for $j_{1}<\cdots<j_{{q}({\xi })}\in I_{\xi }$,
times $\left( -1\right) ^{{q}\left( \xi \right) }$ that is 
\begin{equation*}
\left( \prod_{j\in I_{\xi }}^{{}}\left( -\partial _{j}\right) \right)
g\left( z\right) =\left( -1\right) ^{{q}\left( \xi \right) }\frac{\partial ^{%
{q}({\xi })}}{\partial _{j_{1}}\cdots\partial _{j_{{q}({\xi })}}}g\left(
z\right) .
\end{equation*}

Using the delta-function operator, $\delta \left( z^{j}=a\right) ,$ that
applied to $g\left( z^{1},\cdots ,z^{j},\cdots ,z^{\tilde{q}}\right) $ sets
the $jth$ component to the scalar value $a$, we define the operator $\Delta
_{a,\xi }:$ 
\begin{equation}
\Delta _{a,\xi }g(z)=\prod_{j\in I_{\xi }^{c}}\delta \left( z^{j}=a\right)
\prod_{j\in I_{\xi }}^{{}}\left( -\partial _{j}\right) g(z);  \label{Delta_a}
\end{equation}%
(with often $a=$ $1$ denoting a point on the boundary).

\bigskip We deal with three situations for $\tilde{\psi}\left( W\right) $.
First, the following Lemma \ref{L.phiCondE} provides the expectation of $%
\tilde{\psi}(W)=\tilde{\psi}\left( \frac{\tilde{x}-\tilde{X}}{\tilde{h}}%
\right) $ with respect to the probability measure $P_{\tilde{X}}$ for a
sufficiently differentiable random function $\tilde{\psi}$ on $\mathbb{R}^{%
\tilde{q}}$ with support $\left[ -1,1\right] ^{\tilde{q}}$. The probability
measure $P_{\tilde{X}}$ could be $P_{X}$ or a measure on the product space
with $P_{X}$ marginals. Second, Corollary A.1 to Lemma A.1 details this
result for a symmetric function on $\left[ -1,1\right] ^{\tilde{q}}$. Third,
Lemma A.2 provides the expectation of a sufficiently differentiable function 
$\tilde{\psi}_{+}\left( \frac{\left\Vert \tilde{x}-\tilde{X}\right\Vert }{h}%
\right) $ given on $\left[ 0,1\right] ^{\tilde{q}}$ relative to a measure on 
$\Xi ^{\left[ \tilde{q}\right] }.$ \medskip This is followed by examining
moments for the product $g(\tilde{X})\tilde{\psi}\left( \frac{\tilde{x}-%
\tilde{X}}{\tilde{h}}\right) .$

The moments for $\tilde{\psi}\left( W\right) $ are expressed via sums of
integrals for every $\xi $ over some {$S^{{q}(\xi )}\in $}$\mathbb{R}^{%
\tilde{q}}${\ where usually $S^{{q}(\xi )}=\left[ -1,1\right] ^{{q}(\xi )}$}
or $\left[ 0,1\right] ^{q\left( \xi \right) }.$ The integrals involve
probability measures of some sets, $\widetilde{C},$ $P_{\tilde{X}}(%
\widetilde{C})$, where $\widetilde{C}$ is inside the cuboid $\tilde{C}\left(
x,h\right) $ in $\Xi ^{\left[ \tilde{q}\right] }.$ Given a local function $%
\tilde{\psi}\left( .\right) $ and the operator $\Delta _{a,\xi }$ we define 
\begin{equation}
\mathcal{I}_{\xi }\left( \tilde{x};S^{{q}\left( \xi \right) },P_{\tilde{X}}(%
\widetilde{C}),\Delta _{a,\xi }\tilde{\psi}\right) =\int_{S^{{q}\left( \xi
\right) }}P_{\tilde{X}}(\widetilde{C})\left( \Delta _{a,\xi }\tilde{\psi}%
\left( v\right) \right) dv\left( \xi \right) .  \label{Int def}
\end{equation}

\begin{lemma}
\label{L.phiCondE} Suppose that the function $\tilde{\psi}\left( W\right) $
is sufficiently differentiable with support on $\left[ -1,1\right] ^{\tilde{q%
}\text{ }}$ and $P_{\tilde{X}}$ is the measure associated with $\tilde{X}\in 
\mathbb{R}^{\tilde{q}}$ given by a distribution function $F_{\tilde{X}}$ on
the space $\mathbb{R}^{\tilde{q}}.$ Then with $P_{\tilde{X}}(\widetilde{C})$
expressed as $F_{\tilde{X}}(\tilde{x}-\lambda _{\xi }\circ \tilde{h},\tilde{x%
}+\tilde{h})$ where $\lambda _{\xi }$ is a vector with components:%
\begin{equation}
\left\{ \lambda _{\xi }\right\} ^{j}=v^{j}\text{ if }j\in I_{\xi },\text{
otherwise }\left\{ \lambda _{\xi }\right\} ^{j}=1,  \label{lambda}
\end{equation}%
the expectation is

\begin{eqnarray}
&&E\left( \tilde{\psi}\left( \tfrac{\tilde{x}-\tilde{X}}{\tilde{h}}\right)
\right)  \notag \\
&=& \sum_{\xi =0}^{2^{\tilde{q}}-1}{\int_{\left[-1,1\right]^{{q}(\xi )}}}F_{%
\tilde{X}}(\tilde{x}-\lambda _{\xi }\circ\tilde{h},\tilde{x}+\tilde{h})
\left( \Delta _{1,\xi }\tilde{\psi}\left( v^{1},\cdots,v^{\tilde{q}}\right)
\right) dv(\xi )  \notag \\
&=&\sum_{\xi =0}^{2^{\tilde{q}}-1}\mathcal{I}_{\xi }\left( \tilde{x};\left[
-1,1\right]^{{q}\left( \xi \right) },F_{\tilde{X}}(\tilde{x}-\lambda _{\xi
}\circ\tilde{h},\tilde{x}+\tilde{h}),\Delta _{1,\xi }\tilde{\psi}\right) .
\label{for psi (a)}
\end{eqnarray}
\end{lemma}

\noindent {The notation $F_{\tilde{X}}(\tilde{x}-\lambda _{\xi }\circ \tilde{%
h},\tilde{x}+\tilde{h})$ where $\tilde{X}\in $}$\mathbb{R}^{\tilde{q}}${,
equals 
\begin{equation*}
F_{\tilde{X}}(\tilde{x}-\lambda _{\xi }\circ \tilde{h},\tilde{x}+\tilde{h}%
)=\int_{\tilde{C}}dF_{\tilde{X}}
\end{equation*}%
with $\tilde{C}=\prod_{j\in I_{\xi }}\left[ \tilde{x}^{j}-{v}^{j}\tilde{h}%
^{j},\tilde{x}^{j}+\tilde{h}^{j}\right] \prod_{j\int_{\xi }^{c}}\left[ 
\tilde{x}^{j}-\tilde{h}^{j},\tilde{x}^{j}+\tilde{h}^{j}\right] $, where $F_{%
\tilde{X}}$ is the distribution function of $\tilde{X}$.}\vspace{0.2in}

\noindent \textbf{Proof of Lemma \ref{L.phiCondE}.}

\noindent Note that for a univariate function 
\begin{equation*}
\tilde{\psi}\left( 1\right) =\tilde{\psi}(w)+\int_{w}^{1}\left( \partial
_{v}\right) \tilde{\psi}\left( v\right) dv,
\end{equation*}%
thus 
\begin{equation*}
\tilde{\psi}\left( w\right) =\tilde{\psi}(1)-\int_{w}^{1}\left( \partial
_{v}\right) \tilde{\psi}\left( v\right) dv.
\end{equation*}%
When generalizing to the multivariate case we obtain for $\tilde{\psi}%
(w^{1},\cdots ,w^{\tilde{q}})$ (over $\mathbb{R}^{\tilde{q}})$ 
\begin{align}
& \tilde{\psi}(1,\cdots ,1)+\sum_{\xi =1}^{2^{\tilde{q}}-1}\left\{
\int_{S_{v}^{{q}(\xi )}}\left( \left[ \prod_{j\in I_{\xi }^{c}}^{{}}\delta
\left( v^{j}=1\right) \prod_{j\in I_{\xi }}\left( -\partial _{j}\right) %
\right] \tilde{\psi}\left( v^{1},\cdots,v^{\tilde{q}}\right) \right)
\prod_{j\in I_{\xi }}dv^{j}\right\}  \notag \\
& =\sum_{\xi =0}^{2^{\tilde{q}}-1}\left\{ \int_{S_{v}^{{q}\left( \xi \right)
}}\Delta _{1,\xi }\tilde{\psi}\left( v^{1},\cdots,v^{\tilde{q}}\right)
dv(\xi )\right\}  \label{Sum with v}
\end{align}%
where $S_{v}^{{q}\left( \xi \right) }=\prod_{j\in I_{\xi }}^{{}}\left[
w^{j},1\right] $, $\Delta _{1,\xi }$ is defined in (\ref{Delta_a}), and $%
dv(\xi )=\prod_{j\in I_{\xi }}dv^{j}$.\medskip

To each term of (\ref{Sum with v}) with $v^{j}=\frac{\tilde{x}^{j}-t^{j}}{%
\tilde{h}^{j}}$, we next apply a change of variables: $t^{j}=\tilde{x}%
^{j}-v^{j}\tilde{h}^{j}$ for all $j\in I_{\xi }$. Here 
\begin{equation*}
dv(\xi ):=\prod_{j\in I_{\xi }}dv^{j}=\prod_{j\in I_{\xi }}\left( -\tilde{h}%
^{j}\right) ^{-1}dt^{j}=:(-1)^{{q}(\xi )}\prod_{j\in I_{\xi }}\left( \tilde{h%
}^{j}\right) ^{-1}dt(\xi ).
\end{equation*}%
We replace the integration limits $w^{j}$ in $S_{v}^{{q}(\xi )}$ with $%
W^{j}\equiv \frac{\tilde{x}^{j}-\tilde{X}^{j}}{\tilde{h}^{j}},$ where $%
\tilde{X}^{j}$ denotes the value of a random element in $\mathbb{R}^{\tilde{q%
}}.$ The limits of integrals change with $1\rightarrow \tilde{x}^{j}-\tilde{h%
}^{j};W^{j}\equiv \frac{\tilde{x}^{j}-\tilde{X}^{j}}{\tilde{h}^{j}}%
\rightarrow \tilde{X}^{j}$. Since the function $\tilde{\psi}(\frac{\tilde{x}-%
\tilde{X}}{\tilde{h}})$ is zero outside of the set 
\begin{equation*}
I(\tilde{x}-\tilde{h}\leq \tilde{X}\leq \tilde{x}+\tilde{h})=\prod_{i=1}^{%
\tilde{q}}I(\tilde{x}^{i}-\tilde{h}^{i}\leq \tilde{X}^{i}\leq \tilde{x}^{i}+%
\tilde{h}^{i}),
\end{equation*}%
$\tilde{X}^{j}\geq \tilde{x}^{j}-b^{j}\tilde{h}^{j}$. We obtain

\begin{eqnarray*}
&&\tilde{\psi}\left( \tfrac{\tilde{x}^{1}-\tilde{X}^{1}}{\tilde{h}^{1}}%
,\cdots ,\tfrac{\tilde{x}^{\tilde{q}}-\tilde{X}^{\tilde{q}}}{\tilde{h}^{q}}%
\right) \\
&=&\sum_{\xi =0}^{2^{\tilde{q}}-1}\left\{ \int_{S_{t}^{q\left( \xi \right) }}%
\left[ \Delta _{1,\xi }\tilde{\psi}\left( \frac{\tilde{x}^{1}-t^{1}}{\tilde{h%
}^{1}},\cdots,\frac{\tilde{x}^{\tilde{q}}-t^{\tilde{q}}}{\tilde{h}^{\tilde{q}%
}}\right) \right] \prod_{j\in I_{\xi }}\left( \tilde{h}^{j}\right)
^{-1}dt(\xi )\right\} \\
&&\qquad \times I(\tilde{x}-\tilde{h}\leq \tilde{X}\leq \tilde{x}+\tilde{h}),
\end{eqnarray*}%
where {$S_{t}^{{q}(\xi )}=\prod_{j\in I_{\xi }}[\tilde{x}^{j}-\tilde{h}^{j},%
\tilde{X}^{j}]$}. The $(-1)^{{q}(\xi )}$ term arising from the change of
variables vanishes due to the reversal in the limits integral.

The expectation is 
\begin{eqnarray}
&&E\left( \tilde{\psi}\left(\tfrac{\tilde{x}-\tilde{X}}{h}\right)\right) 
\notag \\
&=&\sum_{\xi =0}^{2^{\tilde{q}}-1}\int_{\mathbb{R}^{\tilde{q}}}\left\{
\int_{S_{t}^{{q}\left( \xi \right) }}\Delta _{1,\xi }\tilde{\psi}\left( 
\frac{\tilde{x}^{1}-{t}^{1}}{\tilde{h}^{1}},\cdots,\frac{\tilde{x}^{\tilde{q}%
}-{t}^{\tilde{q}}}{\tilde{h}^{\tilde{q}}}\right) \prod_{j\in I_{\xi }}\left(
h^{j}\right) ^{-1}dt(\xi )\right\}  \notag \\
&&\qquad \times I(\tilde{x}-\tilde{h}\leq \tilde{X}\leq \tilde{x}+\tilde{h}%
)dP_{\tilde{X}}(\tilde{X})  \notag \\
&=&\sum_{\xi =0}^{2^{\tilde{q}}-1}\int_{\tilde{S}_{t}^{{q}\left( \xi \right)
}}\left[ \Delta _{1,\xi }\tilde{\psi}\left( \frac{\tilde{x}^{1}-t^{1}}{%
\tilde{h}^{1}},\cdots,\frac{\tilde{x}^{\tilde{q}}-t^{\tilde{q}}}{\tilde{h}^{%
\tilde{q}}}\right) \right] \left\{ \int_{S_{X}^{\tilde{q}}\left( t(\xi
)\right) }dP_{\tilde{X}}(\tilde{X})\right\}  \notag \\
&&\times \prod_{j\in I_{\xi }}\left( \tilde{h}^{j}\right) ^{-1}dt(\xi )
\label{E form}
\end{eqnarray}%
where the first equality uses linearity of the expectation operator and the
second equality uses the Fubini theorem and recognizes that the domain of
integration in the curly brackets depends on $t^{j},\ j\in I_{\xi }$. In
particular as for every $j\in I_{\xi }$ 
\begin{equation*}
t^{j}\leq \tilde{X}^{j}\leq \tilde{x}^{j}+\tilde{h}^{j}
\end{equation*}%
we get $S_{X}^{\tilde{q}}\left( t(\xi )\right) =\prod_{j\in I_{\xi }}^{{}}[{t%
}^{j},\tilde{x}^{j}+\tilde{h}^{j}]\prod_{j\in I_{\xi }^{c}}^{{}}\left[ 
\tilde{x}^{j}-\tilde{h}^{j},\tilde{x}^{j}+\tilde{h}^{j}\right]$, which
incorporates the requirement $I(\tilde{x}-\tilde{h}\leq \tilde{X}\leq \tilde{%
x}+\tilde{h})$. This provides 
\begin{eqnarray*}
\int_{S_{X}^{\tilde{q}}\left( t(\xi )\right) }dP_{\tilde{X}}(\tilde{X})
&=&P_{\tilde{X}}\left(\prod_{j\in I_{\xi }}^{{}}[{t}^{j},\tilde{x}^{j}+%
\tilde{h}^{j}]\prod_{j\in I_{\xi }^{c}}^{{}}\left[ \tilde{x}^{j}-\tilde{h}%
^{j},\tilde{x}^{j}+\tilde{h}^{j}\right] \right) \\
&=&F_{\tilde{X}}(\tilde{x}-\lambda _{\xi }(t)\circ \tilde{h},\tilde{x}+%
\tilde{h})
\end{eqnarray*}%
where 
\begin{equation*}
\left\{ \lambda _{\xi }(t)\right\} ^{j}=\frac{t^{j}-\tilde{x}^{j}}{\tilde{h}%
^{j}},\text{ if }j\in I_{\xi },\text{ otherwise }\left\{ \lambda _{\xi
}(t)\right\} ^{j}=1.
\end{equation*}%
For the limits of the integral with respect to $t(\xi )$, we use $\widetilde{%
S}_{t}^{{q}(\xi )}=\prod_{j\in I_{\xi }}[\tilde{x}^{j}-\tilde{h}^{j},\tilde{x%
}^{j}+\tilde{h}^{j}]$ . The last displayed expression (\ref{E form}) then
becomes%
\begin{eqnarray*}
&&\sum_{\xi =0}^{2^{\tilde{q}}-1}\int_{\tilde{S}_{t}^{{q}\left( \xi \right)
}}F_{\tilde{X}}(\tilde{x}-\lambda _{\xi }(t)\circ \tilde{h},\tilde{x}+\tilde{%
h})\left[ \Delta _{1,\xi }\tilde{\psi}\left( \frac{\tilde{x}^{1}-t^{1}}{%
\tilde{h}^{1}},\cdots,\frac{\tilde{x}^{\tilde{q}}-t^{\tilde{q}}}{\tilde{h}^{%
\tilde{q}}}\right) \right] \\
&&\times \prod_{j\in I_{\xi }}\left( \tilde{h}^{j}\right) ^{-1}dt(\xi )
\end{eqnarray*}

After applying a change of variables with $t^{j}=\tilde{x}^{j}-v^{j}\tilde{h}%
^{j}$, this yields 
\begin{equation*}
\sum_{\xi =0}^{2^{\tilde{q}}-1}{\int_{\left[-1,1\right]^{{q}(\xi )}}}\left(
\Delta _{1,\xi }\tilde{\psi}\left( v^{1},\cdots,v^{\tilde{q}}\right) \right)
F_{\tilde{X}}(\tilde{x}-\lambda _{\xi }\circ\tilde{h},\tilde{x}+\tilde{h}%
)dv(\xi )
\end{equation*}%
where 
\begin{equation*}
\left\{ \lambda _{\xi }\right\} ^{j}=v^{j},\text{ if }j\in I_{\xi },\text{
otherwise }\left\{ \lambda _{\xi }\right\} ^{j}=1 .
\end{equation*}
The change of variables uses $dv(\xi )=(-1)^{q(\xi)}( \tilde{h}^{j})
^{-1}dt(\xi )$ together with a reversal in the limits of integration as
before.

Therefore, 
\begin{eqnarray*}
&&E\left( \tilde{\psi}\left( \tfrac{\tilde{x}-\tilde{X}}{\tilde{h}}\right)
\right) \\
&=&\sum_{\xi =0}^{2^{\tilde{q}}-1}\int_{\left[-1,1\right] ^{{q}\left( \xi
\right) }}F_{\tilde{X}}(\tilde{x}-\lambda _{\xi }\circ\tilde{h},\tilde{x}+%
\tilde{h})\prod_{j\in I_{\xi }^{c}}\delta \left( v^{j}=1\right) \prod_{j\in
I_{\xi }}\left( -\partial _{j}\right) \tilde{\psi}\left( v\right) dv(\xi ) .
\end{eqnarray*}
\textcolor{white}{.}\hfill $\blacksquare $

\begin{corollary}
Suppose that the function $\tilde{\psi}\left( W\right) $ is sufficiently
differentiable with support on $\left[ -1,1\right] ^{\tilde{q}}$. If, in
addition to the conditions of Lemma \ref{L.phiCondE}, the function $\tilde{%
\psi} $ is symmetric around zero in every argument, then for $\xi \in I_{\xi
}$ the corresponding $\widetilde{C}=C(\tilde{x},\lambda _{\xi }\circ \tilde{h%
})$ where $\lambda_{\xi}$ a vector with components:%
\begin{equation}
\left\{ \lambda _{\xi }\right\} ^{j}=v^{j}\text{ if }j\in I_{\xi },\text{
otherwise }\left\{ \lambda _{\xi }\right\} ^{j}=1,  \label{lambda1}
\end{equation}
and 
\begin{eqnarray}
&&E\left( \tilde{\psi}\left( \tfrac{\tilde{x}-\tilde{X}}{\tilde{h}}\right)
\right)  \notag \\
&=& \sum_{\xi =0}^{2^{\tilde{q}}-1}{\int_{\left[0,1\right]^{{q}(\xi )}}}F_{%
\tilde{X}}(\tilde{x}-\lambda _{\xi }\circ\tilde{h},\tilde{x}+\lambda _{\xi
}\circ\tilde{h}) \left( \Delta _{1,\xi }\tilde{\psi}\left( v^{1},\cdots,v^{%
\tilde{q}}\right) \right) dv(\xi )  \notag \\
&=&\sum_{\xi =0}^{2^{\tilde{q}}-1}\mathcal{I}_{\xi }\left( \tilde{x};\left[
0,1\right] ^{{q}\left( \xi \right) },P_{\tilde{X}}\left( C(\tilde{x},\lambda
_{\xi }\circ \tilde{h})\right) ,\Delta _{1,\xi }\tilde{\psi}\right).
\label{for psi (a,b)}
\end{eqnarray}
\end{corollary}

\noindent We see that symmetry simplifies the moment expression.\vspace{.2in}

\noindent\textbf{Proof of Corollary A.1.}

\noindent From Lemma A.1 
\begin{equation*}
\sum_{\xi =0}^{2^{\tilde{q}}-1}\int_{\left[ -1,1\right] ^{{q}\left( \xi
\right) }}F_{\tilde{X}}(\tilde{x}-\lambda _{\xi }\circ \tilde{h},\tilde{x}+%
\tilde{h})\prod_{j\in I_{\xi }^{c}}^{{}}\delta \left(
v^{j}=1\right)\prod_{j\in I_{\xi }}^{{}}\left( -\partial _{j}\right) \tilde{%
\psi}\left( v\right) dv(\xi ),
\end{equation*}%
where $dv(\xi )=\prod_{j\in I_{\xi }}dv^{j}$ and 
\begin{equation*}
\left\{ \lambda _{\xi }\right\} ^{j}=v^{j},\text{ if }j\in I_{\xi },\text{
otherwise }\left\{ \lambda _{\xi }(v)\right\} ^{j}=1.
\end{equation*}%
Write 
\begin{equation*}
dv(\xi )=dv^{j_{1}}dv^{j_{2}}\cdots dv^{j_{{q}(\xi )}}.
\end{equation*}%
Consider some $\xi $ and the corresponding multivariate integral. Next, we
integrate such an integral with respect to $v^{j_{1}}$ (meanwhile holding $%
v^{j_{2}},\cdots ,v^{j_{{q}(\xi )}}$ constant) 
\begin{eqnarray}
&&\int_{-1}^{1}F_{\tilde{X}}(\tilde{x}-\lambda _{\xi }\circ \tilde{h},\tilde{%
x}+\tilde{h})\prod_{j\in I_{\xi }^{c}}^{{}}\delta \left( v^{j}=1\right)
\prod_{j\in I_{\xi }}^{{}}\left( -\partial _{j}\right) \tilde{\psi}\left(
v\right) dv^{j_{1}}  \notag \\
&=&\int_{-1}^{0}F_{\tilde{X}}(\tilde{x}-\lambda _{\xi }\circ \tilde{h},%
\tilde{x}+\tilde{h})\prod_{j\in I_{\xi }^{c}}^{{}}\delta \left(
v^{j}=1\right) \prod_{j\in I_{\xi }}^{{}}\left( -\partial _{j}\right) \tilde{%
\psi}\left( v\right) dv^{j_{1}}  \label{IntSum} \\
&&+\int_{0}^{1}F_{\tilde{X}}(\tilde{x}-\lambda _{\xi }\circ \tilde{h},\tilde{%
x}+\tilde{h})\prod_{j\in I_{\xi }^{c}}^{{}}\delta \left( v^{j}=1\right)
\prod_{j\in I_{\xi }}^{{}}\left( -\partial _{j}\right) \tilde{\psi}\left(
v\right) dv^{j_{1}}.  \notag
\end{eqnarray}%
In the first integral, on the rhs of (\ref{IntSum}) apply a change of
variable $v^{j_{1}}=-z^{j_{1}};$ define $\lambda _{\xi }\left( j_{1}\right) $
the same as $\lambda _{\xi }$ for every component, except for $j_{1}^{th}$
where it is $-z^{j_{1}};$ let $v\left( -z^{j_{1}}\right) $ represent $v$
with $v^{j_{1}}$ replaced with $-z^{j_{1}},$ then this term can be written
as 
\begin{equation*}
-\int_{0}^{1}F_{\tilde{X}}(\tilde{x}-\lambda _{\xi }(j_{1})\circ \tilde{h},%
\tilde{x}+\tilde{h})\prod_{j\in I_{\xi }^{c}}^{{}}\delta \left(
v^{j}=1\right) \prod_{j\in I_{\xi }}^{{}}\left( -\partial _{j}\right) \tilde{%
\psi}\left( v\left( -z^{j_{1}}\right) \right) \left( -1\right) dz^{j_{1}}.
\end{equation*}%
where the minus from interchanging the limits of integration and the minus
arising from the change of variables cancel out. By symmetry of $\tilde{\psi}%
\left( \cdot \right) $, $\left( -\partial _{j_{1}}\right) \tilde{\psi}\left(
v\left( -z^{j_{1}}\right) \right) =-\left( -\partial _{j_{1}}\right) \tilde{%
\psi}\left( v\left( z^{j_{1}}\right) \right) $, where $v\left(
z^{j_{1}}\right) $ represents $v$ with $v^{j_{1}}$ replaced with $z^{j_{1}}.$
The first integral then becomes 
\begin{equation*}
-\int_{0}^{1}F_{\tilde{X}}(\tilde{x}-\lambda _{\xi }(j_{1})\circ \tilde{h},%
\tilde{x}+\tilde{h})\prod_{j\in I_{\xi }^{c}}^{{}}\delta \left(
v^{j}=1\right) \prod_{j\in I_{\xi }}^{{}}\left( -\partial _{j}\right) \tilde{%
\psi}\left( v\left( z^{j_{1}}\right) \right) dz^{j_{1}}.
\end{equation*}%
A simple change of notation in the second integral on the rhs of (\ref%
{IntSum}) (replacing $v^{j_{1}}$by $z^{j_{1}}$ in $v$ and denoting the
resulting vector by $v\left( z^{j_{1}}\right) )$ allows us to express the
sum of the two integrals in (\ref{IntSum}) as%
\begin{eqnarray*}
&&\int_{0}^{1}\left\{ F_{\tilde{X}}(\tilde{x}-\lambda _{\xi }\circ \tilde{h},%
\tilde{x}+\tilde{h})-F_{\tilde{X}}(\tilde{x}-\lambda _{\xi }\left(
j_{1}\right) \circ \tilde{h},\tilde{x}+\tilde{h})\right\} \\
&&\qquad \prod_{j\in I_{\xi }^{c}}^{{}}\delta \left(
v^{j}=1\right)\prod_{j\in I_{\xi }}^{{}}\left( -\partial _{j}\right) \tilde{%
\psi}\left( v\left( z^{j_{1}}\right) \right) dz^{j_{1}}.
\end{eqnarray*}%
note that $\lambda _{\xi }$ also has $v_{1}$ replaced with $z_{1}$.

Next, consider the integral with respect to $v^{j_{2}}$ (meanwhile holding $%
v^{j_{3}},\cdots ,v^{j_{q(\xi )}}$ constant). That is, we evaluate 
\begin{eqnarray*}
&&\int_{-1}^{1}\left\{ \int_{0}^{1}\left\{ F_{\tilde{X}}(\tilde{x}-\lambda
_{\xi }\circ \tilde{h},\tilde{x}+\tilde{h})-F_{\tilde{X}}(\tilde{x}-\lambda
_{\xi }\left( j_{1}\right) \circ \tilde{h},\tilde{x}+\tilde{h})\right\}
\right. \\
&&\textcolor{white}{.}\qquad \left. \prod_{j\in I_{\xi
}^{c}}^{{}}\delta\left( v^{j}=1\right)\prod_{j\in I_{\xi }}^{{}}\left(
-\partial _{j}\right) \tilde{\psi}\left( v\left( z^{j_{1}}\right) \right)
dz^{j_{1}}\right\} dv^{j_{2}}.
\end{eqnarray*}%
A similar substitution $z^{j_{2}}=-v^{j_{2}},$ with $v\left(
z^{j_{1}},z^{j_{2}}\right) $, $\lambda _{\xi }\left( j_{2}\right) $ and $%
\lambda _{\xi }\left( j_{1},j_{2}\right) $ similarly defined provides the
integral as 
\begin{eqnarray*}
&&\int_{0}^{1}\int_{0}^{1}\left\{ F_{\tilde{X}}(\tilde{x}-\lambda _{\xi
}\circ \tilde{h},\tilde{x}+\tilde{h})-F_{\tilde{X}}(\tilde{x}-\lambda _{\xi
}\left( j_{2}\right) \circ \tilde{h},\tilde{x}+\tilde{h})\right. \\
&&\qquad \quad \left. +F_{\tilde{X}}(\tilde{x}-\lambda _{\xi }\left(
j_{1},j_{2}\right) \circ \tilde{h},\tilde{x}+\tilde{h})-F_{\tilde{X}}(\tilde{%
x}-\lambda _{\xi }\left( j_{1}\right) \circ \tilde{h},\tilde{x}+\tilde{h}%
)\right\} \\
&&\  \\
&&\times \prod_{j\in I_{\xi }^{c}}\delta \left( v^{j}=1\right) \prod_{j\in
I_{\xi }}\left( -\partial _{j}\right) \tilde{\psi}\left( v\left(
z^{j_{1}},z^{j_{2}}\right) \right) dz^{j_{1}}dz^{j_{2}}.
\end{eqnarray*}%
note that $\lambda _{\xi }$ also has $v_{2}$ also replaced with $z_{2}$.

Continuing this until $z^{j_{q(\xi )}}$ yields 
\begin{eqnarray*}
&&\int_{0}^{1}\cdots \int_{0}^{1}\left\{\underset{j\in I_{\xi }^{c}}{%
\underbrace{\int_{\tilde{x}_{1}-\tilde{h}_{1}}^{\tilde{x}_{1}+\tilde{h}%
_{1}}\int_{\tilde{x}_{2}-\tilde{h}_{2}}^{\tilde{x}_{2}+\tilde{h}_{2}}\cdots }%
}\underset{j\in I_{\xi}}{\underbrace{\int_{\tilde{x}_{j_{1}}-z_{j_{1}}\tilde{%
h}_{j_{1}}}^{\tilde{x}_{j_{1}}+z_{j_{1}}\tilde{h}_{j_{1}}}\cdots\int_{\tilde{%
x}_{j_{q(\xi)}}-z_{j_{q(\xi)}}\tilde{h}_{j_{q(\xi)}}}^{\tilde{x}%
_{j_{q(\xi)}}+z_{j_{q(\xi)}}\tilde{h}_{j_{q(\xi)}}} }}dF_{\tilde{X}}\right\}
\\
&&\  \\
&&\times \prod_{j\in I_{\xi }^{c}}\delta \left( v^{j}=1\right) \prod_{j\in
I_{\xi }}\left( -\partial _{j}\right) \tilde{\psi}\left(
v(z^{j_{1}},z^{j_{2}},\cdots ,z^{j_{q(\xi )}})\right) dz(\xi )
\end{eqnarray*}%
where $dz(\xi )=dz^{j_{1}}dz^{j_{2}}\cdots dz^{j_{q(\xi )}}$. Simply
changing the notation, with $dv(\xi )=dv^{j_{1}}dv^{j_{2}}\cdots
dv^{j_{q(\xi )}}$ and $v(v^{j_{1}},v^{j_{2}},\cdots ,v^{j_{q(\xi )}})=v$,
yields for every $\xi $ 
\begin{eqnarray*}
&&\int_{0}^{1}\cdots \int_{0}^{1}\left\{\underset{j\in I_{\xi }^{c}}{%
\underbrace{\int_{\tilde{x}_{1}-\tilde{h}_{1}}^{\tilde{x}_{1}+\tilde{h}%
_{1}}\int_{\tilde{x}_{2}-\tilde{h}_{2}}^{\tilde{x}_{2}+\tilde{h}_{2}}\cdots }%
}\underset{j\in I_{\xi}}{\underbrace{\int_{\tilde{x}_{j_{1}}-z_{j_{1}}\tilde{%
h}_{j_{1}}}^{\tilde{x}_{j_{1}}+z_{j_{1}}\tilde{h}_{j_{1}}}\cdots\int_{\tilde{%
x}_{j_{q(\xi)}}-z_{j_{q(\xi)}}\tilde{h}_{j_{q(\xi)}}}^{\tilde{x}%
_{j_{q(\xi)}}+z_{j_{q(\xi)}}\tilde{h}_{j_{q(\xi)}}} }}dF_{\tilde{X}}\right\}
\\
&&\  \\
&&\times \prod_{j\in I_{\xi }^{c}}\delta \left( v^{j}=1\right) \prod_{j\in
I_{\xi }}\left( -\partial _{j}\right) \tilde{\psi}\left( v\right) dv(\xi ) \\
&=&\int_{\left[ 0,1\right] ^{q\left( \xi \right) }}F_{\tilde{X}}(\tilde{x}%
-\lambda _{\xi }\circ \tilde{h},\tilde{x}+\lambda _{\xi }\circ \tilde{h}%
)\prod_{j\in I_{\xi }^{c}}\delta \left( v^{j}=1\right) \prod_{j\in I_{\xi
}}\left( -\partial _{j}\right) \tilde{\psi}\left( v\right) dv(\xi ) \\
&=&\int_{\left[ 0,1\right] ^{q\left( \xi \right) }}P_{\tilde{X}}(C(\tilde{x}%
,\lambda _{\xi }\circ \tilde{h})\prod_{j\in I_{\xi }^{c}}\delta \left(
v^{j}=1\right) \prod_{j\in I_{\xi }}\left( -\partial _{j}\right) \tilde{\psi}%
\left( v\right) dv(\xi ) \\
&=&\int_{\left[ 0,1\right] ^{q\left( \xi \right) }}P_{\tilde{X}}(C(\tilde{x}%
,\lambda _{\xi }\circ \tilde{h})\Delta _{1,\xi }\tilde{\psi}\left( v\right)
dv(\xi )
\end{eqnarray*}
This concludes the proof.\hfill$\blacksquare $\vspace{.2in}

The lemma below applies to a functional or metric product space with $\Xi ^{%
\left[ \tilde{q}\right] \text{ }}$ given by $\Xi ^{\lbrack q]}$ or $\Xi
^{\lbrack q]}\times \Xi ^{\lbrack q]}.$\medskip

\begin{lemma}
\label{L.phiCondE2}Suppose that the function $\tilde{\psi}\left( W\right) =%
\tilde{\psi}_{+}\left( \frac{\left\Vert \tilde{x}^{1}-\tilde{X}%
^{1}\right\Vert _{1}}{\tilde{h}^{1}},\cdots,\frac{\left\Vert \tilde{x}^{q}-%
\tilde{X}^{\tilde{q}}\right\Vert _{q}}{h^{\tilde{q}}}\right) $ where $\tilde{%
\psi}_{+} $ defined on $\left[ 0,1\right] ^{\tilde{q}}$ is sufficiently
differentiable {and $P_{\tilde{X}}$ is a probability measure defined on $\Xi
^{\left[ \tilde{q}\right] }$} and $\widetilde{C}=C(\tilde{x},\lambda _{\xi
}\circ \tilde{h})$ with $\lambda _{\xi }$ defined in (\ref{lambda1}) then 
\begin{eqnarray}
&&E\left( \tilde{\psi}_{+}\left( \tfrac{\left\Vert \tilde{x}^{1}-\tilde{X}%
^{1}\right\Vert _{1}}{\tilde{h}^{1}},\cdots,\tfrac{\left\Vert \tilde{x}^{%
\tilde{q}}-\tilde{X}^{\tilde{q}}\right\Vert _{\tilde{q}}}{\tilde{h}^{\tilde{q%
}}}\right) \right)  \notag \\
&=&\sum_{\xi =0}^{2^{\tilde{q}}-1}\mathcal{I}_{\xi }\left( \tilde{x};\left[
0,1\right] ^{{q}\left( \xi \right) },P_{\tilde{X}}\left( C(\tilde{x},\lambda
_{\xi }\circ \tilde{h})\right) ,\Delta _{1,\xi }\tilde{\psi}\right) .
\label{for psi (c)}
\end{eqnarray}
\end{lemma}

\vspace{.2in}

\noindent \textbf{Proof of Lemma A.2.}

\noindent Consider $\Xi ^{\lbrack \tilde{q}]}$. When $\Xi ^{\lbrack \tilde{q}%
]}=\Xi ^{\lbrack q]}$, the vector $\tilde{x}=x,$ but when $\Xi ^{\lbrack 
\tilde{q}]}=\Xi ^{\lbrack q]}\times \Xi ^{\lbrack q]},$ $\tilde{x}=\left(
x,x\right) $. Given $\tilde{x}$\ the probability measure $P_{\tilde{X}}$ on $%
\Xi ^{\lbrack \tilde{q}]}$ defines a distribution $F_{\tilde{Z}_{+}}$ in $%
\mathbb{R}^{\tilde{q}}$ with support on $\mathbb{R}_{+}^{\tilde{q}},$ a
non-negative multivariate quadrant, given by the measurable mapping $\Xi
^{\lbrack \tilde{q}]}\rightarrow \mathbb{R}_{+}^{\tilde{q}}\,$\ with $\tilde{%
X}=\left( X^{1},\cdots,X^{\tilde{q}}\right) $ mapped into a random vector 
\begin{equation*}
\tilde{Z}_{+}=\left( \tilde{Z}_{+}^{1},\cdots,\tilde{Z}_{+}^{\tilde{q}%
}\right) ;\text{ }\tilde{Z}_{+}^{i}=\left\Vert \tilde{x}^{i}-\tilde{X}%
^{i}\right\Vert _{i}.
\end{equation*}%
Then $\tilde{X}=\tilde{x}$ transforms into $\tilde{Z}_{+}=0$. The $F_{\tilde{%
Z}}$ measure of the cuboid $C(0,r)$, with $r\in \mathbb{R}_{+}^{\tilde{q}}$,
is concentrated in the non-negative quadrant and is given by 
\begin{equation*}
F_{\tilde{Z}_{+}}\left( C(0,r)\right) =P_{\tilde{X}}(C\left( \tilde{x}%
,r\right) )=P_{\tilde{X}}\left( \prod_{i=1}^{\tilde{q}}B(\left\Vert \tilde{x}%
^{i}-\tilde{X}^{i}\right\Vert _{i}\leq r^{i})\right) .
\end{equation*}%
We can consider the symmetric function $\tilde{\psi} $ on $\mathbb{R}^{%
\tilde{q}}$ with support in $\left[ -1,1\right] ^{\tilde{q}},$ based on the
given $\tilde{\psi}_{+} $ defined as $\tilde{\psi}\left( v^{1},\cdots,v^{%
\tilde{q}}\right) =\tilde{\psi}_{+}\left( \left\Vert v^{1}\right\Vert
,\cdots,\left\Vert v^{\tilde{q}}\right\Vert \right) .$ Thus the symmetric
function $\tilde{\psi}\left( \frac{\tilde{x}-\tilde{X}}{\tilde{h}}\right) $
of Corollary A.1 can be written as $\tilde{\psi}\left( \frac{\tilde{Z}}{%
\tilde{h}}\right) $ and its values are given by $\tilde{\psi}\left( \frac{%
\tilde{Z}}{\tilde{h}}\right) =\tilde{\psi}_{+}\left( \frac{\tilde{Z}_{+}}{%
\tilde{h}}\right) .$ The result from Corollary A.1 then written with $\tilde{%
x}=0$ and the probability measure corresponding to the distribution $F_{%
\tilde{Z}_{+}}$, 
\begin{equation*}
P_{\tilde{Z}_{+}}\left( C(0,\lambda _{\xi }\circ \tilde{h})\right) =F_{%
\tilde{Z}_{+}}\left( C(0,\lambda _{\xi }\circ \tilde{h})\right),
\end{equation*}%
becomes 
\begin{equation*}
E\left( \tilde{\psi}\left( \tfrac{\tilde{Z}}{\tilde{h}}\right) \right)
=\sum_{\xi =0}^{2^{\tilde{q}}-1}\mathcal{I}_{\xi }\mathcal{(}0;\left[ 0,1%
\right] ^{q\left( \xi \right) },F_{\tilde{Z}_{+}}\left( C(0,\lambda _{\xi
}\circ \tilde{h})\right) ,\Delta _{1,\xi }\tilde{\psi}).
\end{equation*}%
Recognizing that $E_{\tilde{Z}_{+}}\left( \tilde{\psi}_{+}\left( \frac{%
\tilde{Z}_{+}}{\tilde{h}}\right) \right) =E_{\tilde{Z}_{+}}\left( \tilde{\psi%
}\left( \frac{\tilde{Z}_{+}}{\tilde{h}}\right) \right) $ and transforming
back to the original probability measure we get%
\begin{equation*}
E\left( \tilde{\psi}\left( \tfrac{\left\Vert \tilde{x}^{1}-\tilde{X}%
^{1}\right\Vert _{1}}{\tilde{h}^{1}},\cdots,\tfrac{\left\Vert \tilde{x}^{%
\tilde{q}}-\tilde{X}^{\tilde{q}}\right\Vert _{\tilde{q}}}{\tilde{h}^{\tilde{q%
}}}\right) \right) =\sum_{\xi =0}^{2^{\tilde{q}}-1}\mathcal{I}_{\xi }%
\mathcal{(}\tilde{x};\left[ 0,1\right] ^{q\left( \xi \right) },P_{\tilde{X}%
}\left( C(\tilde{x},\lambda _{\xi }\circ \tilde{h})\right) ,\Delta _{1,\xi }%
\tilde{\psi})
\end{equation*}%
as required.\hfill $\blacksquare $\medskip

A simplified expression for functions that take zero value on the boundary
is provided in the next corollary. We say that a function $\tilde{\psi}$ in
Lemmas A.1, A.2 and Corollary A.1 is zero on the boundary if for any $w$
with at least one $w^{l}$ with $\left\vert w^{l}\right\vert =1$ the value $%
\tilde{\psi}\left( w\right) \,\ $is zero. In this case the only non-zero
term in the sums in (\ref{for psi (a)}, \ref{for psi (a,b)}, \ref{for psi
(c)}) corresponds to $\xi =2^{\tilde{q}}-1$.

The notation $\frac{\left\Vert \tilde{x}-\tilde{X}\right\Vert }{\tilde{h}}$
can be used as an argument of $\tilde{\psi}$ under either the conditions of
the Corollary A.1 to Lemma A.1 or satisfies the conditions of Lemma A.2.

\begin{corollary}
\label{C.oneterm} If $\tilde{\psi}\left( W\right) $ satisfies Corollary A.1
to Lemma A.1 or satisfies Lemma A.2 and is zero at the boundary, then 
\begin{equation}
E\left( \tilde{\psi}\left( \tfrac{\left\Vert \tilde{x}-\tilde{X}\right\Vert 
}{\tilde{h}}\right) \right) =\int_{\left[ 0,1\right] ^{\tilde{q}}}P_{X}(C(%
\tilde{x},\tilde{h}\circ v))(-1)^{\tilde{q}}\partial \psi \left( v\right) dv
\label{one term}
\end{equation}
\end{corollary}

\noindent\textbf{Proof of Corollary A.2.}

\noindent (\ref{one term}) arises immediately from the sums in the
expressions for the expectation since any term $\xi$ for which $I_{\xi
}^{c}\neq \varnothing $ is zero. \hfill$\blacksquare $\vspace{.2in}

The lemma below gives a general expression for the upper bound which depends
on $h$ via the small cube probability.

\begin{lemma}
\label{C.phiUBound} Under Corollary A.1 of Lemma \ref{L.phiCondE} or Lemma %
\ref{L.phiCondE2} 
\begin{equation*}
\left\vert E\left( \tilde{\psi}\left( \tfrac{\left\Vert \tilde{x}-\tilde{X}%
\right\Vert }{\tilde{h}}\right) \right) \right\vert \leq P_{\tilde{X}}(C(%
\tilde{x},\tilde{h}))M_{E\psi }\left( \tilde{x}\right) ,
\end{equation*}%
where 
\begin{equation*}
M_{E\psi }\left( x\right) =2^{\tilde{q}}\underset{0\leq \xi \leq 2^{\tilde{q}%
}-1}{\max }\int_{[0,1]^{q\left( \xi \right) }}\left\vert \Delta _{1,\xi }%
\tilde{\psi}\left( v\right) \right\vert dv(\xi ) .
\end{equation*}
\end{lemma}

The expressions simplify for functions that take zero value on the boundary.%
\vspace{.2in}

\noindent\textbf{Proof of Lemma A.3.}

\noindent Since the components of $\lambda _{\xi }\circ \tilde{h}$ are
between zero and the components of $\tilde{h},$ 
\begin{equation*}
P_{\tilde{X}}\left( C(\tilde{x},\lambda _{\xi }\circ \tilde{h})\right) \leq
P_{\tilde{X}}\left( C(\tilde{x},\tilde{h})\right) .
\end{equation*}%
We detail the bound for the expression of Lemma A.1; similar derivations
provide it under Lemma A.2. Consider the expression

\begin{equation*}
\sum_{\xi =0}^{2^{\tilde{q}}-1}\int_{\left[ -1,1\right] ^{q\left( \xi
\right) }}F_{\tilde{X}}(\tilde{x}-\lambda _{\xi }\circ \tilde{h},\tilde{x}+%
\tilde{h})\prod_{j\in I_{\xi }^{c}}^{{}}\delta \left( v^{j}=-1\right)
\prod_{j\in I_{\xi }}^{{}}\left( -\partial _{j}\right) \tilde{\psi}\left(
v\right) dv\left( \xi \right) ;
\end{equation*}%
this can be bounded by 
\begin{eqnarray*}
&&\sum_{\xi =0}^{2^{\tilde{q}}-1}\int_{\left[ -1,1\right] ^{\tilde{q}\left(
\xi \right) }}F_{\tilde{X}}(\tilde{x}-\tilde{h},\tilde{x}+\tilde{h}%
)\left\vert \prod_{j\in I_{\xi }^{c}}^{{}}\delta \left( v^{j}=-1\right)
\prod_{j\in I_{\xi }}^{{}}\left( -\partial _{j}\right) \tilde{\psi}\left(
v\right) \right\vert dv\left( \xi \right) \\
&\leq &F_{\tilde{X}}(\tilde{x}-\tilde{h},\tilde{x}+\tilde{h})2^{\tilde{q}}%
\underset{0\leq \xi \leq 2^{\tilde{q}}-1}{\max }\int_{\left[ -1,1\right]
^{q\left( \xi \right) }}\left\vert \prod_{j\in I_{\xi }^{c}}^{{}}\delta
\left( v^{j}=-1\right)\prod_{j\in I_{\xi }}^{{}}\left( -\partial _{j}\right) 
\tilde{\psi}\left( v\right) \right\vert dv\left( \xi \right) .
\end{eqnarray*}%
Recall that $F_{\tilde{X}}(\tilde{x}-\tilde{h},\tilde{x}+\tilde{h})=P_{%
\tilde{X}}(C(\tilde{x},\tilde{h})).$ Thus under Lemma A.1 or Lemma A.2 
\begin{equation*}
\left\vert E\left( \tilde{\psi}\left( \tfrac{\left\Vert \tilde{x}-\tilde{X}%
\right\Vert }{\tilde{h}}\right) \right) \right\vert \leq P_{\tilde{X}}(C(%
\tilde{x},\tilde{h}))M_{E\psi }\left( \tilde{x}\right) ,
\end{equation*}%
where

\begin{equation*}
M_{E\psi }\left( \tilde{x}\right) =2^{\tilde{q}}\underset{0\leq \xi \leq 2^{%
\tilde{q}}-1}{\max }\int_{[0,1]^{q\left( \xi \right) }}\left\vert \Delta
_{1,\xi }\tilde{\psi}\left( v\right) \right\vert dv(\xi ).
\end{equation*}

\textcolor{white}{.}\hfill $\blacksquare $\medskip

In part (a) of the next lemma we provide a condition on the probability
measure that ensures the existence of a positive lower bound for such
functions; the bound is expressed via the small cube probability. For
functions in Lemmas A.1, A.2 and Corollary A.1, that are not zero on the
boundary, no restrictions on the probability measure are required; lower
bounds in (b) do not require any additional conditions on the probability
measure.\medskip

\begin{lemma}
\label{L.phiBounds} If $\tilde{\psi}\left( w\right) $ is non-increasing for $%
w>0$ and $\tilde{\psi}\left( w\right) $ satisfies Corollary A.1 to Lemma A.1
or satisfies Lemma A.2

\noindent (a) {If $\tilde{\psi}$ is zero at the boundary and the measure
satisfies 
\begin{equation*}
\frac{P_{\tilde{X}}(C(\tilde{x},\tilde{h}))}{P_{\tilde{X}}(C(\tilde{x}%
,\varepsilon \tilde{h}))}<C_{\varepsilon }<\infty ,
\end{equation*}%
for some $0<\varepsilon <1$ then there is the lower bound}: 
\begin{equation*}
E\left( \tilde{\psi}\left( \tfrac{\left\Vert \tilde{x}-\tilde{X}\right\Vert 
}{\tilde{h}}\right) \right) \geq P_{\tilde{X}}(C(\tilde{x},\tilde{h}%
))L_{E\psi }
\end{equation*}%
where 
\begin{equation*}
L_{E\psi }=\frac{1}{C_{\varepsilon }}\int_{\left[ \varepsilon ,1\right] ^{%
\tilde{q}}}(-1)^{\tilde{q}}\partial \tilde{\psi}\left( v\right) dv>0.
\end{equation*}

\noindent (b) {If $\psi $ satisfies 
\begin{equation*}
\psi \left( 1,\cdots ,1\right) >0,
\end{equation*}%
then there is the lower bound}: 
\begin{equation*}
E\left( \tilde{\psi}\left( \tfrac{\left\Vert \tilde{x}-\tilde{X}\right\Vert 
}{\tilde{h}}\right) \right) \geq P_{\tilde{X}}(C(\tilde{x},\tilde{h}%
))L_{E\psi }
\end{equation*}%
where 
\begin{equation*}
L_{E\psi }=\tilde{\psi}\left( 1,\cdots ,1\right) >0
\end{equation*}
\end{lemma}

\vspace{.2in}

\noindent \textbf{Proof of Lemma A.4.}\newline
\noindent (a) The lower bound follows after substituting the lower bound, $%
C_{\varepsilon }P_{\tilde{X}}(C(\tilde{x},\tilde{h}))$ for $P_{\tilde{X}}(C(%
\tilde{x},\varepsilon \tilde{h}))$ in (\ref{one term}).\newline
\noindent (b) The term $\tilde{\psi}\left( 1,..,1\right) >0$ appears in the
moment expression for the expectation which appears for $\xi =0$ with $%
I_{0}=\emptyset .$ As all other terms are non-negative, this term is
sufficient for the lower bound.$\hfill \blacksquare $\medskip

Next, consider moments for a product of the local random function $\tilde{%
\psi}\left( w\right) $ with some continuous function $\tilde{g}:C( \tilde{x},%
\tilde{h}) \rightarrow \mathbb{R}.$ Consider here the sets $\widetilde{C}$
that are either small cubes or unions of small cubes in $C( \tilde{x},\tilde{%
h}) $. Define for a continuous function $\tilde{g} $ (bounded on $\tilde{C})$%
\begin{equation}
\Omega _{\tilde{g}}(\widetilde{C})=\int\limits_{\widetilde{C}}\tilde{g}%
(z)dP_{\tilde{X}}(z)  \label{omega2}
\end{equation}

\noindent The next lemma gives the expression for the moment of the product
and bounds on the moment. The upper bound for this moment is expressed as a
multiple of the small cube probability. The lower bound can be defined
similarly, and need not be positive.

\begin{lemma}
\label{L.gKm bounds} (a) Under the conditions of Corollary A.1 of Lemma A.1
or Lemma A.2 for $\tilde{\psi}$, for a bounded continuous function $\tilde{g}
$ the moment 
\begin{equation*}
E\left[ \tilde{g}(\tilde{X})\tilde{\psi}\left( \tfrac{\left\Vert \tilde{x}-%
\tilde{X}\right\Vert }{\tilde{h}}\right) \right] =\sum_{\xi =0}^{2^{\tilde{q}%
}-1}\mathcal{I}_{\xi }\mathcal{(}\tilde{x};\left[ 0,1\right] ^{{q}\left( \xi
\right) },\Omega _{\tilde{g}}\left( C(\tilde{x},\lambda _{\xi }\circ \tilde{h%
})\right) ,\Delta _{1,\xi }\tilde{\psi})
\end{equation*}%
with $\Omega _{\tilde{g}}\left( C(\tilde{x},\lambda _{\xi }\circ \tilde{h}%
)\right) $ given by (\ref{omega2}) with $\widetilde{C}=C(\tilde{x},\lambda
_{\xi }\circ \tilde{h})$

\noindent (b) The moment is bounded 
\begin{equation*}
\left\vert E\left[ \tilde{g}(\tilde{X})\tilde{\psi}\left( \tfrac{\left\Vert 
\tilde{x}-\tilde{X}\right\Vert }{\tilde{h}}\right) \right] \right\vert \leq
M_{Eg\psi }P_{\tilde{X}}\left( C(\tilde{x},\tilde{h})\right) ,
\end{equation*}%
with 
\begin{equation*}
M_{E{g}{\psi }}={\underset{x\in C(\tilde{x},\tilde{h})}{\sup }}\left\vert 
\tilde{g}\left( x\right) \right\vert M_{E\psi };\qquad
\end{equation*}

\noindent (c) Under the conditions of (a) or (b) of Lemma A.4 
\begin{equation*}
\left\vert E\left[ \tilde{g}(X)\tilde{\psi}\left( \tfrac{\left\Vert \tilde{x}%
-\tilde{X}\right\Vert }{\tilde{h}}\right) \right] \right\vert \geq L_{Eg\psi
}P_{\tilde{X}}\left( C(\tilde{x},\tilde{h})\right) ,
\end{equation*}%
with 
\begin{equation*}
L_{Eg\psi }=\max \left\{ 0,\left( {\underset{x\in C(\tilde{x},\tilde{h})}{%
\inf }}\tilde{g}\left( {x}\right) \right) L_{E\psi }\right\}
\end{equation*}%
where for (a)%
\begin{equation*}
L_{E\psi }=\frac{1}{C_{\varepsilon }}\int_{\left[ \varepsilon ,1\right] ^{%
\tilde{q}}}(-1)^{\tilde{q}}\partial \tilde{\psi}\left( v\right) dv>0;
\end{equation*}%
under (b) 
\begin{equation*}
L_{E\psi }=\tilde{\psi}\left( 1,\cdots ,1\right) >0
\end{equation*}
\end{lemma}

When {${\underset{x \in C(\tilde{x},\tilde{h})}{\inf}} \tilde{g}\left(
x\right) >0,$ the lower bound is positive.}\vspace{.2in}

\noindent \textbf{Proof of Lemma A.5.}\newline
\noindent (a) The derivation is identical to that in Lemma A.1 with the only
difference that $dF_{\tilde{X}}$ or $dP_{\tilde{X}}$ is replaced by $d\Omega
_{\tilde{X}}$ in all the derivations providing the result.

\noindent (b) The upper bound follows from the boundedness of the function $%
g(\tilde{X})$ around $\tilde{x}.$

\noindent (c) The lower bound in the case of positive $g(\tilde{X})$
follows; when $g(\tilde{X})$ can take non-positive values, the lower bound
on the absolute value of the moment is zero.\hfill$\blacksquare $

\section{Proofs of lemmas and theorems}

In this appendix we provide proofs of the results from the paper and a
remark on bandwidth selection with mass points.

\subsection{Some preliminaries: moments and moment bounds}

\renewcommand{\theequation}{B.\arabic{equation}} \setcounter{equation}{0}%
\renewcommand{\thelemma}{B.\arabic{lemma}} \setcounter{lemma}{0} %
\renewcommand{\thecorollary}{B.\arabic{corollary}} \setcounter{corollary}{0} %
\renewcommand{\theproposition}{B.\arabic{proposition}} %
\setcounter{proposition}{0}

The proofs of the lemmas and theorems apply the results in Appendix A.%
\newline

The next lemma and its corollary provide bounds used in the proofs below,
which implement the results {from} Appendix A.

\begin{lemma}
\label{L.Km bounds} {Given Assumption 1, 2(a-c) and 3}, the moment for $\tilde{K}\left( \frac{\left\Vert 
\tilde{x}-\tilde{X}\right\Vert }{\tilde{h}}\right) $ (with $\tilde{K}$ given
by either $K,$ or $K\times K$) satisfies 
\begin{equation*}
L_{EK^{m}}P_{\tilde{X}}\left( C(\tilde{x},\tilde{h})\right) \leq E\left( {%
\tilde{K}}^{m}\left( \tfrac{\left\Vert \tilde{x}-\tilde{X}\right\Vert }{%
\tilde{h}}\right) \right) \leq M_{EK^{m}}P_{\tilde{X}}\left( C(\tilde{x},%
\tilde{h})\right) .
\end{equation*}%
with

\begin{enumerate}
\item[(a)] 
\begin{equation*}
M_{EK^{m}}=2^{\tilde{q}}\underset{0\leq \xi \leq 2^{\tilde{q}-1}}{\max }%
\int_{\left[ 0,1\right] ^{q\left( \psi \right) }}\left\vert \Delta _{1,\xi }{%
\tilde{K}}^{m}\left( v\right) \right\vert dv\left( \xi \right) ;
\end{equation*}

\item[(b)] {when Assumption 5 holds, for some $0<\varepsilon <1$} 
\begin{equation*}
L_{EK^{m}}=\frac{1}{C_{\varepsilon }}\int_{[\varepsilon ,1]^{\tilde{q}%
}}(-1)^{\tilde{q}}\partial {\tilde{K}}^{m}\left( v\right) dv
\end{equation*}

\item[(c)] when instead, the kernel is type I (add Assumption 2%
(d)) 
\begin{equation*}
L_{EK^{m}}={\tilde{K}}^{m}\left( \iota \right) >0.\qquad \qquad \qquad
\end{equation*}%
with $\iota =\left( 1,\cdots,1\right) ^{T}.$ 
\end{enumerate}
\end{lemma}

For odd $m$ the sign of $\partial {K}^{m}\left( v\right) $ is $(-1)^{\tilde{q%
}}$ for any $v\geq 0,$ thus the lower bound $L_{EK^{m}}$ is always positive.
This is important since such functions appear in the denominator of the
estimator.\vspace{.2in}

\noindent \textbf{Proof of Lemma \ref{L.Km bounds}.}\newline
(a) The upper bound is obtained by substituting ${\tilde{K}}^{m} $ for $%
\tilde{\psi} $ in Lemma A.3. \newline
(b) The bound follows from Lemma A.4 (a). \newline
(c) The bound follows from Lemma A.4 (b).\hfill$\blacksquare $\medskip

Recall, for some bounded continuous function $\tilde{g}( \tilde{X}) $, we
defined 
\begin{equation}
\Omega _{\tilde{g}}(C(\tilde{x},\tilde{h}))=\int\limits_{C(\tilde{x},\tilde{h%
})}\tilde{g}(z)dP_{\tilde{X}}(z).  \label{omega}
\end{equation}

\begin{corollary}
\label{C.gKm bounds} {Under the conditions of Lemma \ref{L.Km bounds} for a
bounded function $\tilde{g}$ }the moment%
\begin{equation*}
E\left[ \tilde{g}(\tilde{X})\tilde{K}^{m}\left( \tfrac{\left\Vert \tilde{x}-%
\tilde{X}\right\Vert }{\tilde{h}}\right) \right] =\int_{\left[ 0,1\right] ^{%
\tilde{q}}}\Omega _{\tilde{g}}\left( C(\tilde{x},\tilde{h}\circ v)\right)
(-1)^{\tilde{q}}\partial \tilde{K}^{m}\left( v\right) dv ,
\end{equation*}%
%
with $\tilde{h}\circ v=(\tilde{h}^1v^1,\cdots,\tilde{h}^qv^q)$, is bounded
as 
\begin{equation*}
L_{EgK^{m}}P_{\tilde{X}}\left( C(\tilde{x},\tilde{h})\right) \leq \left\vert
E\left[ \tilde{g}(\tilde{X}){\tilde{K}}^{m}\left( \tfrac{\left\Vert \tilde{x}%
-\tilde{X}\right\Vert }{\tilde{h}}\right) \right] \right\vert \leq
M_{EgK^{m}}P_{\tilde{X}}\left( C(\tilde{x},\tilde{h})\right) ,
\end{equation*}%
where 
\begin{equation*}
L_{EK^{m}g}=\max \{0,{\underset{x\in C(\tilde{x},\tilde{h})}{\inf }}\tilde{g}%
(x)L_{EK^{m}}\};\ M_{EK^{m}}={\underset{x\in C(\tilde{x},\tilde{h})}{\sup }}%
\left\vert \tilde{g}\left( x\right) M_{EK^{m}}\right\vert .
\end{equation*}

\noindent If $\tilde{g}\left( z\right) $ is a continuous function that is
positive at all $x$ in support of $P_{X}$ the lower bound $L_{EK^{m}g}$ is
positive.
\end{corollary}

\noindent The corollary follows directly from Lemma {\ref{L.Km bounds}. }%
Note that here the lower bound does not have to be positive. \medskip

Next, we provide preliminary results on moments of $B(x)$ and $A(x)$ of (5).

\begin{lemma}
\label{PrelMom} (Preliminary for moments) {Under either of the following
sets of conditions (i) Assumptions 1-4 and 6 or (ii) Assumptions 1, 2(a-c), 3-6}

\begin{eqnarray}
&&L_{EK}P_{X}\left( C\left( x,h\right) \right) \leq EB\left( x\right) \leq
M_{EK}P_{X}\left( C\left( x,h\right) \right) ;  \label{Bounds EB} \\
&&varB\left( x\right) =\frac{1}{n}{E\left[ {K}^{2}\left( \frac{\left\Vert
x-X\right\Vert }{h}\right) \right] }\left( 1+o\left( 1\right) \right); 
\notag
\end{eqnarray}%
\begin{equation}
0<{\frac{1}{n}}L_{varK}P_{X}\left( C\left( x,h\right) \right) \leq
varB\left( x\right) \leq {\frac{1}{n}}M_{varK}P_{X}\left( C\left( x,h\right)
\right) <\infty  \label{Var B}
\end{equation}%
and 
\begin{eqnarray}
EA\left( x\right) &=&E\left[ m\left( X\right) {K}\left( \frac{\left\Vert
x-X\right\Vert }{h}\right) \right] \leq M_{EmK}P_{X}\left( C\left(
x,h\right) \right) ;  \label{EA} \\
varA\left( x\right) &=&\frac{1}{n}E\left[ {K}^{2}\left( \frac{\left\Vert
x-X\right\Vert }{h}\right) \left( \mu _{2}\left( X\right) +m\left( X\right)
^{2} \right) \right] \left( 1+o\left( 1\right) \right);  \notag
\end{eqnarray}%
\begin{equation}
0<{\frac{1}{n}}L_{varA}P_{X}\left( C\left( x,h\right) \right) \leq
varA\left( x\right) \leq {\frac{1}{n}}M_{varA}P_{X}\left( C\left( x,h\right)
\right) <\infty  \label{var A}
\end{equation}
\end{lemma}

\vspace{.2in}

\noindent\textbf{Proof of Lemma \ref{PrelMom}}.

\noindent The results for the first moments follow immediately from Lemma {%
\ref{L.Km bounds}} and Corollary {\ref{C.gKm bounds}}. To establish the
results for the variances, consider a generic random variable $Q_{i},$ that
could be either {$Q_{i}\left( 1\right) ={K}\left( \frac{\left\Vert
x-X_i\right\Vert }{h}\right) ,$ or $Q_{i}\left( 2\right) ={K}\left( \frac{%
\left\Vert x-X_i\right\Vert }{h}\right) Y_{i}.$} Our assumptions imply that $%
\left\{ Q_{i}\left( \cdot \right) \right\} $ is a stationary mixing sequence
with the mixing coefficient $\alpha \left( s\right) .$ We show 
\begin{equation}
E\left\vert Q\right\vert ^{2+\zeta }\leq M_{\left\vert Q\right\vert
^{2+\zeta }}P_{X}\left( C\left( x,h\right) \right) .  \label{EQ-i}
\end{equation}%
Indeed for $Q_{i}\left( 2\right) $ first $E\left\vert Q_{i}\left( 2\right)
\right\vert ^{2+\zeta }=$ 
\begin{equation*}
E\left( \left\vert {K}\left( \frac{\left\Vert x-X_{i}\right\Vert }{h}\right)
Y_{i}\right\vert ^{2+\zeta }\right) = E \left( \left\vert K \left( \frac{%
\left\Vert x-X_{i}\right\Vert }{h} \right) \left( m(X_i) + u_i \right)
\right\vert ^{2+\zeta } \right) . 
\end{equation*}%
which is bounded by 
\begin{equation*}
2^{1+\zeta} \left[ E \left( K^{2+\zeta} \left( \frac{\left\Vert
x-X_{i}\right\Vert }{h} \right) \left\vert m(X_i) \right\vert ^{2+\zeta }
\right) + E \left( K^{2+\zeta} \left( \frac{\left\Vert x-X_{i} \right\Vert }{%
h} \right) \left\vert u_i \right\vert^{2+\zeta} \right) \right] 
\end{equation*}
using the $C_r$ inequality. Then by Assumption 3(c) the bound
follows from the results developed in Lemma A.5. For $Q_{i}\left( 1\right) $
(\ref{EQ-i}) follows immediately.\medskip

Write 
\begin{eqnarray*}
var\left( \frac{1}{n}\sum_{i=1}^{n}Q_{i}\right) &=&\frac{1}{n^{2}}%
\sum_{i=1}^{n}varQ_{i}+\frac{2}{n^{2}}{\sum_{i=1}^{n-1}}%
\sum_{s=1}^{n-i}cov(Q_{i},Q_{i+s}) \\
&=&\frac{1}{n}varQ_{1}+\frac{2}{n^{2}}\sum_{s=1}^{n-1}\left( n-s\right)
cov(Q_{1},Q_{1+s}).
\end{eqnarray*}%
By the results from Lemma {\ref{L.Km bounds}.} and Corollary {\ref{C.gKm
bounds}} we have that 
\begin{equation*}
0<\frac{1}{n}L_{varQ}P_{X}\left( C\left( x,h\right) \right) <\frac{1}{n}%
varQ_{1}<\frac{1}{n}M_{varQ}P_{X}\left( C\left( x,h\right) \right) ,
\end{equation*}%
where the positivity of the lower bound follows from the assumptions,
including $\mu _{2}\left( x\right) >0.$ Thus $\frac{1}{n}varQ_{1}\approx
O\left( \frac{1}{n}P_{X}\left( C\left( x,h\right) \right) \right) .$\medskip

It remains to show that the expression with the covariances goes to zero
faster. Following the usual approach (e.g., as in Masry, 2005) consider some
integer $u_{n}$ between $1$ and $n$ and {use the partitioned sum} 
\begin{equation*}
\frac{1}{n^{2}}\sum_{s=1}^{u_{n}-1}\left( n-s\right) cov(Q_{1},Q_{1+s})+%
\frac{1}{n^{2}}\sum_{s=u_{n}}^{n-1}\left( n-s\right)
cov(Q_{1},Q_{1+s})=E_{1}+E_{2} .
\end{equation*}%
Using Assumption 6, $E_{1}$ can be bounded 
\begin{equation*}
E_{1}\leq \frac{u_{n}}{n}M_{varQ}P_{\tilde{X}}\left( C\left( x,h\right)
\times C\left( x,h\right) \right) \leq \frac{u_{n}}{n}%
M_{varQ}M_{FF}P_{X}(C(x,h))^{2},
\end{equation*}%
which by appropriate selection of $u_{n}$ specified below is of smaller
order. To bound $E_{2}$ we utilize Davydov's Lemma {(1968)} which provides 
\begin{equation*}
\left\vert cov(Q_{1},Q_{1+s})\right\vert \leq 8\left( E\left\vert
Q\right\vert ^{2+\zeta }\right) ^{\frac{2}{2+\zeta }}s^{-\kappa \frac{\zeta 
}{2+\zeta }}.
\end{equation*}%
As $E_{2}$ can then be bounded by a geometric progression, it is dominated
by $\left\vert n\;cov(Q_{1},Q_{1+u_{n}})\right\vert $ and hence 
\begin{eqnarray*}
E_{2} &\leq &nu_{n}^{-k\frac{\zeta }{2+\zeta }}8\left( E\left\vert
Q\right\vert ^{2+\zeta }\right) ^{\frac{2}{2+\zeta }} \\
&\leq &M_{covQ}nu_{n}^{-k\frac{\zeta }{2+\zeta }}P_{X}\left( C\left(
x,h\right) \right) ^{\frac{2}{2+\zeta }}.
\end{eqnarray*}

By setting $u_{n}=\left[ P_{X}\left( C\left( x,h\right) \right) \right] ^{-%
\frac{2(2+\zeta )}{\kappa \zeta }}$ we can bound the expression with the
covariances by 
\begin{eqnarray*}
&&\frac{2}{n^{2}}\sum_{s=1}^{n-1}\left( n-s\right) cov(Q_{1},Q_{1+s}) \\
&\leq &M\left\{ \frac{1}{n}\left[ P_{X}\left( C\left( x,h\right) \right) %
\right] ^{2-\frac{2(2+\zeta )}{\kappa \zeta }}+\frac{1}{n}P_{X}\left(
C\left( x,h\right) \right) ^{\frac{2}{2+\zeta }+2}\right\}
\end{eqnarray*}%
and since $2-\frac{2(2+\zeta )}{\kappa \zeta }>1$ the first component is $%
o\left( \frac{1}{n}P_{X}\left( C\left( x,h\right) \right) \right) ;$ the
second component is of smaller order.

Thus for $varA\left( x\right) $ we have that 
\begin{equation*}
varA\left( x\right) {=\frac{1}{n}var\left( {K}\left( \frac{\left\Vert
x-X_1\right\Vert }{h}\right) Y_{1}\right)} \left( 1+o\left( 1\right) \right).
\end{equation*}%
Consider $\frac{1}{n}var\left( {K}\left( \frac{\left\Vert x-X_1\right\Vert }{%
h}\right) Y_{1}\right) .$ For $g\left( x\right) =m\left( x\right) ^{2}+\mu
_{2}\left( x\right) $ by the rates from {Lemma \ref{L.Km bounds} }we have 
\begin{eqnarray*}
&&var\left( {K}\left( \frac{\left\Vert x-X_1\right\Vert }{h}\right)
Y_{1}\right) =var\left( {K}\left( \frac{\left\Vert x-X_1\right\Vert }{h}%
\right) (u_1+m\left( X_1\right) \right) \\
&=&E\left( {K}^2\left( \frac{\left\Vert x-X_1\right\Vert }{h}\right)
(u_1+m\left( X_1\right) )^2 \right)-\left[ E{K}\left( \frac{\left\Vert
x-X_1\right\Vert }{h}\right) (u_1+m\left( X_1\right) )\right] ^{2} \\
&=&E\left( K^2 \left( \frac{\left\Vert x-X_1\right\Vert }{h}\right)u_1^2
\right) +E\left( K^2 \left( \frac{\left\Vert x-X_1\right\Vert }{h}\right)
m^2\left( X_1\right) \right)-\left[ E\left( m\left( X_1\right) K \left( 
\frac{\left\Vert x-X_1\right\Vert }{h}\right)\right) \right] ^{2} \\
&=& \mu_2(x) E\left( K^2 \left( \frac{\left\Vert x-X_1\right\Vert }{h}%
\right) \right) + m(x)^2 var \left( K \left( \frac{\left\Vert
x-X_1\right\Vert }{h}\right) \right);
\end{eqnarray*}%
by the corollary to {Lemma \ref{L.Km bounds} the leading term of }$%
varA\left( x\right) $ then {has a positive lower bound (}with some $%
0<\varepsilon <1)$ {as }$n\rightarrow \infty ,h\rightarrow 0$.{\ }%
\begin{equation*}
0<{\frac{1}{n}}L_{K^{2}g}P_{X}\left( C\left( x,h\right) \right) \left(
1-\varepsilon \right) \leq varA\left( x\right) \leq {\frac{1}{n}}%
M_{K^{2}g}P_{X}\left( C\left( x,h\right) \right) \left( 1-\varepsilon
\right) <\infty .
\end{equation*}%
\textcolor{white}{.}\hfill $\blacksquare $

\bigskip

\subsection{Proof of results from Section 2}

Following Ferraty et al. (2007, p 270) write for all $\varepsilon >0$ 
\begin{equation*}
\tau _{h}\left( \varepsilon \right) =\frac{P_{X}\left( B\left( x,\varepsilon
h\right) \right) }{P_{X}\left( B\left( x,h\right) \right) }.
\end{equation*}%
{Assumption $H_{3}$ in Ferraty et al. (2007) requires that as $h\rightarrow
0 $ the functions }${\tau }${$_{h}\left( \varepsilon \right) $ converge: 
\begin{equation}
\tau _{h}\left( \varepsilon \right) \rightarrow \tau _{0}\left( \varepsilon
\right) \quad \text{for every }\varepsilon >0,  \label{Fetal
cond}
\end{equation}%
where the limit could be a regular function or a delta-function $\delta
\left( \varepsilon =1\right) .$ Proposition 1 of that paper provides
examples of distributions for which $\tau _{0}\left( \varepsilon \right) >0.$
A similar condition (Lemma 4.4) in Ferraty and Vieu (2006) is that 
\begin{equation}
\int_{0}^{v}P_{X}\left( B\left( x,u\right) \right) du>CvP_{X}\left( B\left(
x,v\right) \right)  \label{FV cond}
\end{equation}%
holds for some $C>0,$ $v_{0}>0$ and any $v<v_{0}.$ } The lemma below shows
that condition (6) holds whenever (\ref{Fetal cond}) or (\ref{FV cond}) hold.

\begin{lemma}
\label{L.CompareFV} Condition (6) of Assumption 5
is a necessary condition (a) for (\ref{Fetal cond}) with $\tau _{0}\left(
\varepsilon \right) >0$ and (b) for (\ref{FV cond}) to hold.
\end{lemma}

\vspace{.2in}

\noindent \textbf{Proof of Lemma \ref{L.CompareFV}.}

\noindent (a) Given (\ref{Fetal cond}) with $\tau _{0}\left( \varepsilon
\right) =C\left( \varepsilon \right) >0$ we have that for small enough $h$ 
\begin{equation*}
P_{X}\left( B\left( x,\varepsilon h\right) \right) >C_{\varepsilon
}P_{X}\left( B\left( x,h\right) \right) ,
\end{equation*}%
where $\tilde{C}=C\left( \varepsilon \right) +\delta ,$ $\delta >0$ and (6) holds.\medskip

\noindent (b) Since $P_{X}\left( B\left( x,u\right) \right) $ is
non-decreasing and continuous in $u$ 
\begin{equation*}
\int_{0}^{v}P_{X}\left( B\left( x,u\right) \right) du=P_{X}\left( B\left( x,%
\tilde{s}v\right) \right) v,
\end{equation*}%
where $\tilde{s}<1.$ From (\ref{FV cond}) it follows that $P_{X}\left(
B\left( x,\tilde{s}v\right) \right) v>CvP_{X}\left( B\left( x,v\right)
\right) ,$ thus for $\varepsilon =\tilde{s}$ (6)
holds.\hfill$\blacksquare$\vspace{0.2in}

The following proposition shows that (7) applies to a finite
mixture of $L$ Ahlfors regular distributions that each satisfy (7) with some constant $s_{l},l=1,...,L$ and bounds $L_{s_{l}}$ and $%
M_{s_{l}}.$

\begin{proposition}
\label{mixture} [Mixture] Suppose that $P_{X}$ corresponds to a finite
mixture of probability measures that are A-r for different constants $%
s_{l}:P_{X}=\sum_{l=1}^{d}\alpha _{l}P_{0}\left( x_{l}\right)
+\sum_{l=d+1}^{L}\alpha _{l}P_{s_{l}}$ with $P_{0}\left( x_{l}\right) $
representing a mass point at $x_{l}$ and $s_{l}=0$ for $\ l=1,...d;$ $%
0<s_{d+1}\leq ...\leq s_{L}\leq 1,\ \alpha _{l}\geq 0,$ and $%
\sum_{l=1}^{L}\alpha _{l}=1$ and with $0<L_{s_{l}}<M_{s_{l}}<\infty ,$ for
each $P_{s_{l}}.$ Then given $x$ in the support of $P_{X}$ there exists
\thinspace $H\left( x\right) ,L_{P}\left( x\right) ,M_{P}\left( x\right) $
such that condition (7) holds.
\end{proposition}

\noindent \textbf{Proof of Proposition \ref{mixture}.}

\noindent Since $x$ is assumed to be a point of support of $P_{X}$ there is
a subset of $K\subset \left\{ 1,...,L\right\} $ such that for $l\in K$ and
any $h$ with $\underline{h}>0$ we have that $P_{s_{l}}\left( C\left(
x,h\right) \right) >0.$

Denote by $s\left( x\right) $ the value $\underset{l\in K}{\min }\left\{
s_{l}\right\} .$ Consider two possibilities: (a) $s\left( x\right) =0$ and
(b) $s\left( x\right) >0.$

\noindent (a) Then $l\leq d$ and $x=x_{l}.$ Set $H\left( x\right) <\underset{%
l^{\prime }\in \left\{ 1,...,d\right\} ;l^{\prime }\neq l;1\leq i\leq q}{%
\min }\left\{ \left\Vert x_{l}^{i}-x_{l^{\prime }}^{i}\right\Vert \right\} .$
Then with $L_{P}\left( x\right) =\alpha _{l}>0$ and $M_{P}\left( x\right)
=\Sigma $ $_{l=1}^{L}M_{s_{l}}\left( x_{l}\right) $ and $s\left( x\right) =0$
we obtain (7).

\noindent (b) Consider the subset $\left\{ l_{1},...,l_{K}\right\} $ of $%
\left\{ d+1,...,L\right\} $ such that $P_{s_{l_{t}}}\left( C\left(
x,h\right) \right) >0$ for every $h$ if and only if $l_{t}\in \left\{
l_{1},...,l_{K}\right\} .$ Define $s\left( x\right) =\underset{l_{t}\in
\left\{ l_{1},...,l_{K}\right\} }{\min }\left\{ s_{lt}\right\} .$ We show
that there exists $H:P_{s_{l_{t}}}\left( C\left( x,H\iota \right) \right) >0$
such that for any $\tilde{x}\in C\left( x,H\iota \right) $ if $P_{s^{\prime
}}\left( C\left( \tilde{x},h\right) \right) >0$ for every $h,$ then $%
s^{\prime }\geq s\left( x\right) .$ The proof is by contradiction.

Suppose that for any $\tilde{H},$ no matter how small $( \tilde{H}%
\rightarrow 0) $ we get some $\tilde{x}( \tilde{H}) \in C( x,\tilde{H}\iota
) $ with $s\left( \tilde{x}\right) <s\left( x\right) .$ Without loss of
generality assume that $s\left( \tilde{x}\right) $ is constant (as the set
of values it can take is finite).

Define $\tilde{h}=\left( \tilde{h}^{1},...,\tilde{h}^{q}\right) $ with $%
\tilde{h}^{i}=\min \left\{ \left\Vert x^{i}-\tilde{x}^{i}\right\Vert
,\left\Vert \tilde{x}^{i}-\tilde{H}\right\Vert \right\} .$ Then \thinspace $%
C( x,\tilde{H}\iota) \supset C( \tilde{x},\tilde{h}) $ and since by
assumption $P_{s\left( \tilde{x}\right) }(C( \tilde{x},\tilde{h}) ) >0$ it
follows that $P_{s\left( \tilde{x}\right) }( C( x,\tilde{H}\iota)) >0$ for
any $\tilde{H},$ which contradicts the assumption on $s\left( x\right) .$

We set $H\left( x\right) $ such that $C\left( x,H\iota \right) $ does not
contain any $\tilde{x}:s\left( \tilde{x}\right) <s\left( x\right) .$ Then 
\begin{equation*}
P_X\left( C\left( x,H\iota \right) \right) =\sum_{s_{l}=s\left( x\right)
}\alpha _{s_{l}}P_{s_{l}}\left( C\left( x,H\iota \right) \right)
+\sum_{s_{l}>s\left( x\right) }\alpha _{s_{l}}P_{s_{l}}\left( C\left(
x,H\iota \right) \right) .
\end{equation*}

$L_{P}\left( x\right) =\sum _{s_{l=s\left( x\right)
}}\alpha_{s_{l}}L_{s_{l}}I\left( P_{s_{l}}\left( C\left( x,H\iota \right)
\right)>0\right) ;$ this bound is positive since at least one of the $%
P_{s_{l}}\left( C\left( x,H\iota \right) \right) >0.$ Finally, $%
M_{P}\left(x\right) =\Sigma _{l=d+1}^{L}M_{s_{l}};$ indeed with $%
H^{s_{l}q}\leq H^{s\left( x\right) q}$ for $s_{l}\geq s\left( x\right) $ 
\begin{equation*}
P_X\left( C\left( x,H\iota \right) \right) \leq
\sum_{l=d+1}^{L}M_{s_{l}}H^{s\left( x\right) q}.
\end{equation*}%
Condition (7) is then satisfied for $s\left( x\right).$%
\hfill $\blacksquare $

\subsection{Example where the limit for $\frac{EK(W_i(x))}{P_X(C(x,h))}$ as $%
h \rightarrow 0$ does not exist.}

\noindent Without loss of generality assume a univariate setting and $x=0.$

Consider a sequence $a_{i}=2^{-i}$ for $i\in \mathbb{N};$ $f\left(
a_{i}\right) =\exp \left( -\frac{1}{a_{i}}\right) =\exp \left( -2^{i}\right) 
$ and a sequence of $h:h_{n}=2^{-n+\varepsilon },$ $0<\varepsilon <1.$

Suppose that the distribution is such that on $\left[ 0,1\right] $ the
measure in the denominator is $P_{X}\left( C(0,h)\right) =\sum_{a_{i}\leq
h}f_X\left( a_{i}\right) ;$ then for $h_{n}$ we get 
\begin{eqnarray*}
P_{X}\left(C(0,h_{n})\right) &=& \sum_{i=n}^{\infty }f_X\left( a_{i}\right)
\\
&=&\exp \left( -2^{n}\right)\left( 1+\sum_{l=1}^{\infty }\exp \left( -{2^{n}}%
\left(1-2^{-l}\right) \right) \right) \\
&=&\exp \left( -2^{n}\right) \left( 1+o\left( 1\right) \right) .
\end{eqnarray*}

Consider now for $h_{n}$ the numerator $EK\left( W_{i}\left( 0\right)
\right)=$%
\begin{eqnarray*}
\sum_{i=n}^{\infty }K\left( \frac{a_{i}}{h_{n}}\right) f_X\left(a_{i}\right)
&=&\exp \left( -2^{n}\right) \left( K\left( 2^{-\varepsilon}\right)
+\sum_{l=1}^{\infty }\exp \left( -{2^{n}}\left(1-2^{-l}\right) \right)
K\left( 2^{l-\varepsilon }\right) \right) \\
&=&K\left( 2^{-\varepsilon }\right) \exp \left( -2^{n}\right)
\left(1+o\left( 1\right) \right) .
\end{eqnarray*}%
Then the leading term of the ratio is $K\left( 2^{-\varepsilon }\right) .$
Thus the limit depends on $\varepsilon $ and on $K.$ As long as $K$ is not a
constant function, say $K\left( \frac{1}{\sqrt{2}}\right) \neq K\left(\frac{1%
}{\sqrt[4]{2}}\right) $, we can consider two bandwidth sequences $h_{n}$
(with $\varepsilon _{1}=\frac{1}{2}$ and $\varepsilon _{2}=\frac{1}{4}$)
where the limits of the ratio will differ.

\subsection{Proofs of the main results from Section 3}

Here we prove the asymptotic normality results of the NW estimator, $%
\widehat{m}\left( x\right) $ defined as: 
\begin{eqnarray*}
\widehat{m}\left( x\right) &=&B_{n}^{-1}\left( x\right) A_{n}\left( x\right)
,  \label{NWx} \\
B_{n}\left( x\right) &=&\frac{1}{n}\sum_{i=1}^{n}K\left( W_{i}(x)\right) ;%
\text{ }A_{n}\left( x\right) =\frac{1}{n}\sum_{i=1}^{n}K\left(
W_{i}(x)\right) Y_{i}.\;\;\;\;\;\;\;  \label{B, Ax}
\end{eqnarray*}

\noindent We first provide useful preliminary results in the form of two
lemmas and a proposition together with their proofs. To simplify exposition
we sometimes denote $A_{n}\left( x\right) $ by $A,$ $B_{n}\left( x\right) $
by $B,$ $B_{n}(x)-EB_{n}(x)$ by $\mathcal{\zeta };$ $P_{X}\left( C\left(
x,h\right) \right) $ by $P,$ also, $K\left( W_{i}\right) $ by $K_{i}.$
Denote by $o_{m.s}\left( 1\right) $ convergence to zero in mean square; this
of course implies convergence in probability to zero. If some $Q=O\left(
r\right) $ and $Q^{-1}=O\left( r^{-1}\right) $ for $r\rightarrow 0,$ we
write $Q\simeq O\left( r\right) .$ Part (b) of the lemma below examines the
bias, $bias\left( \widehat{m}\left( x\right) \right) .$ The variance is in
part (c). Denote $A_{n}\left( x\right) -m\left( x\right) B_{n}\left(
x\right) $ by $A^{c}\left( x\right) .$

\begin{lemma}
\label{L.mhat mtilde}{Under either of the following conditions (i)
Assumptions 1-4 or (ii) 1, 2(a-c), 3-5}\newline
(a)%
\begin{equation*}
\dfrac{B_{n}(x)-EB_{n}(x)}{EB_{n}(x)}=\frac{\mathcal{\zeta }}{EB_{n}\left(
x\right) }=O_{m.s.}\left( (nP_{X}\left( C\left( x,h\right) \right)
^{-1/2}\right) =o_{m.s.}\left( 1\right) ;
\end{equation*}%
(b) 
\begin{equation*}
bias\left( \widehat{m}\left( x\right) \right) =O\left( \bar{h}^{\delta
}\right) +O\left( (nP_{X}\left( C\left( x,h\right) \right) ^{-1}\right) ;
\end{equation*}%
(c)\ 
\begin{eqnarray*}
&& \widehat{m}\left( x\right) -m\left( x\right) =\frac{A_{n}^{c}\left(
x\right) }{EB_{n}\left( x\right) }(1+o_{p}\left( 1\right) ); \\
&&\frac{varA_{n}^{c}\left( x\right) }{\left( EB_{n}\left( x\right) \right)
^{2}} =\frac{\frac{1}{n}\mu _{2}\left( x\right) E\left[ K^2\left(
h^{-1}\left\Vert x-X\right\Vert \right) \right]}{\left( E\left[ K\left(
h^{-1}\left\Vert x-X\right\Vert \right) \right] \right)^2}\left( 1+o\left(
1\right) \right) \simeq O\left( (nP_{X}\left( C\left( x,h\right) \right)
^{-1}\right) .
\end{eqnarray*}
\end{lemma}

\vspace{.2in}

\noindent \textbf{Proof of Lemma \ref{L.mhat mtilde}.}

\noindent (a) follows immediately by making use of the lower and upper
bounds in (\ref{Bounds EB}) and the upper bound from (\ref{Var B}) which
provide%
\begin{eqnarray*}
E\left( \frac{B_{n}(x)-EB_{n}(x)}{EB_{n}(x)}\right) ^{2} &=&\frac{%
E(B_{n}\left( x\right) -E\left( B_{n}\left( x\right) \right) ^{2}}{\left(
EB_{n}(x)\right) ^{2}}; \\
\frac{L_{varB}P_{X}\left( x,h\right) }{n\left( M_{K}P_{X}\left( x,h\right)
\right) ^{2}} &\leq &\frac{E(B_{n}\left( x\right) -E\left( B_{n}\left(
x\right) \right) ^{2}}{\left( EB_{n}(x)\right) ^{2}}\leq \frac{%
M_{varB}P_{X}\left( x,h\right) }{n\left( L_{K}P_{X}\left( x,h\right) \right)
^{2}}.
\end{eqnarray*}%
Thus 
\begin{equation*}
\frac{B_{n}(x)-EB_{n}(x)}{EB_{n}(x)}\simeq O_{m.s}\left( \left( \frac{1}{%
nP_{X}\left( C\left( x,h\right) \right) }\right) ^{1/2}\right) .
\end{equation*}

\noindent (b) For any constant $C_{\zeta }:0<C_{\zeta }<1$%
\begin{eqnarray}
\widehat{m}\left( x\right) -m\left( x\right) &=&\frac{A_{n}^{c}\left(
x\right) }{EB_{n}\left( x\right) }\left( 1-\frac{\mathcal{\zeta }%
/EB_{n}\left( x\right) }{\left( 1+\zeta /EB_{n}\left( x\right) \right) }%
\right) I\left( \left\vert \zeta \right\vert \leq C_{\zeta }EB_{n}\left(
x\right) \right)  \notag \\
&&+\frac{A_{n}^{c}\left( x\right) }{EB_{n}\left( x\right) \left( 1+\zeta
/EB_{n}\left( x\right) \right) }I\left( \left\vert \zeta \right\vert
>C_{\zeta }EB_{n}\left( x\right) \right) .
\end{eqnarray}%
Represent $E\left( \widehat{m}\left( x\right) -m\left( x\right) \right)
=T_{1}+T_{2}$ with%
\begin{eqnarray*}
T_{1} &=&E\left[ \frac{A_{n}^{c}\left( x\right) }{EB_{n}\left( x\right) }%
\left( 1-\frac{\mathcal{\zeta }/EB_{n}\left( x\right) }{\left( 1+\zeta
/EB_{n}\left( x\right) \right) }\right) I\left( \left\vert \zeta \right\vert
\leq C_{\zeta }EB_{n}\left( x\right) \right) \right] ; \\
T_{2} &=&E\left[ \frac{A_{n}^{c}\left( x\right) }{EB_{n}\left( x\right)
\left( 1+\zeta /EB_{n}\left( x\right) \right) }I\left( \left\vert \zeta
\right\vert >C_{\zeta }EB_{n}\left( x\right) \right) \right] .
\end{eqnarray*}

Then 
\begin{equation*}
\left\vert T_{1}\right\vert \leq \left\vert E\frac{A_{n}^{c}\left( x\right) 
}{EB_{n}\left( x\right) }\right\vert +E\left\vert \frac{A_{n}^{c}\left(
x\right) }{EB_{n}\left( x\right) }\frac{\zeta }{EB_{n}\left( x\right) }\frac{%
1}{1-C_{z}}\right\vert .
\end{equation*}%
Also, $\frac{1}{1+\zeta /EB_{n}\left( x\right) }\leq \frac{1}{1+C_{gz}},$ so 
\begin{equation*}
\left\vert T_{2}\right\vert \leq E\left[ \frac{\left\vert A_{n}^{c}\left(
x\right) \right\vert }{EB_{n}\left( x\right) }\frac{1}{1+C_{\zeta }}I\left(
\left\vert \zeta \right\vert >C_{\zeta }EB_{n}\left( x\right) \right) \right]
.
\end{equation*}

For $T_{1}$, recalling the bounds on the moments, we have 
\begin{eqnarray*}
&& \left\vert E\frac{A_{n}^{c}\left( x\right) }{EB_{n}\left( x\right) }%
\right\vert \leq \bar{h}^{\delta }M_{\Delta }; \\
&&E\left\vert \frac{A_{n}^{c}\left( x\right) }{EB_{n}\left( x\right) }\frac{%
\zeta }{EB_{n}\left( x\right) }\frac{1}{1-C_{z}}\right\vert \leq \frac{%
\left( varA_{n}^{c}\left( x\right) var\zeta \right) ^{1/2}}{\left( EB\right)
^{2}}\frac{1}{1-C_{\zeta }} \leq M_{T_{1}}\frac{1}{nP}
\end{eqnarray*}%
with some bound $M_{T_{1}}<\infty .$

For $T_{2}$%
\begin{eqnarray*}
E\left\vert T_{2}\right\vert &\leq &\frac{\left( varA_{n}^{c}\left( x\right)
\right) varI\left( \left\vert \zeta \right\vert >C_{\zeta }EB_{n}\left(
x\right) \right) ^{1/2}}{EB}\frac{1}{1+C_{\zeta }} \\
&\leq &M_{T_{2}^{\prime }}\frac{1}{\left( nP\right) ^{1/2}}\left[ \Pr \left(
\left\vert \zeta \right\vert >C_{\zeta }EB_{n}\left( x\right) \right)
-\left( \Pr \left( \left\vert \zeta \right\vert >C_{\zeta }EB_{n}\left(
x\right) \right) \right) ^{2}\right] ^{1/2}\leq M_{T_{2}}\frac{1}{nP}
\end{eqnarray*}%
with some $M_{T_{2}}<\infty .$ Combining we get the result (b).\medskip

\noindent (c) We have that 
\begin{eqnarray*}
\widehat{m}\left( x\right) -m\left( x\right) &=&\frac{A_{n}^{c}\left(
x\right) }{EB_{n}\left( x\right) \left( 1+\zeta /EB_{n}\left( x\right)
\right) } \\
&=&\frac{A_{n}^{c}\left( x\right) }{EB_{n}\left( x\right) }\left( 1-\frac{%
\zeta /EB_{n}\left( x\right) }{\left( 1+\zeta /EB_{n}\left( x\right) \right) 
}\right)
\end{eqnarray*}%
and since $\zeta /EB_{n}\left( x\right) =o_{p}\left( 1\right) $ from (a) we
get that the limit distribution of $\widehat{m}\left( x\right) -m(x)$ is the
same as for $\frac{A_{n}^{c}\left( x\right) }{EB_{n}\left( x\right) }.$

The variance of $\frac{A_{n}^{c}\left( x\right) }{EB_{n}\left( x\right) }$
is $\frac{var\left( A_{n}^{c}\left( x\right) \right) }{\left( EB_{n}\left(
x\right) \right) ^{2}}$; for $A_{n}^{c}\left( x\right)=A(x)-m(x)B(x) $ the
variance is obtained from Lemma B.2 and provides 
\begin{equation*}
varA_{n}^{c}\left( x\right) =\frac{1}{n}\mu _{2}\left( x\right)E \left[
K^2\left( \frac{\left\Vert x-X \right\Vert}{h}\right) \right] (1+o\left(
1)\right)
\end{equation*}%
then dividing by $\left( EB\right) ^{2}=\left( E K^2\left( \frac{\left\Vert
x-X \right\Vert}{h}\right) \right) $ we get 
\begin{equation*}
\frac{varA_{n}^{c}\left( x\right) }{\left( EB_{n}\left( x\right) \right) ^{2}%
}=\frac{\frac{1}{n}\mu _{2}\left( x\right) E\left[ K^2\left( \frac{%
\left\Vert x-X \right\Vert}{h}\right)\right]}{\left( E\left[ K\left( \frac{%
\left\Vert x-X \right\Vert}{h}\right)\right]\right) ^{2}}\left( 1+o\left(
1\right) \right) \simeq O\left( (nP_{X}\left( C\left( x,h\right) \right)
^{-1}\right) .
\end{equation*}%
\textcolor{white}{.}\hfill $\blacksquare $\vspace{0.2in}

Next, consider the numerator of $\widehat{m}\left( x\right) -m\left(
x\right) $: 
\begin{equation*}
A_{n}^{c}(x)=\frac{1}{n}\sum_{i=1}^{n}\zeta _{in}\text{ with }\zeta
_{in}=K\left( W_{i}(x)\right) (Y_{i}-m\left( x)\right) .
\end{equation*}

Each term, $\zeta _{in},$ is a function of $x,$ but to simplify we suppress
in notation this dependence. The next lemma then expresses the moments of $%
\zeta _{in}.$

\begin{lemma}
\label{L.zeta moms} {Under the conditions of Lemma \ref{L.mhat mtilde}}

\begin{itemize}
\item[(a)] 
\begin{equation*}
E\zeta _{in}=O\left( \bar{h}^{\delta }\right) EK=O\left( \bar{h}^{\delta
}P\right) ;\qquad
\end{equation*}
\end{itemize}

\begin{itemize}
\item[(b)] 
\begin{eqnarray*}
var\zeta _{in} &=&\int_{\left[ 0,1\right] ^{q}}\Omega _{\mu _{2}}\left(
C\left( x,h\circ v\right) \right) (-1)^{q}\partial K^{2}\left( v\right) dv \\
&=&\mu _{2}(x)EK^{2}(1+o\left( 1\right) )
\end{eqnarray*}
\end{itemize}
\end{lemma}

\vspace{.2in}

\noindent \textbf{Proof of Lemma \ref{L.zeta moms}. }

\noindent (a) We have 
\begin{eqnarray}
E\zeta _{in} &=&E\left( E((Y_{i}-m(x))K_{i}|X_{i}\right) )  \label{E-1} \\
&=&E\left[ K\left( h^{-1}\left\Vert x-X_{i}\right\Vert \right) \left(
m\left( X_{i}\right) -m\left( x\right) +E(u_{i}|X_{i})\right) \right]  \notag
\\
&=&\int K\left( h^{-1}\left\Vert x-X\right\Vert \right) (m\left( X\right)
-m\left( x\right) )dF_{X}  \label{g=m(x)} \\
&=&O\left( \bar{h}^{\delta }\right) EK=O\left( \bar{h}^{\delta }P\right)
\end{eqnarray}%
where we use $E(u|X)=0$ and Assumption {3.}

\noindent (b) For variance similarly%
\begin{eqnarray*}
E(\zeta _{in}-E\zeta )^{2} &=&E\left( K\left( W_{i}(x)\right) (Y_{i}-m\left(
x)\right) \right) ^{2}-\left( E\zeta _{in}\right) ^{2} \\
&=&E\left( K_{i}\left( u_{i}+(m\left( X_{i}\right) -m\left( x\right) \right)
)\right) ^{2}-\left( E\zeta _{in}\right) ^{2} \\
&=&\mu _{2}\left( x\right) EK^{2}_i+E\left( K_{i}(m\left( X_{i}\right)
-m\left( x\right) )-EK_{i}(m\left( X_{i}\right) -m\left( x\right) )\right)
^{2}-\left( E\zeta _{in}\right) ^{2} \\
&=&\mu _{2}\left( x\right) EK^{2}_i+O\left( \bar{h}^{2\delta }\right)
varK_i-\left( E\zeta _{in}\right) ^{2}=\mu _{2}\left( x\right)
EK^{2}_i\left( 1+o\left( 1\right) \right) .
\end{eqnarray*}
\textcolor{white}{.}\hfill $\blacksquare $\vspace{.2in}

Next, define 
\begin{equation}
\widetilde{\xi }_{in}=\frac{1}{\sqrt{n}}\frac{\xi _{in}-E\xi _{in}}{\sqrt{%
var\xi _{in}}}.  \label{zeta-tilde}
\end{equation}

The proposition below shows asymptotic normality of the sum $\sum_{i=1}^{n}%
\widetilde{\xi }_{in}.$

\begin{proposition}
\label{P.zeta moms} {Under the conditions of Lemma \ref{L.zeta moms} and
Assumption 6 } as $n\rightarrow \infty ,$ $h\rightarrow 0$
\begin{equation*}
\sum_{i=1}^{n}\widetilde{\xi }_{in}\rightarrow _{d}Z\sim N\left( 0,1\right) .
\end{equation*}
\end{proposition}

\noindent \textbf{Proof of Proposition \ref{P.zeta moms}.} 

\noindent Our proof proceeds similarly to the proof of Theorem 4 in Masry
(2005, pp.172--176). For convenience of the reader we list the main steps in
the notation of our paper: we replace the $\tilde{Z}_{ni}$ of that paper by
scaled $\widetilde{\xi }_{in}$ from (\ref{zeta-tilde}) and denote $\sqrt{n}%
\widetilde{\xi }_{in}$ by $V_{in}$. We consider $S_{n}=\sum_{i=1}^{n}V_{in}$
in place of their $\sum_{i=1}^{n}\tilde{Z}_{in}$; our standardization
implies that for us $\sigma ^{2}\left( x\right) =1.$ Thus we need to show 
\begin{equation*}
\frac{1}{\sqrt{n}}S_{n}\rightarrow ^{L}N\left( 0,1\right) .
\end{equation*}%
We follow Masry (2005) in partitioning $\left\{ 1,\cdots ,n\right\} $ into
alternating big blocks of size $u_{n}$ and small blocks of size $v_{n},$
respectively. The block sizes satisfy the following conditions as $%
n\rightarrow \infty $:

\begin{itemize}
\item[(a)] $\left\{ v_{n}\right\} :$ $v_{n}\rightarrow \infty ,$ but $%
v_{n}=o\left( \left( nP_{X}\left( x,h\right) \right) ^{1/2}\right) ;\left( 
\frac{n}{P_{X}\left( x,h\right) }\right) ^{1/2}\alpha \left( v_{n}\right)
\rightarrow 0.$ This implies that there exists a sequence of positive
integers $\left\{q_{n}\right\} $ such that $q_{n}\rightarrow \infty $ but $%
q_{n}v_{n}=o\left(\left( nP_{X}\left( x,h\right) \right) ^{1/2}\right) $ and 
$q_{n}\left( \frac{n}{P_{X}\left( x,h\right) }\right) ^{1/2}\alpha \left(
v_{n}\right) \rightarrow 0.$

\item[(b)] $\left\{ u_{n}\right\} : u_{n}\rightarrow \infty , \
u_{n}=\lfloor \left( nP_{X}\left( x,h\right) \right) ^{1/2}/q_{n}\rfloor ,$
where $\lfloor .\rfloor $ stands for integer value here and below.
\end{itemize}

Next, denote by $\eta _{j}$ the $j$th sum over a large block:%
\begin{equation*}
\eta _{j}=\sum_{i=j\left( u+v\right) +1}^{j\left( u+v\right) +u}V_{in};
\end{equation*}%
by $\xi _{j}$ the $j$th sum over a small block: 
\begin{equation*}
\xi _{j}=\sum_{i=j\left( u+v\right) +u+1}^{(j+1)\left( u+v\right) }V_{in}.
\end{equation*}%
Here $j$ varies from $0$ to $k-1$ with $k=\lfloor \frac{n}{u_{n}+v_{n}}%
\rfloor .$ Finally, after $k$ large blocks, followed by a small block each,
there may be a remainder part in the total sum:%
\begin{equation*}
\zeta _{k}=\sum_{i=k\left( u+v\right) +1}^{n}V_{in}.
\end{equation*}
The sum $S_{n}$ is represented as 
\begin{eqnarray*}
S_{n} &=&S_{n}^{I}+S_{n}^{II}+S_{n}^{III} \\
S_{n}^{I} &=&\sum_{j=0}^{k-1}\eta _{j}; \ S_{n}^{II}=\sum_{j=0}^{k-1}\xi
_{j}; \ S_{n}^{III}=\zeta _{k}.
\end{eqnarray*}

Identically to Masry we can show that 
\begin{eqnarray*}
&&\frac{1}{n}E\left( S_{n}^{II}\right) ^{2}\rightarrow 0; \\
&&\frac{1}{n}E\left( S_{n}^{III}\right) ^{2}\rightarrow 0; \\
&&E\left( \exp (itn^{-1/2}S_{n}^{I}\right) -\prod_{j=0}^{k-1}E\left( \exp
(itn^{-1/2}\eta _{j}\right) \rightarrow 0; \\
&& \frac{1}{n}\sum_{j=0}^{k-1}E\eta _{j}^{2} \rightarrow 1 \quad ;\frac{1}{n}%
\sum_{j=0}^{k-1}E\left[ \eta _{j}^{2}I\left\{ \left\vert \eta
_{j}\right\vert >\varepsilon \sqrt{n}\right\} \right] \rightarrow 0
\end{eqnarray*}%
for any $\varepsilon >0$. The results require similar evaluation of the
second order moments using the conditions on the relative sizes of the
blocks in the sums and the strong mixing condition; for the last limit Masry
(2005) employs a truncation argument to deal with the fact that the response
variable $Y_i$ is not necessarily bounded.

\noindent The limits show that only $S_{n}^{I}$ matters in the limit and
that it satisfies the Lindeberg-Feller theorem.\hfill $\blacksquare $\vspace{%
0.2in}

\noindent \textbf{Proof of Theorem 1.}

\noindent (a) The asymptotic normality result in Proposition \ref{P.zeta
moms} provides the asymptotic normality for the numerator, $A_{n}^{c}(x).$
To establish asymptotic normality of $\widehat{m}\left( x\right) =\frac{%
A_{n}\left( x\right) }{B_{n}\left( x\right) }$ first recall that by Lemma %
\ref{L.mhat mtilde} $\widehat{m}\left( x\right) -m\left( x\right) =\frac{%
A_{n}^{c}\left( x\right) }{EB_{n}\left( x\right) }(1+o_{p}\left( 1\right) ),$
the variance of $\frac{A_{n}^{c}\left( x\right) }{EB_{n}\left( x\right) }$
is $\frac{varA_{n}^{c}\left( x\right) }{\left( EB_{n}\left( x\right) \right)
^{2}}=\frac{\frac{1}{n}\mu _{2}\left( x\right) EK^{2}}{\left( EK\right) ^{2}}%
=O\left( \frac{1}{nP}\right) ,$ thus we have convergence%
\begin{equation*}
\frac{\sqrt{n}E\left[K\left( h^{-1}\left\Vert x-X\right\Vert \right)\right] 
}{\sqrt{ \mu _{2}\left( x\right) E\left[K^{2}\left( h^{-1}\left\Vert
x-X\right\Vert \right)\right] }}\left( \widehat{m}\left( x\right) -m\left(
x\right) -bias\left( \widehat{m}\left( x\right) \right) \right) \rightarrow
_{d}Z\sim N\left( 0,1\right)
\end{equation*}%
and obtain (a) of Theorem 1. \hfill

\noindent (b) Lemma \ref{L.mhat mtilde} (b) gives rates for the bias as $%
O\left( \bar{h}^{\delta }\right) +O\left( \frac{1}{nP}\right) ,$ for $\frac{1%
}{n}\mu _{2}\left( x\right) EK^{2}$ as $O\left( \left( nP\right)
^{-1}\right) ,$ so that $\frac{\sqrt{n}EK}{\sqrt{\mu _{2}\left( x\right)
EK^{2}}}$ is $O\left( \left( nP\right) ^{1/2}\right) .$

\noindent (c) If $\bar{h}^{2\delta }=o\left( \frac{1}{nP}\right) $ then $%
\frac{\sqrt{n}EK}{\sqrt{\mu _{2}\left( x\right) EK^{2}}}bias\left( \widehat{m%
}\left( x\right) \right) \rightarrow 0.$ From (a) it then follows that 
\begin{equation*}
\frac{\sqrt{n}E\left[K\left( h^{-1}\left\Vert x-X\right\Vert \right)\right] 
}{\sqrt{ \mu _{2}\left( x\right) E\left[K^{2}\left( h^{-1}\left\Vert
x-X\right\Vert \right)\right] }}(\widehat{m}\left( x\right) -m\left(
x\right) )\rightarrow _{d}Z\sim N\left( 0,1\right) .
\end{equation*}

\textcolor{white}{.}\hfill $\blacksquare $ \vspace{0.2in}

\noindent\textbf{Proof of Theorem 2.} The result follows by
substituting the limits in Assumption 7 into the result of Theorem %
1.$\hfill \blacksquare $ \vspace{0.2in}

\subsection{Cross-validated bandwidth from Section 4}

%

The consistency of the bandwidth result extends to singular distributions
that satisfy the following assumption: \renewcommand{\theassumption}{CV}

\begin{assumption}
The support of the probability measure $P_{X}$ is a subset $\Omega $\ of an
affine subspace $V\left( r\right) \subset \mathbb{R}^{q}.$ Restricted to $\Omega $,
the probability measure, the regression function and the conditional error
variance, the bandwidth, and kernel function satisfy the Assumptions of
Theorem 2.1 in Hall et al. (2007) or the Assumptions of Theorem 1 in Li et
al. (2009).
\end{assumption}

The proposition below shows the consistency result for the continuous
relevant regressors.

\renewcommand{\theproposition}{CV}

\begin{proposition}
Under the conditions of Theorem 1, Assumption CV and assuming that $d
$ regressors are continuous and relevant 
\begin{equation*}
n^{\frac{1}{4+rd}}h_{cv}\rightarrow _{p}a^{o}
\end{equation*}%
in probability.
\end{proposition}

No knowledge of $V\left( r\right) ,$ or $r$ itself is assumed for this
result. This implies, for example, that for functional dependence in the
regressors knowledge of the number of factors is not required for
consistency of the cross-validated bandwidth. The result could be extended
to account for irrelevant regressors as well. We conjecture (with Hall et
al., 2007) that in many other cases the cross-validation procedure could
provide a consistent approximation of the optimal bandwidth.\medskip

\noindent \textbf{Proof of Proposition CV. }

\noindent Since the support of the measure is $V\left( r\right) $ with $v\in
V\left( r\right) $ represented in the coordinates of $\mathbb{R}^{q}$ as $%
v=Ax$ the WIMSE can be expressed as%
\begin{eqnarray*}
&&\int E\left( \widehat{m}^{r}\left( v\right) -m^{r}\left( v\right) \right)
^{2}M^{r}\left( v\right) dF_X; \\
&& \widehat{m}^{r}\left( v\right) =\widehat{m}\left( Ax\right) ,m^{r}\left(
v\right) =m\left( Ax\right) ,M^{r}\left( v\right) =M\left( Ax\right) ,
\end{eqnarray*}%
with a similar substitution in the CV criterion. The values of WMISE and CV
expressed via $x$ are identical to those expressed via $v$, thus if $h\left(
x\right) =\left( h^{1},\dots ,h^{q}\right) $ provides some value for WIMSE
or CV expressed via $x,$ then $h\left( v\right) =Ah\left( x\right) =\left(
h_{v}^{1},\dots ,h_{v}^{r}\right) $ gives the same value for those functions
expressed via $v.$ Assumption CV implies that a cross-validated bandwidth
restricted to considering $V$ is consistent there. Due to the identity of
values of the WMISE and CV functions, the CV procedure over $\mathbb{R}^{q}$
is consistent. \hfill$\blacksquare$\vspace{0.2in}

\noindent \textbf{Remarks.}

\begin{enumerate}

\item Consistency holds without assuming knowledge of $V\left( r\right) $ (or
knowledge of $r);$ just existence is assumed.

\item If there are irrelevant regressors in $V\left( r\right) ,$ then
correspondingly, some of the cross-validated bandwidths over $\mathbb{R}^{q}$
will exceed the range. Assuming that there are $r_{1}$ relevant regressors
among those in $V\left( r\right) $ the optimal bandwidth rate will be $n^{-%
\frac{1}{4+r_{1}}}.$

\end{enumerate}

\subsection{\textbf{Bandwidth selection with mass points.}}

Suppose that we have a mixture of mass points $\left( s=0\right) $ with
continuous distributions of other degrees of singularity $\left( 1\geq
s>0\right) .$ Then $P_{X}=\alpha _{0}P_{0}+(1-\alpha _{0})P_{1}$, a mixture
of an absolutely continuous $(s=1)$ distribution with mass points $%
\left(s=0\right) .$ More generally we could have a mixture of a singular
distribution such as one that satisfies Assumption CV with support $\Omega ,$
with mass points supported on a discrete set $MP$. Assume that the set $MP$
is finite. A mass point $x_{0}$ is associated with $P_{0}$ even when it is
also a point of support for the continuous distribution.

\begin{itemize}
\item[(a)] The bandwidth can be selected by ``naive'' application of the
cross validation statistical packages that do not take account of the
concentration of the observations at a mass point. Such a procedure takes
account of the variance where the contribution of mass points is relatively
small (parametric rather than non-parametric rate) and the squared bias
where the mass point observations contribute on par with those outside of
mass points.

\item[(b)] The bandwidth selection could be adapted to the presence of mass
points. We propose a two step adaptive procedure:

\begin{itemize}
\item Step 1. With known mass points eliminate all the observations at each
mass point and estimate the regression function at the other values based on
a cross-validated bandwidth over the subsample after exclusion. It is also
possible to apply the standard adaptive bandwidth for this continuous
component of the distribution.

\item Step 2. Compute a cross-validated bandwidth for the mass points only
and estimate the function at those points. Since variance convergence is
fast at the mass points and reduction of bias for the finite sample in the
estimation of a continuous function can benefit from adding some points in
the vicinity, a cross-validated bandwidth for those points could provide
better estimate than a zero bandwidth.
\end{itemize}
\end{itemize}

We evaluate these procedures in the simulations section and apply it in the
empirical study.

\pagebreak

\section{Simulations Supplement}

\renewcommand{\theequation}{C.\arabic{equation}} \setcounter{equation}{0}%
\renewcommand{\thelemma}{C.\arabic{lemma}} \setcounter{lemma}{0}%
\renewcommand{\thetheorem}{C.\arabic{theorem}} \setcounter{theorem}{0}%
\renewcommand{\thecorollary}{C.\arabic{corollary}} \setcounter{corollary}{0} %
\renewcommand{\thetable}{C.\arabic{table}} \setcounter{table}{0}%
\renewcommand{\thefigure}{C.\arabic{figure}} \setcounter{figure}{0}

In this supplement we provide more details of the simulations and insights
they provide. Simulations were conducted in R, and make use of the NP
package in R (Hayfield and Racine, 2008) where suitable. In Section C.1, we
consider the univariate (point mass) setting discussed in Section 5.1.
Section C.2 presents a bivariate setting in the presence of a singularity
that was not presented in the main text. Sections C.3-C.5 provide additional
details related to Sections 5.2-5.3.

\subsection{Univariate (Point mass example)}


Illustrative graphs of the density of the regressor distribution considered
in Section 5.1 are displayed in Figure C.1. In this supplement, details are
provided for point mass distribution for different values of $p$, in
particular $p=0, 0.1,$ and $p=0.2$, to allow us to see the impact of
changing the probability of mass points. The trinormal mixture, considered
in Kotlyarova et al. (2016), is given by the following density 
\begin{equation*}
f_X(x)=0.5\phi (x+0.767)+3\phi \left(\frac{x+0.767-0.8}{0.1}\right)+2\phi
\left(\frac{x+0.767-1.2}{0.1}\right) ,
\end{equation*}%
where $\phi$ denotes the standard Gaussian density function.

\begin{figure}[h]
\caption{Density of regressor}%
\begin{equation*}
\text{Mass point}\hspace{1.25in}\text{Trinormal}\qquad \qquad \vspace{-.1in} 
\end{equation*}
\includegraphics[scale=.55]{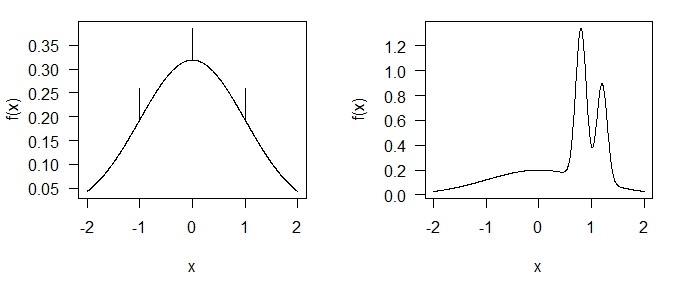}
\end{figure}

From Table C.1, we observe that the mean cross-validated bandwidth across
the 500 simulation show a decrease in the presence of mass points. The
bandwidth under the trinormal distribution with $n=1,000$ is comparable to
that of the distribution with mass points with $p=0.2$, which is why we
focused on this setting in the main text. 

The performance indicators of the NW estimator (mean absolute error (MAE)
and root mean squared error (RMSE)) reveal an improved aggregate performance
arising from the presence of mass points, as does an increased sample size
and larger signal to noise ratio. Details hereof are provided in Table C.2. 

\pagebreak

\begin{table}[H]
\caption{Cross validated bandwidths for the NW estimator by regressor
distribution, sample size ($n$), and signal to noise ratio (snr).}%
\begin{tabular}{lp{1cm}p{1.2cm}p{1cm}p{1.2cm}p{1cm}p{1cm}}
\  &  &  &  &  &  &  \\ 
& \multicolumn{6}{c}{\textbf{snr=1}} \\ 
& \multicolumn{2}{c}{\textbf{n=100}} & \multicolumn{2}{c}{\textbf{n=500}} & 
\multicolumn{2}{c}{\textbf{n=1,000}} \\ 
\cmidrule(lr){2-3} \cmidrule(lr){4-5} \cmidrule(lr){6-7} & Mean & Std & Mean
& Std & Mean & Std \\ 
F$_{mass}(p=0)$ & 0.407 & 0.123 & 0.305 & 0.069 & 0.267 & 0.057 \\ 
F$_{mass}(p=0.1)$ & 0.380 & 0.118 & 0.278 & 0.070 & 0.243 & 0.050 \\ 
F$_{mass}(p=0.2)$ & 0.261 & 0.115 & 0.250 & 0.067 & 0.222 & 0.050 \\ 
F$_{trinormal}$ & 0.348 & 0.112 & 0.256 & 0.065 & 0.216 & 0.054 \\ 
\  &  &  &  &  &  &  \\ 
& \multicolumn{6}{c}{\textbf{snr=2}} \\ 
& \multicolumn{2}{c}{\textbf{n=100}} & \multicolumn{2}{c}{\textbf{n=500}} & 
\multicolumn{2}{c}{\textbf{n=1,000}} \\ 
\cmidrule(lr){2-3} \cmidrule(lr){4-5} \cmidrule(lr){6-7} & Mean & Std & Mean
& Std & Mean & Std \\ 
F$_{mass}(p=0)$ & 0.351 & 0.101 & 0.265 & 0.061 & 0.236 & 0.046 \\ 
F$_{mass}(p=0.1)$ & 0.330 & 0.093 & 0.237 & 0.059 & 0.212 & 0.044 \\ 
F$_{mass}(p=0.2)$ & 0.310 & 0.088 & 0.211 & 0.054 & 0.189 & 0.041 \\ 
F$_{trinormal}$ & 0.304 & 0.092 & 0.218 & 0.054 & 0.178 & 0.044%
\end{tabular}%
\end{table}
\begin{table}[H]
\caption{Aggregate performance (MAE and RMSE) by regressor distribution, sample size ($n$), and signal to noise ratio (snr); cross-validated
bandwidth.}%
\begin{tabular}{lp{1cm}p{1.2cm}p{1cm}p{1.2cm}p{1cm}p{1cm}}
\  &  &  &  &  &  &  \\ 
& \multicolumn{6}{c}{\textbf{snr=1}} \\ 
& \multicolumn{2}{c}{\textbf{n=100}} & \multicolumn{2}{c}{\textbf{n=500}} & 
\multicolumn{2}{c}{\textbf{n=1,000}} \\ 
\cmidrule(lr){2-3} \cmidrule(lr){4-5} \cmidrule(lr){6-7} & MAE & RMSE & MAE
& RMSE & MAE & RMSE \\ 
F$_{mass}(p=0)$ & 0.591 & 0.739 & 0.572 & 0.717 & 0.569 & 0.713 \\ 
F$_{mass}(p=0.1)$ & 0.577 & 0.722 & 0.558 & 0.700 & 0.555 & 0.695 \\ 
F$_{mass}(p=0.2)$ & 0.562 & 0.703 & 0.543 & 0.681 & 0.540 & 0.677 \\ 
F$_{trinormal}$ & 0.541 & 0.677 & 0.525 & 0.658 & 0.521 & 0.653 \\ 
\  &  &  &  &  &  &  \\ 
& \multicolumn{6}{c}{\textbf{snr=2}} \\ 
& \multicolumn{2}{c}{\textbf{n=100}} & \multicolumn{2}{c}{\textbf{n=500}} & 
\multicolumn{2}{c}{\textbf{n=1,000}} \\ 
\cmidrule(lr){2-3} \cmidrule(lr){4-5} \cmidrule(lr){6-7} & MAE & RMSE & MAE
& RMSE & MAE & RMSE \\ 
F$_{mass}(p=0)$ & 0.423 & 0.530 & 0.406 & 0.509 & 0.403 & 0.506 \\ 
F$_{mass}(p=0.1)$ & 0.414 & 0.518 & 0.396 & 0.497 & 0.393 & 0.493 \\ 
F$_{mass}(p=0.2)$ & 0.404 & 0.506 & 0.386 & 0.484 & 0.383 & 0.481 \\ 
F$_{trinormal}$ & 0.389 & 0.487 & 0.373 & 0.467 & 0.369 & 0.463%
\end{tabular}%
\end{table}

\pagebreak

The MAE typically is smaller than the RMSE which is a reflection of the
variability associated with the NW estimator across simulations.\newline

In Figure C.2, we evaluate the NW nonparametric fit locally. The figure
shows the root mean squared error (RMSE) over the 500 replications with $n=1,000$ . 
\begin{figure}[H]
\caption{Local RMSE of the NW estimator by regressor distribution, n=1,000;
cross-validated bandwidth.}\includegraphics[scale=.6]{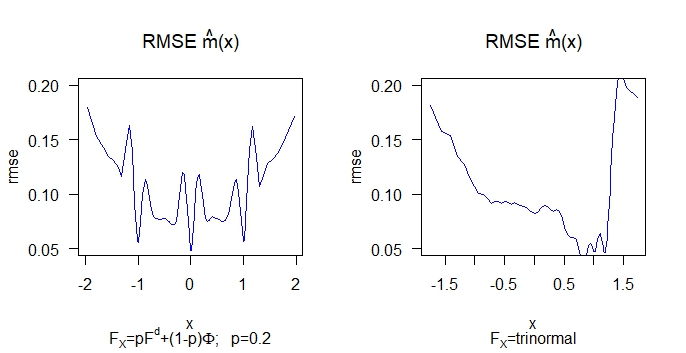}
\end{figure}

In order to assess the sensitivity of the results to the selected bandwidth,
Figure C.3 displays the results for different bandwidth choices. The RMSE
results are presented together with the bias and standard deviation.

\begin{figure}[H]
\caption{Bias, standard deviation and RMSE of the NW estimator by regressor
distribution and bandwidth (n=1,000).}%
\includegraphics[scale=.6]{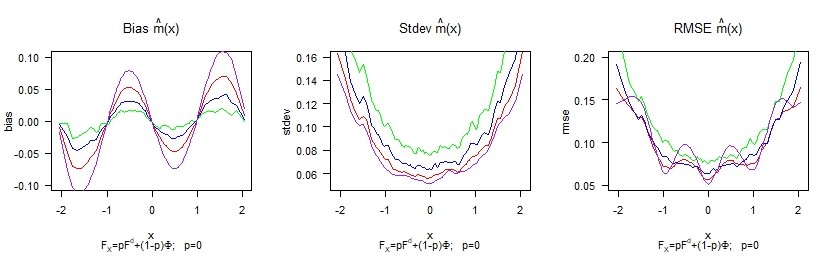} %
\includegraphics[scale=.6]{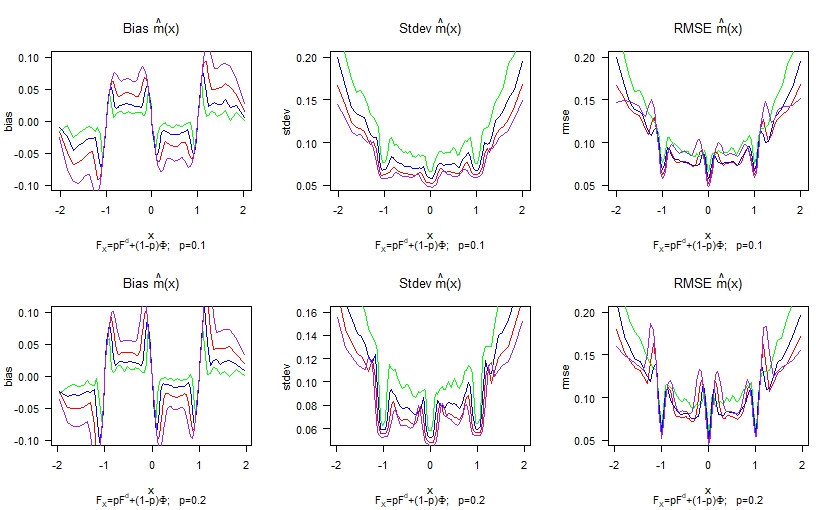} %
\includegraphics[scale=.6]{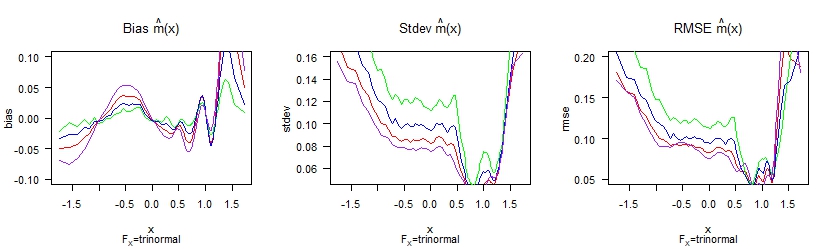} 
\begin{minipage}{1.0\textwidth}{Note: The top panel is the base setting where $F_X$  is standard gaussian, in the panels 2-3 $F_X$ has mass points (with increasing probability), and in the 4th panel $F_X$ is the trinormal distribution. The green and blue lines present results with $h=0.5h_{cv}$ and $h=0.75h_{cv}$ respectively), the purple line use $h=1.25h_{cv}$, and the red line presents the cross-validated result.}
\end{minipage}
\end{figure}

As expected, smaller bandwidths result in a smaller bias and larger variance
at all points, where the standard deviation increases at points where the
distribution is more sparse. The cross-validated bandwidth provides
reasonable performance in that this bias is small.

The impact of increasing the probability of mass points is marked. While, as
expected, the bias is close to zero at the mass points (where a parametric
rate of convergence can be obtained), when evaluating points slightly
above/below these mass points its impact is non-negligible. The impact on
the standard deviation around the point mass points also becomes more
pronounced.

Interesting patterns are also observed when we consider the trinormal
mixture. In regions where the derivative of the trinormal mixture is large,
small standard deviations are observed as well as larger fluctuations in the
bias; the bias and standard deviation are fairly stable when $x$ takes
values from [-1,0.5] where the distribution is not sparse and does not have
large derivatives, while the standard deviation increases again where the
distribution is more sparse.

Finally, in Table C.3 we provide a more extensive set of results regarding
the pointwise convergence rates. The results obtained with different
probabilities of mass points is similar, the pointwise convergence rates
estimates close to masspoints improve when a smaller bandwidth is chosen and
notably when using our adaptive bandwidth procedure. 
\begin{table}[H]
\caption{Empirical Rates by regressor distribution and bandwidth.}%
\begin{tabular}{lp{1.25cm}p{1.25cm}p{1.25cm}p{1.25cm}}
&  &  &  &  \\ 
&  & \multicolumn{3}{c}{\textbf{Mass Point, p=0.2}} \\ 
\cmidrule(lr){3-5} & \multicolumn{1}{l}{\textbf{X}} & \multicolumn{1}{c}{$%
.75h_{cv}$} & \multicolumn{1}{c}{$h_{cv}$} & \multicolumn{1}{c}{$1.25h_{cv}$}
\\ 
& 0.0 & -0.456 & -0.455 & -0.455 \\ 
& 0.1 & -0.246 & -0.186 & -0.166 \\ 
& 0.2 & -0.435 & -0.411 & -0.315 \\ 
& 0.3 & -0.433 & -0.465 & -0.474 \\ 
& 0.4 & -0.436 & -0.460 & -0.480 \\ 
& 0.8 & -0.398 & -0.390 & -0.326 \\ 
& 0.9 & -0.265 & -0.212 & -0.199 \\ 
& 1.0 & -0.469 & -0.465 & -0.461%
\end{tabular}
\begin{tabular}{lp{1.25cm}p{1.25cm}p{1.25cm}}
&  &  &  \\ 
& \multicolumn{3}{c}{\textbf{Mass Point, p=0.2}} \\ 
\cmidrule(lr){2-4} & \multicolumn{1}{c}{$.75h_{cv,a}$} & \multicolumn{1}{c}{$%
h_{cv,a}$} & \multicolumn{1}{c}{$1.25h_{cv,a}$} \\ 
& -0.516 & -0.515 & -0.515 \\ 
& -0.434 & -0.443 & -0.443 \\ 
& -0.444 & -0.438 & -0.423 \\ 
& -0.438 & -0.430 & -0.413 \\ 
& -0.437 & -0.425 & -0.404 \\ 
& -0.439 & -0.443 & -0.435 \\ 
& -0.447 & -0.449 & -0.454 \\ 
& -0.524 & -0.526 & -0.523%
\end{tabular}
\newline
\begin{tabular}{lp{1.25cm}p{1.25cm}p{1.25cm}p{1.25cm}}
&  &  &  &  \\ 
&  & \multicolumn{3}{c}{\textbf{Mass Point, p=0.1}} \\ 
\cmidrule(lr){3-5} & \multicolumn{1}{l}{\textbf{X}} & \multicolumn{1}{c}{$%
.75h_{cv}$} & \multicolumn{1}{c}{$h_{cv}$} & \multicolumn{1}{c}{$1.25h_{cv}$}
\\ 
& 0.0 & -0.453 & -0.451 & -0.451 \\ 
& 0.1 & -0.318 & -0.280 & -0.371 \\ 
& 0.2 & -0.445 & -0.421 & -0.335 \\ 
& 0.3 & -0.455 & -0.470 & -0.460 \\ 
& 0.4 & -0.441 & -0.461 & -0.465 \\ 
& 0.8 & -0.433 & -0.412 & -0.350 \\ 
& 0.9 & -0.335 & -0.301 & -0.295 \\ 
& 1.0 & -0.457 & -0.454 & -0.453%
\end{tabular}%
\begin{tabular}{lp{1.25cm}p{1.25cm}p{1.25cm}}
&  &  &  \\ 
& \multicolumn{3}{c}{\textbf{Mass Point, p=0.1}} \\ 
\cmidrule(lr){2-4} & \multicolumn{1}{c}{$.75h_{cv,a}$} & \multicolumn{1}{c}{$%
h_{cv,a}$} & \multicolumn{1}{c}{$1.25h_{cv,a}$} \\ 
& -0.509 & -0.508 & -0.507 \\ 
& -0.445 & -0.448 & -0.445 \\ 
& -0.446 & -0.442 & -0.430 \\ 
& -0.447 & -0.438 & -0.421 \\ 
& -0.443 & -0.437 & -0.417 \\ 
& -0.448 & -0.439 & -0.429 \\ 
& -0.440 & -0.440 & -0.442 \\ 
& -0.508 & -0.506 & -0.505%
\end{tabular}%
\newline
\begin{tabular}{lp{1.25cm}p{1.25cm}p{1.25cm}p{1.25cm}}
&  &  &  &  \\ 
&  & \multicolumn{3}{c}{\textbf{Mixture of normal}} \\ 
\cmidrule(lr){3-5} & \multicolumn{1}{l}{\textbf{X}} & \multicolumn{1}{c}{$%
.75h_{cv}$} & \multicolumn{1}{c}{$h_{cv}$} & \multicolumn{1}{c}{$1.25h_{cv}$}
\\ 
& 0.00 & -0.444 & -0.449 & -0.451 \\ 
& 0.50 & -0.376 & -0.381 & -0.421 \\ 
& 0.75 & -0.432 & -0.416 & -0.413 \\ 
& 1.00 & -0.399 & -0.413 & -0.436 \\ 
& 1.25 & -0.421 & -0.397 & -0.390 \\ 
& 1.50 & -0.397 & -0.356 & -0.295 \\ 
\  &  &  &  &  \\ 
&  &  &  & 
\end{tabular}
\newline
\begin{minipage}{1.0\textwidth}{Note: For the mass point distribution the usual, $h_{cv}$, and adaptive, $h_{cv,a }$, bandwidth selection are considered.}
\end{minipage}
\end{table}

\pagebreak
 
\subsection{Bivariate (with singularity: $X_1$ and $X_2$ on unit circle)}

In this section we consider a bivariate regression model in the presence of
a singularity. In particular, we consider the setting where the regressors
lie on a unit circle, that is 
\begin{equation*}
X_1^2 + X_2 ^2 = 1 .
\end{equation*}
This example may find its origin in Hotelling's (1929) spatial model of
horizontal differentiation which assumes that each consumer has an `ideal'
variety identified by his location on the unit circle (bounded product
space). See also Desmet and Parente (2010).\medskip

Subject to this singularity, we simulated 500 random samples $%
\{(y_i,X_{1i},X_{2i}\}_{i=1}^n$ from the model 
\begin{align*}
Y_i & = X_{1i} + X_{2i} + \sigma \varepsilon_i ,
\end{align*}
for different sample sizes. We assume $F_X(x_1,x_2)$ is uniform on the unit
circle. To obtain our random sample we draw $\phi$ from $U[0,2\pi]$ and use $%
(X_1,X_2)=(\cos(\phi),\sin(\phi))$. The error $\{\varepsilon_i\}_{i=1}^n$ is
drawn independently of the regressors and has a standard Gaussian
distribution; $\sigma$ is selected to yield a given signal to noise ratio.

The conditional mean function under consideration is displayed below. 
\begin{figure}[H]
\caption{Bivariate conditional mean with singularity.}%
\includegraphics[scale=.6]{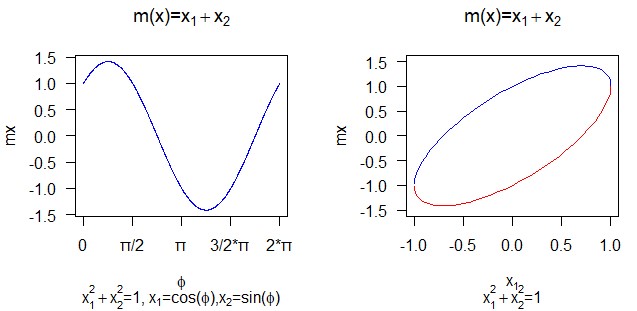}\newline
\textcolor{white}{.}\hspace{1in}(a)\hspace{1.75in}(b)\newline
\ \newline
\begin{minipage}{1.0\textwidth}{Notes: (a) $m(x)$ with $\phi$ on the horizontal axis, (b) $m(x)$ with $x_1$ on the horizontal axis; the blue line represents the setting where $x_2 \leq 0$, the red line where $x_2 \geq 0$, with $x_2=0$ the lines intersect.}
\end{minipage}
\end{figure}

The display on the left uses $\phi \in [0,2\pi]$ on the horizontal axis with 
$(X_1,X_2)=(\cos(\phi),\sin(\phi))$; the display on the right has $X_1 \in
[-1,1]$ on the horizontal axis.\newline

From Table C.4 we can see that the leave-one-out cross validated bandwidths
for the NW estimator are comparable for both arguments $X_1$ and $X_2$ given
its symmetric formulation. 
\begin{table}[H]
\caption{: Cross validated bandwidth for the NW estimator by sample size ($n$%
), and signal to noise ratio (snr).}%
\begin{tabular}{lcccccc}
\  &  &  &  &  &  &  \\ 
& \multicolumn{3}{c}{\textbf{snr=1}} & \multicolumn{3}{c}{\textbf{snr=2}} \\ 
\cmidrule(lr){2-4} \cmidrule(lr){5-7} & n=100 & n=500 & n=1,000 & n=100 & 
n=500 & n=1,000 \\ 
&  &  &  &  &  &  \\ 
$h_{1,cv}$ & 0.861 & 0.604 & 0.523 & 0.735 & 0.529 & 0.456 \\ 
& (.324) & (.182) & (.141) & (.267) & (.155) & (.112) \\ 
$h_{2,cv}$ & 0.879 & 0.611 & 0.528 & 0.748 & 0.530 & 0.459 \\ 
& (.314) & (.180) & (.141) & (.261) & (.150) & (.114) \\ 
&  &  &  &  &  & 
\end{tabular}%
\end{table}

In Table C.5 the performance indicators of the NW estimator (mean absolute
error (MAE) and root mean squared error (RMSE)) reveal an improvement when
the signal to noise ratio is larger or the sample size increases. 
\begin{table}[H]
\caption{: Performance indicators of the NW estimator by sample size ($n$),
and signal to noise ratio (snr); cross-validated bandwidth.}%
\begin{tabular}{lcccccc}
\  &  &  &  &  &  &  \\ 
& \multicolumn{3}{c}{\textbf{snr=1}} & \multicolumn{3}{c}{\textbf{snr=2}} \\ 
\cmidrule(lr){2-4} \cmidrule(lr){5-7} & n=100 & n=500 & n=1,000 & n=100 & 
n=500 & n=1,000 \\ 
&  &  &  &  &  &  \\ 
\quad MAE & 0.810 & 0.804 & 0.801 & 0.576 & 0.569 & 0.567 \\ 
\quad RMSE & 1.014 & 1.007 & 1.003 & 0.721 & 0.713 & 0.710 \\ 
\  &  &  &  &  &  &  \\ 
&  &  &  &  &  & 
\end{tabular}
\begin{minipage}{1.0\textwidth}{Note: The MAE is given by $\sum_{i=1}^{n}\left|y_i-\widehat{m}\left(x_i\right)\right|/n$ and the RMSE is given by $\sqrt{\sum_{i=1}^{n}\left(y_i-\widehat{m}\left(x_i\right)\right)^2/n}$ where the leave-one-out estimator for $m(x_i)$ is used. }
\end{minipage}
\end{table}

We evaluate the average RMSE over 500 simulations at a grid of 100 $%
(x_1,x_2) $ points based on equidistant values of $\phi$ on $[0,2\pi]$ for $%
n=1,000$. In Figure C.5 we present the average RMSE over the 500
simulations, together with the bias and standard deviation for the NW
estimator when $n=1,000$. The graphs reveal the sensitivity to the bandwidth
choice. 
\begin{figure}[H]
\caption{Bias, standard deviation and RMSE of the NW estimator by bandwidth
(n=1,000).}%
\begin{equation*}
\includegraphics[scale=.5]{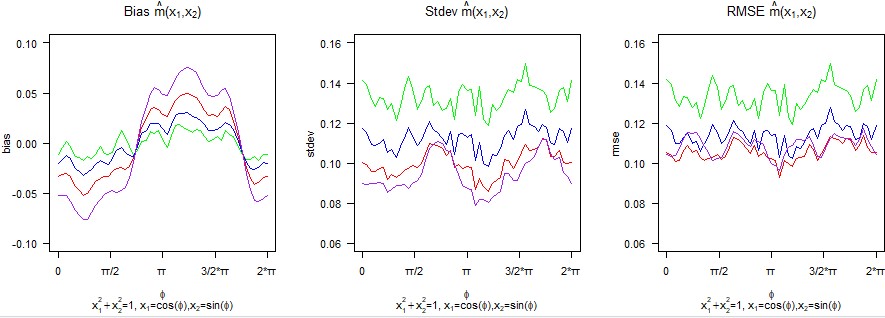} 
\end{equation*}
\begin{minipage}{1.0\textwidth}{Note: The green and blue lines present results with $h=0.5h_{cv}$ and $h=0.75h_{cv}$ respectively, the purple line uses $h=1.25h_{cv}$, and the red line presents the cross-validated bandwidth.}
\end{minipage}
\end{figure}

When evaluating the pointwise convergence rate, we find that at points on
the unit circle, the NW estimator typically achieves a faster empirical rate
of convergence than $-0.4$. Illustrative results hereof are provided in
Table C.6. For comparison we consider simulations with $X_1$ and $X_2$ drawn
independently on $U[-1,1]$. There, as expected, the rate of convergence is
slower. The results support the rate reduction due to singularity. 
\begin{table}[H]
\caption{Empirical rate of convergence of the NW estimator at various points
on the unit circle by distribution of regressors.}%
\begin{tabular}{lp{2cm}p{1.75cm}p{1.75cm}}
\  &  &  &  \\ 
&  & \multicolumn{2}{c}{Distribution $(X_1,X_2)$} \\ 
\cmidrule(lr){3-4} & $\phi$ & \multicolumn{1}{l}{Singular} & 
\multicolumn{1}{l}{Non-Singular} \\ 
\  &  &  &  \\ 
& 0 & -0.444 & -0.306 \\ 
& $\pi/2$ & -0.448 & -0.297 \\ 
& $\pi$ & -0.451 & -0.291 \\ 
& $3\pi/2$ & -0.465 & -0.203 \\ 
\  &  &  &  \\ 
&  &  & 
\end{tabular}%
\newline
\begin{minipage}{1\textwidth}{Note: $(X_1,X_2)=(\cos(\phi),\sin(\phi))$. For our singular setting $\phi$ is drawn randomly from $U[0,2\pi]$; for the non-singular setting $X_1$ and $X_2$ are drawn randomly from $U[-1,1]$.}
\end{minipage}
\end{table}

\subsection{Bivariate (with effective dimension 1)}

The conditional mean function under consideration in Section 5.2 is
displayed in Figure C.6. 

\begin{figure}[H]
\caption{Bivariate conditional mean with reduced dimensionality.}%
\begin{equation*}
\includegraphics[scale=.5]{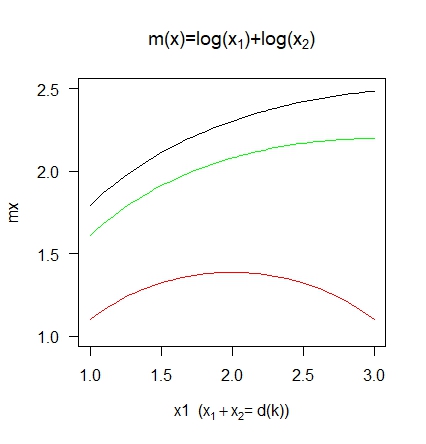} 
\end{equation*}
\begin{minipage}{1.0\textwidth}{Note: The red line represents $d(1)=3$, the green line $d(2)=6$ and the black line $d(3)=7$.}
\end{minipage}
\end{figure}

In Table C.7, the overall performance indicators (MAE and MSE) are given for
the NW estimator at different sample sizes. For the ordered discrete
variable in NW.d, we use both the Epanechnikov and the geometric weights
proposed by Wang et al. (1981), with 
\begin{equation*}
W(h,i,j)=\left\lbrace 
\begin{array}{ll}
\frac{1}{2}(1-h)h^{|i-j|} & |i-j|\geq 1 \\ 
1-h & i=j%
\end{array}
\right.
\end{equation*}
$h\in [0,1]$, where $i$ and $j$ denote values taken by the ordered discrete
variable. NW.d with the Epanechnikov kernel is labeled ``Unordered''; NW.d
with the geometric weights for $d(k)$ is labeled ``Ordered''.

For the smaller sample size, $n=100$ the use of the geometric weights
(``Ordered'') in place of the Epanechnikov kernel (``Unordered'') may be
beneficial for the NW.d estimator. The overall performance of the NW.d
estimator suggest only a slight improvement over NW.c for small samples. At
the cross validated bandwidth, $n=500$, and $srn=1$, the MAE on average
equals 0.354 and 0.351 for the NW.c, and NW.d approach respectively; the
RMSE respectively 0.443 and 0.440. 
\begin{table}[H]
\caption{Aggregate performance of the NW estimator with cross validated
bandwidth by sample size (n) and snr=1.}%
\begin{tabular}{p{1.5cm}p{1.25cm}p{1.25cm}p{1.25cm}p{1.25cm}p{1.25cm}p{1.25cm}}
\  &  &  &  &  &  &  \\ 
& \multicolumn{2}{c}{NW.c} & \multicolumn{4}{c}{NW.d} \\ 
& \multicolumn{2}{c}{$(X_1,X_2)$} & \multicolumn{4}{c}{$(X_1,d(k))$} \\ 
\cmidrule(lr){2-3} \cmidrule(lr){4-7} & \multicolumn{2}{c}{} & 
\multicolumn{2}{c}{Unordered} & \multicolumn{2}{c}{Ordered} \\ 
\cmidrule(lr){4-5} \cmidrule(lr){6-7} n & \textbf{MAE} & \textbf{RMSE} & 
\textbf{MAE} & \textbf{RMSE} & \textbf{MAE} & \textbf{RMSE} \\ 
\  &  &  &  &  &  &  \\ 
100 & 0.360 & 0.450 & 0.359 & 0.449 & 0.350 & 0.438 \\ 
500 & 0.354 & 0.443 & 0.354 & 0.443 & 0.351 & 0.440 \\ 
1,000 & 0.352 & 0.442 & 0.352 & 0.442 & 0.351 & 0.440 \\ 
\  &  &  &  &  &  &  \\ 
&  &  &  &  &  & 
\end{tabular}
\begin{minipage}{1.0\textwidth}{Note: The MAE is given by $\sum_{i=1}^{n}\left|y_i-\widehat{m}\left(x_i\right)\right|/n$ and the RMSE is given by $\sqrt{\sum_{i=1}^{n}\left(y_i-\widehat{m}\left(x_i\right)\right)^2/n}$ where the leave-one-out estimator for $m(x_i)$ is used. }
\end{minipage}
\end{table}

In Figure C.7, we present the bias, standard deviation, and RMSE of the NW.c
and NW.d (unordered) estimators over a grid of values for $x_{1}$ separately
for each sub-population with $n=1,000$. The standard deviation and RMSE
reveal patterns across the range of $X_1$ which are not dissimilar for
different values of $d(k)$ and reveal increases at the boundary values with $%
X_1$ taking values close to 1 or 3. 
\begin{figure}[H]
\caption{Bias, Standard deviation and RMSE of the NW estimator at
cross-validated bandwidth, n=1,000 (separately d(k), k=1,2,3).}%
\includegraphics[scale=.6]{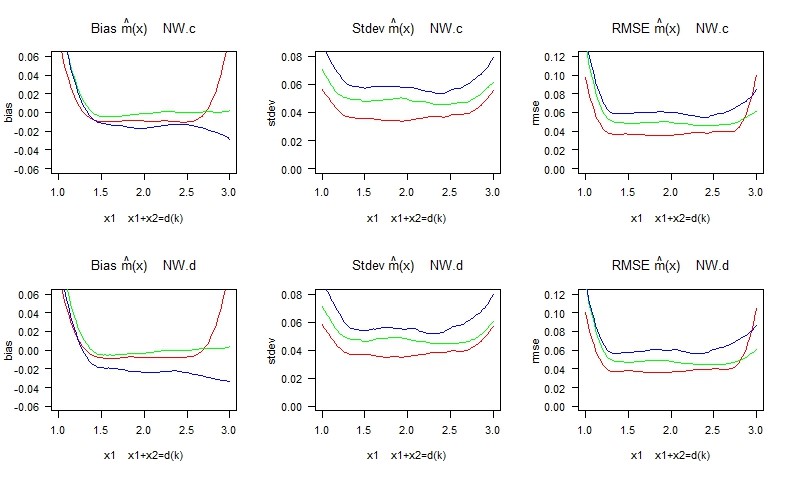}\newline
\begin{minipage}{1.0\textwidth}{Note: Red lines are for d(1)=4, blue lines for d(2)=6, and green lines for d(3)=7..}
\end{minipage}
\end{figure}

\pagebreak

\subsection{Bivariate (in the presence of a functional regressor)}

In Figure C.8 (a) we display a random sample of 20 random curves of the
functional regressor $X_1$ (and $Z$) used in Section 5.4 and in (b) we
display the associated $m_1(X_1)$ realizations used in defining the
conditional mean. 

\begin{figure}[H]
\caption{A sample of 20 random curves and associated function used for
conditional mean $m(X)=m_1(X_1)+X_2$ .}%
\includegraphics[scale=.6]{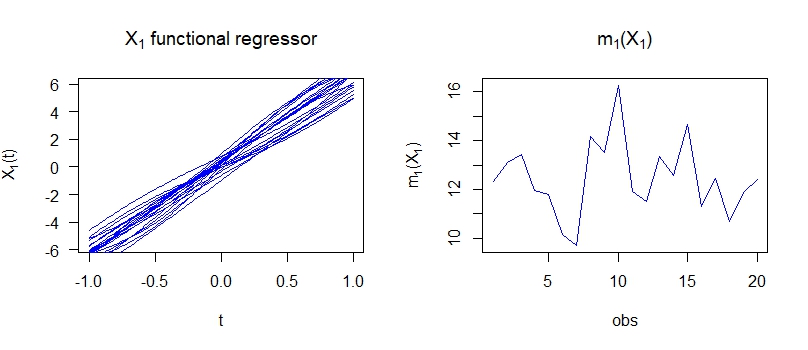}\newline
\textcolor{white}{.}\hspace{1.25in}(a)\hspace{2.5in}(b)
\end{figure}

The cross-validated bandwidths for $h_{1,cv}$ and $h_{2,cv}$ were obtained
using a grid search on $[h_{1,low},h_{1,up}]\times[h_{2,low},h_{2,up}]$; we
set the lower bound equal to the minimal distance ensuring each regressor
has at least one neighboring observation and the upper bound equals half the
maximum distance for each regressor, $j=1,2$.\newline

In Table C.8 we present details of the cross-validated bandwidths obtained
for the models considered in Section 5.3. The columns specify the different
additional control $X_2$ considered, either standard normal, or $m_1(Z)$
where $Z$ is a functional regressor like $X_1$ with correlation of its
parameters $(a,b,w)$ given by $\rho$. The attained residual sum of squares
achieved is around 90\% of the sum of squares residual of the Oracle OLS.
Ignoring either the functional or scalar regressor when implementing the NW
estimator deteriorates the residual sum of squares. When $X_2$ is given by
an independently drawn N(0,1) regressor, the loss of ignoring $X_2$ is more
severe than dropping the functional regressor (1.616 versus 1.170). When
using $X_2=m_1(Z)$ as additional regressor, the loss of dropping either
regressor is stronger when $\rho=0$ than $\rho=0.8$. 
\begin{table}[H]
\caption{Some details of the cross validated bandwidths for the NW estimator
in the presence of functional regressor $X_1$ and associated residual sum of
squares, n=250.}
\begin{tabular}{lccc}
\  &  &  &  \\ 
& \multicolumn{1}{c}{$X_2 = N(0,1)$} & \multicolumn{2}{c}{$X_2 = m_1(Z)$} \\ 
\cmidrule(lr){3-4} &  & $\rho =0.0$ & $\rho=0.8$ \\ 
\  &  &  &  \\ 
$h_{1,cv}$ & \quad 1.369 (0.242) & \quad 1.583 (0.288) & \quad 1.746 (0.334)
\\ 
$h_{2,cv}$ & \quad 1.513 (0.310) & \quad 2.075 (0.414) & \quad 2.321 (0.546)
\\ 
\  &  &  &  \\ 
mean(CV) & 4.670 & 7.189 & 12.020 \\ 
\  &  &  &  \\ 
\multicolumn{4}{l}{\textbf{Oracle:}} \\ 
\quad OLS/CV & 0.890 & 0.891 & 0.916 \\ 
\  &  &  &  \\ 
\multicolumn{4}{l}{\textbf{Misspecification:}} \\ 
\quad CV$_1$/CV & 1.170 & 1.409 & 1.280 \\ 
\quad CV$_2$/CV & 1.616 & 1.377 & 1.178 \\ 
\  &  &  &  \\ 
&  &  & 
\end{tabular}
\begin{minipage}{1.0\textwidth}{Note: CV$_1$ stands for the mean(CV) achieved when only including the functional regressor $X_1$ and CV$_2$ stands for the mean(CV) achieved when only including $X_2$.}
\end{minipage}
\end{table}

In Figure C.9 representative scatter plots are presented for the NW
estimates against $m(x)$ in a simulation where we include both regressors
(black), or omit either the functional (green) or other regressor (blue). In
the top panel $X_2$ is N(0,1), in the bottom panel $X_2=m_1(Z)$ with $\rho=0$%
. The scatter plots clearly reveal the loss in fit with either of the
regressors omitted. 
\begin{figure}[H]
\caption{NW fit against $m(x)$ under potential misspecification.}%
\begin{equation*}
m(X)=m_1(X_1) + X_2, \quad X_2=N(0,1) 
\end{equation*}
\textcolor{white}{.}\hspace{.7in}$X_1,X_2$\hspace{1.2in}$X_2$\hspace{1.3in}$%
X_1$\newline
\includegraphics[scale=.8]{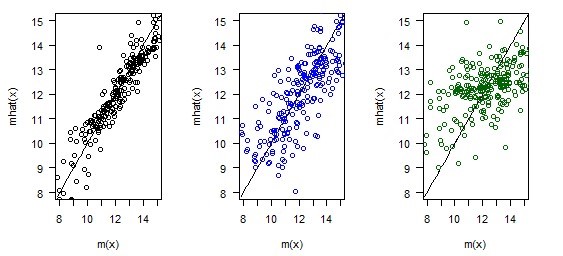}\newline
\begin{equation*}
m(X)=m_1(X_1) + X_2, \quad X_2=m_1(Z), \rho=0 
\end{equation*}
\textcolor{white}{.}\hspace{.7in}$X_1,X_2$\hspace{1.2in}$X_2$\hspace{1.3in}$%
X_1$\newline
\includegraphics[scale=.8]{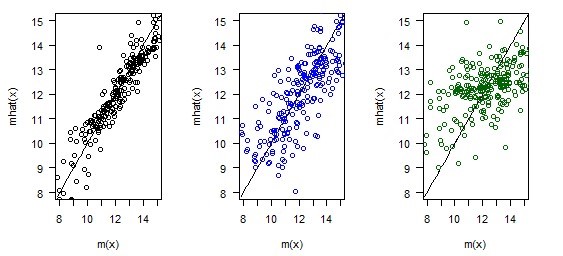} 
\begin{minipage}{1.0\textwidth}{Note: The scatter plot in black indicates the correctly specified NW fit, the blue indicates the fit including $X_2$ only, the green including $X_1$ only.}
\end{minipage}
\end{figure}

In Figure C.10 we present the pointwise RMSE at out-of-sample observations
at the cross-validated bandwidth, where the true value of $m(x)$ of
out-of-sample observations is on the horizontal axis 
\begin{figure}[H]
\caption{Pointwise RMSE of the NW estimator in the presence of functional
regressor as a function of the number of neighbors, $n=250$ (out-of-sample).}%
\includegraphics[scale=.4]{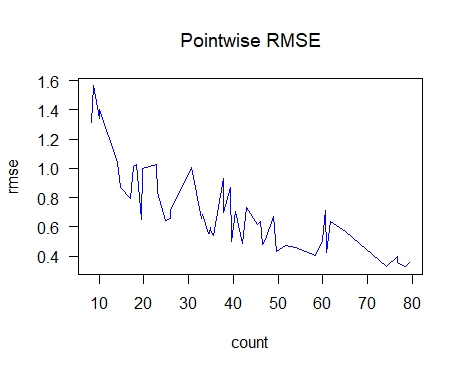}%
\includegraphics[scale=.4]{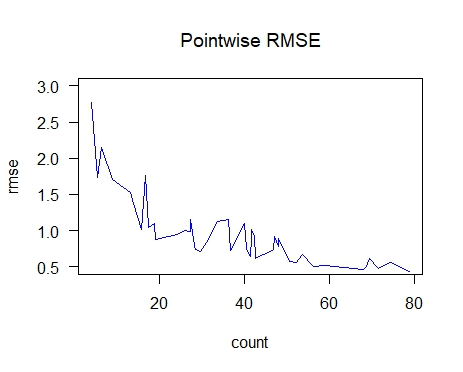}%
\includegraphics[scale=.4]{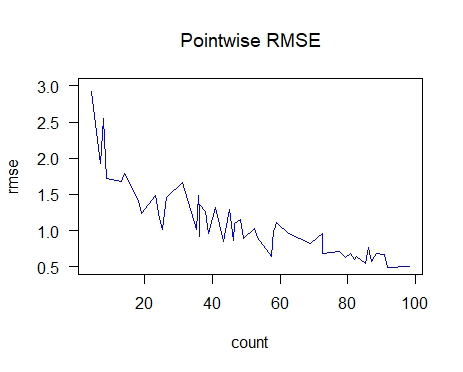}
\end{figure}

Details about the pointwise bias, standard deviation and RMSE of the NW
estimator (not shown) reveal that smaller bandwidths result in a smaller
bias and larger variance at all points. The standard deviation and RMSE is
large where the density for an out-of-sample observation, as represented by
the number of neighbors, is small, as in Figure C.11. 
\begin{figure}[H]
\caption{Pointwise RMSE of the NW estimator in presence of functional
regressor as a function of the true value $m(x)$, $n=250$ (50 out-of-sample
observations).}\includegraphics[scale=.4]{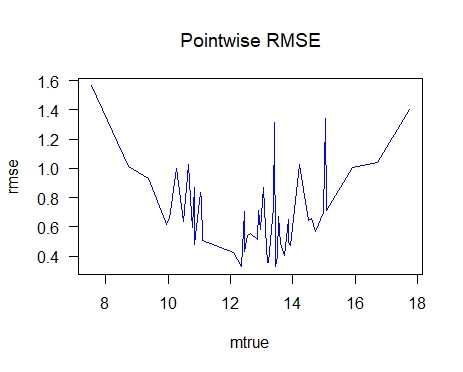}%
\includegraphics[scale=.4]{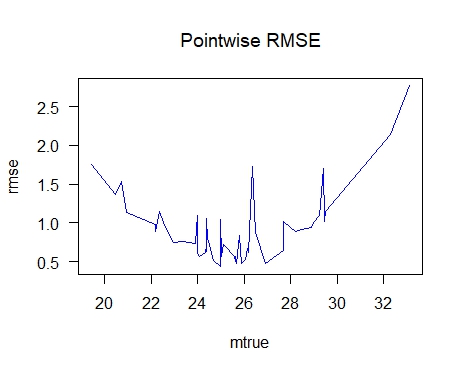}%
\includegraphics[scale=.4]{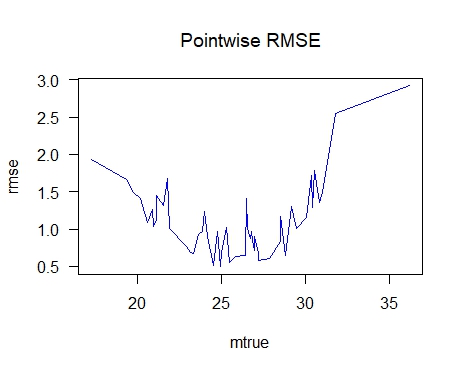}
\end{figure}

\pagebreak

\section{Empirical Study Supplement}

\renewcommand{\theequation}{D.\arabic{equation}} \setcounter{equation}{0}%
\renewcommand{\thelemma}{D.\arabic{lemma}} \setcounter{lemma}{0}%
\renewcommand{\thetheorem}{D.\arabic{theorem}} \setcounter{theorem}{0}%
\renewcommand{\thecorollary}{D.\arabic{corollary}} \setcounter{corollary}{0} %
\renewcommand{\thetable}{D.\arabic{table}} \setcounter{table}{0}%
\renewcommand{\thefigure}{D.\arabic{figure}} \setcounter{figure}{0}

In this supplement we provide more details of the empirical implementation.
The data can be obtained from %
\url{https://users.nber.org/~rdehejia/nswdata2.html} or %
\url{https://github.com/xuyiqing/lalonde/blob/main/data/lalonde}. For
reference to the data and descriptive statistics, we refer the reader to
LaLonde (1986) and the recent survey by Imbens et al. (2024).

We make use of the np package in R (Hayfield and Racine, 2008), which
permits the implementation of both the NW and Local Linear LL kernel based
estimator. For the local linear kernel based regression it implements a
ridge factor correction to achieve non-singularity in the kernel estimate
regression. To obtain the random forest based CATT, we use the grf package
in R (Tibshirani et al. 2024). These estimates are obtained through the
augmented inverse propensity weighting (AIPW) with propensity scores
estimated by generalized random forest (GRF). See also Imbens et al. (2024).

\subsection{Cross validation results}

The leave-one-out cross-validation procedure was used to determine the
bandwidth. We considered both a grid search and the npregbw procedure with
repeated restarts equal to 30 to avoid local minima. The cross-validated NW
kernel bandwidths together with the obtained cross-validation value for the
kernels considered for $(T,X)$ are presented in Table D.1. The top half
relates to the bivariate model, the bottom half to the multivariate model.
We denote the kernel with two arguments: the first argument denotes the
kernel applied to all binary regressors (treat, u75, nodegree, black,
hispanic, and married) and the second argument denotes the kernel applied to
the other regressors (re75, educ, and age). 
\begin{table}[H]
\caption{Cross validated (cv) bandwidth}%
\begin{tabular}{lcccccccccl}
&  &  &  &  &  &  &  &  &  &  \\ 
\  &  &  &  &  &  &  &  &  &  &  \\ 
\multicolumn{1}{c}{kernel} & \multicolumn{6}{c}{regressor} &  &  &  &  \\ 
\cmidrule(lr){2-10} & treat & re75$^*$ & u75 & educ & nodegree & black & 
hispanic & age & married & c.v. \\ 
(e,e) & $\leq 1$ & 11.606 & - & - & - & - & - & - & - & 38.060 \\ 
(d,e) & 0.000 & 12.291 & - & - & - & - & - & - & - & 38.066 \\ 
(d,u) & 0.000 & 7.866 & - & - & - & - & - & - & - & 37.985 \\ 
\  &  &  &  &  &  &  &  &  &  &  \\ 
(e,e) & $\leq 1$ & 23.608 & B & 1.842 & B & $\leq 1$ & B & 25.999 & B & 
36.911 \\ 
(d,e) & 0.001 & 24.584 & 0.396 & 1.747 & 0.500 & 0.002 & 0.500 & 26.000 & 
0.500 & 36.876 \\ 
(d,u) & 0.009 & 22.267 & 0.399 & 1.700 & 0.500 & 0.018 & 0.500 & 20.190 & 
0.500 & 36.742 \\ 
\  &  &  &  &  &  &  &  &  &  &  \\ 
&  &  &  &  &  &  &  &  &  & 
\end{tabular}
\begin{minipage}{1.25\textwidth}{Note: For binary regressors, the Epanechnikov kernel with bandwidths $\leq 1$ indicate no smoothing is implemented; for the discrete kernel this is achieved when the cv bandwidth is zero.  For the Epanechnikov kernel, extremely large bandwidths are denoted with a B; here cv smoothes these irrelevant variables automatically out; for the discrete kernel this is achieved when the cv bandwidth is 0.5;  c.v. stands for cross validated objective function. \\$^*$ The variable re75 is expressed in '000\$s. }
\end{minipage}
\end{table}

For the bivariate regression model, the cross validated bandwidths confirm
that we should not smooth across treated and untreated observations. The
bandwidth on pre-treatment earnings is comparable across the kernel choices
considered, somewhat smaller when using the uniform kernel in place of the Epanechnikov kernel.

For the multivariate regression model extremely large bandwidths are
represented with a B, here cross-validation smoothes out irrelevant
variables; with the discrete kernel this is achieved when the cross
validated bandwidth is 0.5. The multivariate bandwidth results indicate that the variables nodegree, hispanic and married are irrelevant. The bandwidth
on pre-treatment earnings is still relevant and is much larger than in the
baseline model for all kernels, suggesting a reduction of the heterogeneous
impact with respect to pre-treatment earnings. At the same time, the
bandwidths for education and age imply a heterogeneous impact associated
with those characteristics, although the size of the bandwidth for age is
fairly large.

\subsection{Nonparametric fit and CATT}

\subsubsection{Bivariate Model}

In Figure D.1 the bivariate nonparametric fit of the conditional expectation
by pre-treatment earnings and treatment status is given for the kernels
under consideration. 
\begin{figure}[h]
\caption{Nonparametric fit of the conditional expectation by pre-treatment
earnings, treatment status, kernel, and bandwidth.}%
\begin{equation*}
h_{cv} 
\end{equation*}
\begin{equation*}
(e,e)\hspace{1.3in} (d,e) \hspace{1.3in} (d,u)\vspace{-.1in} 
\end{equation*}
\includegraphics[scale=.6]{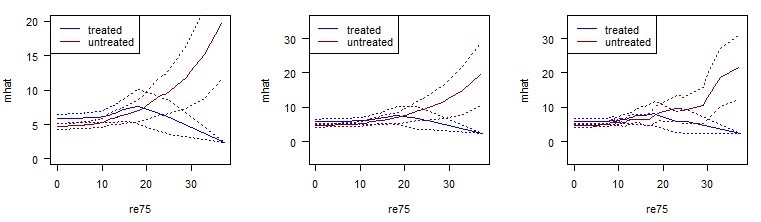} 
\begin{equation*}
1.5h_{cv} 
\end{equation*}
\begin{equation*}
(e,e)\hspace{1.3in} (d,e) \hspace{1.3in} (d,u)\vspace{-.1in} 
\end{equation*}
\includegraphics[scale=.6]{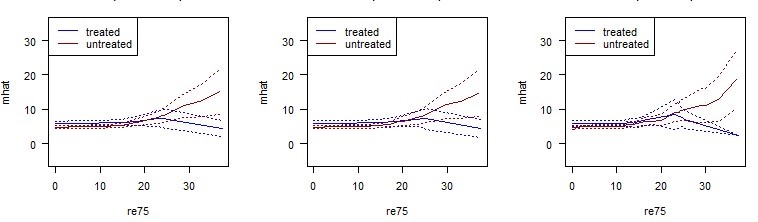} 
\begin{minipage}{1\textwidth}{Note: The values are rescaled and are denoted in '000\$. The top panel applies the cross-validated bandwidth, the bottom panel uses a bandwidth on the pre-treatment earnings that is 50\% larger.}
\end{minipage}
\end{figure}
The nonparametric fit yields a correlation between the post-treatment
outcome and its fit which exceeds that for OLS. The correlation is 0.210 for
the kernels considered, against a correlation of 0.170 using OLS. The graphs clearly
reveal a heterogeneity in the resulting treatment effect which is fairly
stable across kernel. Gains of treatment arise where the confidence band
around the estimated nonparametric fit $\widehat{m}(x,1)$ exceeds that of $%
\widehat{m}(x,0)$. When we increase the bandwidth of the pre-treatment
earnings (resulting in a drop in the correlation to 0.198) the estimates as
well as the associated confidence bandwidth are more attenuated. The
Box-plot of CATT estimates for the bivariate model in Figure D.2, similarly,
shows an attenuation of the CATT effects when increasing the bandwidth on
the pre-treatment earnings. The use of the adaptive bandwidth also has a
notable impact on the CATT estimates and renders the interquartile range
similar to that of the random forest based result. 
\begin{figure}[H]
\caption{Box-plot of CATT estimates in bivariate model with different
bandwidths on pre-treatment earnings.}%
\begin{equation*}
\qquad h_{cv}\hspace{1.3in} 1.5h_{cv} \hspace{1.2in} h_{cv,a}\vspace{-.1in} 
\end{equation*}
\includegraphics[scale=.55]{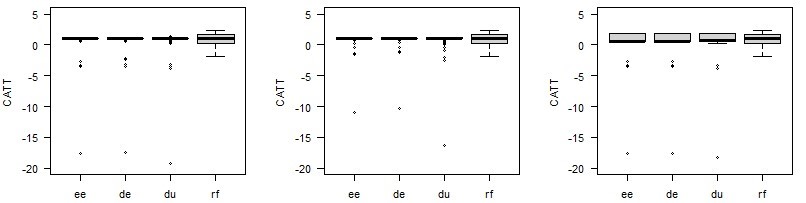} 
\begin{minipage}{1\textwidth}{Note: The values are rescaled and are denoted in '000\$. In the first graph, $h_{cv}$, we provide the cross-validated NW based estimates CATT estimates against the random forest; the  second graph, $1.5h_{cv}$, increases the bandwidth for pre-treatment earnings with 50\%; the final graph, $h_{cv,a}$ uses the adaptive NW estimates. }
\end{minipage}
\end{figure}

Table D.2 presents details of these results for the bivariate model.
Adaptive NW based estimates at mass points equal \$1,784 $(\overline{re78}_1-\overline{re78}_0$). 
\begin{table}[H]
\caption{CATT results (mean (stde) and range) for bivariate model.}%
\begin{tabular}{llccrccrc}
&  &  &  &  &  &  &  &  \\ 
\  &  &  &  &  &  &  &  &  \\ 
& \multicolumn{2}{c}{$h_{cv}$} &  & \multicolumn{2}{c}{$1.5h_{cv}$} &  & 
\multicolumn{2}{c}{$h_{cv,a}$} \\ 
\cmidrule(lr){2-3} \cmidrule(lr){5-6} \cmidrule(lr){8-9} NW(e,e) & 931 (68)
& [-17,558--1,095] &  & 999 (43) & [-11,042--1,098] &  & 920 (76) & 
[-17,558--1,794] \\ 
NW(d,e) & 941 (67) & [-17,378--1,097] &  & 1,005 (41) & [-10,356--1,097] & 
& 920 (76) & [-17,558--1,794] \\ 
NW(d,u) & 908 (75) & [-19,175--1,354] &  & 1,008 (62) & [-16,311--1,164] & 
& 906 (80) & [-18,194--1,794] \\ 
\  &  &  &  &  &  &  &  &  \\ 
RF & 848 (90) & [-1,893--2,304] &  &  &  &  &  &  \\ 
&  &  &  &  &  &  &  &  \\ 
&  &  &  &  &  &  &  & 
\end{tabular}
\begin{minipage}{1.2\textwidth}{Note: The descriptive statistics are provided for the NW based estimates of CATT for different bandwidths (cross-validated, 50\% increase in bandwidth for re75, and adaptive bandwidth) together with the random forest, RF, based estimates.}
\end{minipage}
\end{table}

The NW kernel based CATT results are more disperse than those obtain using
the random forest approach, an approach that can be seen as an adaptive
nearest neighbor estimator. The average of the NW kernel based CATT results
are comparable to those obtained using random forest approach. The local
linear kernel based estimates yield comparable results for the bivariate
model and equal \$834 (75) with the (e,e) kernel with estimates ranging from -\$17,268 to \$1,257.\footnote{%
The LL results with the (d,e) kernel is \$837 (86) [-17,268--1,257]; with
the (d,u) kernel \$793 (90) [-20,876--1,241].} 

\subsubsection{Multivariate Model}

The correlation between the post-treatment earnings and its fit increases
when additional regressors are included. The in-sample correlation after
including u75, educ, nodegree, black, hispanic, age and married equals
0.338, 0.354 and 0.326 for the (e,e), (d,e), and (d,u) kernels on $(T,X)$
respectively. For comparison, the correlation when we use OLS and add age
squared as an additional control equals 0.209.

In Figure D.3 we provide the multivariate regression fit of the conditional
expectation by years of education and ethnicity for an individual with
median age (23), median pre-treatment earnings (\$936), and u75=0 together
with the bootstrap confidence bounds for all kernels we consider. The value
of the dummy variables nodegree, hispanic and married are irrelevant The
graph suggests an improved performance of the Epanechnikov kernel for all
regressors relative to the use of discrete and/or uniform kernels as
reflected by the larger confidence bands displayed in those settings. While
the fit itself suggests gains from treatment for individuals identified as
black with more than 10 years of education for all kernels, the difference
in only statistically significant using the (e,e) kernel. 
\begin{figure}[H]
\caption{Nonparametric fit of the conditional expectation by years of
education, `black' indicator, and treatment status with median pre-treatment
earnings, median age, and kernel.}%
\begin{equation*}
\text{black=0} 
\end{equation*}
\begin{equation*}
(e,e)\hspace{1.3in} (d,e) \hspace{1.3in} (d,u)\vspace{-.1in} 
\end{equation*}
\includegraphics[scale=.55]{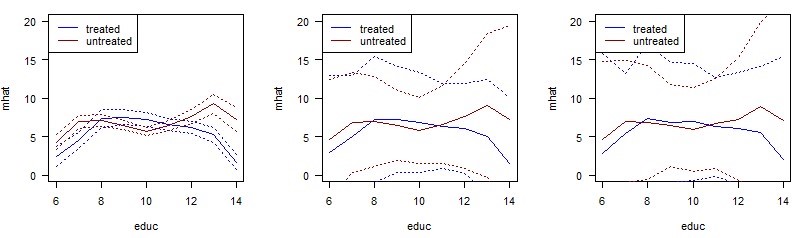}\newline
\begin{equation*}
\text{black=1} 
\end{equation*}
\begin{equation*}
(e,e)\hspace{1.3in} (d,e) \hspace{1.3in} (d,u)\vspace{-.1in} 
\end{equation*}
\includegraphics[scale=.55]{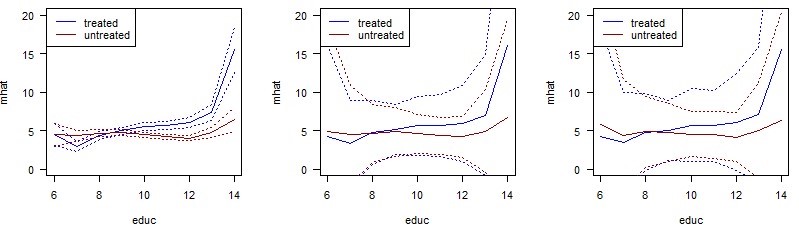}\newline
\begin{minipage}{1\textwidth}{Note: The median pre-treatment earnings equals $\$936$ and the median age is $23$. The estimates are rescaled and are denoted in '000\$. }
\end{minipage}
\end{figure}

Further support for the heterogeneous treatment effect is given in Table
D.3, where we report the average CATT estimates obtained using the NW kernel
regression estimates by race for various educational attainment groups. 
\begin{table}[H]
\caption{CATT for multivariate model by educational category and `black'
indicator.}%
\begin{tabular}{r r r r r r } \\ \ \\
& \multicolumn{5}{c}{Treated with Black=1}\\
\cmidrule(lr){2-6} 
\multicolumn{1}{c}{Educ}  & n & \multicolumn{1}{c}{ee}   & \multicolumn{1}{c}{de} & \multicolumn{1}{c}{du} & \multicolumn{1}{c}{rf}  \\  
$<10$ & 63 & 339 (240)    & 330  (240)    & 265   (230)  & 587 (\ 76) \\
$10$  & 49 & 1,042 (\ \ 8)& 1,014 (\ \ 9) & 1,097 (\ 12) & 566 (\ 79) \\
$11$  & 64 & 1,414 (\ 32) & 1,410 (\ 34)  & 1,228 (\ 27)& 1,105 (125) \\
$>11$ & 62 & 2,651 (252)  & 2,630 (262)   & 2,706 (257) & 1,188 (157)  \\
& \multicolumn{5}{c}{Treated with Black=0}\\
\cmidrule(lr){2-6} 
\multicolumn{1}{c}{Educ}  & n & \multicolumn{1}{c}{ee}   & \multicolumn{1}{c}{de} & \multicolumn{1}{c}{du} & \multicolumn{1}{c}{rf}  \\  
$<10$ & 20 & 566   (564)  & 584 (533) & 642 (493)  & 294 (106) \\
$10$  & 8  & 1,402 (\ \ 9)& 1,041 (190) & 1,144 (137) & 494 (287) \\
$11$  & 13 &  -160 (133)  & -270  (134) & -468  (118)& 280 (177) \\
$>11$ & 18 & -2,081 (259) & -2,096 (239) & -1,670 (262) & 744 (266)  
\end{tabular}
\end{table}
For individuals identified as black we see a statistically significant
increase in the mean CATT with increasing years of education, when using
random forest based estimates there is no statistically significant increase
with over 11 years of education.

As Table D.4 indicates, the multivariate model reduces the variability of
the NW kernel based CATT estimates. 
\begin{table}[h]
\caption{CATT results (mean (stde) and range) for bivariate and multivariate
model.}%
\begin{tabular}{llccrc}
&  &  &  &  &  \\ 
\  &  &  &  &  &  \\ 
\multicolumn{1}{l}{Method} & \multicolumn{2}{c}{Bivariate Model} &  & 
\multicolumn{2}{c}{Multivariate Model} \\ 
\cmidrule(lr){2-3} \cmidrule(lr){5-6}  & \multicolumn{2}{c}{$h_{cv}$} &  & 
\multicolumn{2}{c}{$h_{cv}$} \\ 
(e,e) & 931 (68) & [-17,558--1,095] &  & 1,045 (107) & [-5,657--9,181] \\ 
(d,e) & 941 (67) & [-17,378--1,097] &  & 931 (108) & [-4,828--9,545] \\ 
(d,u) & 908 (75) & [-19,175--1,354] &  & 931 (104) & [-4,484--9,237] \\ 
\  &  &  &  &  &  \\ 
RF & 848 (90) & [-1,893--2,304] &  & 794 (\ 54) & [-1,209--3,437] \\ 
&  &  &  &  &  \\ 
&  &  &  &  & 
\end{tabular}%
\end{table}

The Box-plots of the CATT estimates in Figure D.4 reveal again the
similarity in the interquartile range of CATT estimates. 
\begin{figure}[H]
\caption{Box-plot of CATT estimates in bivariate and multivariate model.}%
\begin{equation*}
\qquad \qquad \text{Bivariate model} \hspace{1.25in} \qquad \text{%
Multivariate model} 
\end{equation*}
\includegraphics[scale=.6]{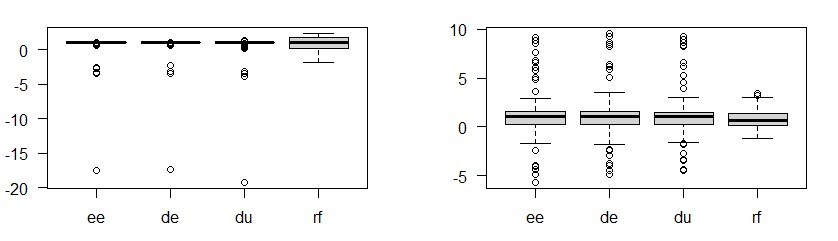}\newline
\begin{minipage}{1\textwidth}{Note: The estimates are rescaled and are denoted in '000\$. }
\end{minipage}
\end{figure}


\begin{thebibliography}{99}
\bibitem{} Ackerberg, D.A., K. Caves, and G. Frazer (2015) ``Identification
properties of recent production function estimators,'' \textit{Econometrica}%
, \textbf{83}, 2411--2451.

\bibitem{} Ahlfors, Lars (1966) \textit{Lectures on quasiconformal mappings}%
,\ Princeton University Press.

\bibitem{} Aitchison, J. and C.G.G. Aitkin (1976) ``Multivariate binary
discrimination by the kernel method,'' \textit{Biometrika}, \textbf{63},
413--420.

\bibitem{} Angrist, J.D. and J.-S. Pischke (2009) \textit{Mostly harmless
econometrics: An empiricist's companion}, Princeton University Press.

\bibitem{} Arulampalam, W., V. Corradi, and D. Gutknecht (2017) ``Modeling
heaped duration data: An application to neonatal mortality,'' \textit{%
Journal of Econometrics}, \textbf{200}, 363--377.

\bibitem{} Bai, J., S. Ng (2006) ``Confidence intervals for diffusion index forecarsts and inference with factor-augmented regressors,'' \textit{Econometrica}, \textbf{74}, 1133--1150.

\bibitem{} Caldeira J.F., R. Gupta, H.S. Torrent (2020) ``Forecasting U.S.
aggregate stock market excess return: Do functional data analysis add
economic value?,'' \textit{Mathematics}, \textbf{8}, 2042.
https://doi.org/10.3390/math8112042 .


%





\bibitem{} Dehejia R.H. and S.Wahba (1999) ``Causal effect in
nonexperimental studies: Reevaluating the evaluation of training programs,'' 
\textit{Journal of the American Statistical Association}, \textbf{94},
1053--1062.

\bibitem{} Dehejia R.H. and S.Wahba (2002) ``Propensity score-matching
methods for nonexperimental causal studies,'' \textit{The Review of
Economics and Statistics}, \textbf{84}, 151--161.

\bibitem{} Demir, S. and \"{O}. Toktamis (2010) ``On the adaptive
Nadaraya-Watson kernel regression estimators,'' \textit{Hacettepe Journal of
Mathematics and Statistics}, \textbf{39}, 429--437.

\bibitem{} Desmet, K. and S.L. Parente (2010) ``Bigger is better: Market
size, demand elasticity, and innovation,'' \textit{International Economic
Review}, \textbf{51}, 319--333.

\bibitem{} Donkers, A.C. and M.M.A. Schafgans (2008) ``Estimation and
specification of semiparametric index models,'' \textit{Econometric Theory}, 
\textbf{24}, 1584--1606.

\bibitem{} Fan, J. and I. Gijbels (1996) \textit{Local polynomial modelling
and its applications}, Chapman and Hall.


\bibitem{} Ferraty F., A. Mas, and P. Vieu (2007) ``Nonparametric regression
on functional data: Inference and practical aspects,'' \textit{Australian
and New Zealand Journal of Statistics}, \textbf{49}, 267--286.

\bibitem{} Ferraty, F. and S. Nagy (2022) ``Scalar-on-function local linear
regression and beyond,''\ \textit{Biometrika}, \textbf{109}, 439--455.


\bibitem{} Ferraty, F. and P. Vieu (2004) ``Nonparametric models for
functional data, with application in regression, time series prediction and
curve discrimination,'' \textit{Journal of Nonparametric Statistics}, 
\textbf{16}, 111--125.

\bibitem{} Ferraty F. and P. Vieu (2006) \textit{Nonparametric functional
data analysis: Theory and Practice}, Springer, New York.

\bibitem{} Gasser, T., P. Hall, and B. Presnell (1998) ``Nonparametric
estimation of the mode of a distribution of random curves,'' \textit{Journal
of Royal Statistical Society, Series B}, \textbf{60}, 681--691.

\bibitem{} Geenens, G. (2015) ``Moments, errors, asymptotic normality and
large deviation principle in nonparametric functional regression,'' \textit{%
Statistics and Probability Letters}, \textbf{107}, 369--377.

\bibitem{} Gy$\ddot{o}$rfi, L., M. Kohler, A. Krzyzak, and H. Walk (2002) 
\textit{A distribution-free theory of nonparametric regression}, Springer,
New York.

\bibitem{} Hall, P. and J. Horowitz (2013) ``A simple bootstrap method for
constructing nonparametric confidence bands for functions,'' \textit{Annals
of Statistics}, \textbf{41}, 1892--1921.

\bibitem{} Hall, P., Q. Li, and J.S. Racine (2007) ``Nonparametric
estimation of regression functions in the presence of irrelevant
regressors,'' \textit{The Review of Economics and Statistics}, \textbf{89},
784--789.

\bibitem{} Hall, P. and J.S. Racine (2015) ``Infinite order cross-validated
local polynomial regression,'' \textit{Journal of Econometrics}, \textbf{185}%
, 510--525.



\bibitem{} Heckman, J.J., H. Ichimura, and P.E. Todd (1997) ``Matching as an
econometric evaluation estimator: Evidence from evaluating a job training
programme,'' \textit{Review of Economic Studies}, \textbf{64}, 605--654.

\bibitem{} Heckman, J.J., H. Ichimura, and P.E. Todd (1998) ``Matching as an
econometric evaluation estimator,'' \textit{Review of Economic Studies}, 
\textbf{65}, 261--294.

\bibitem{} Hong,S. and O. Linton (2020) ``Nonparametric estimation of
infinite order regression and its application to the risk-return tradeoff,'' 
\textit{Journal of Econometrics}, \textbf{219}, 389--424.

\bibitem{} Hotelling, H. (1929) ``Stability in competition,'' \textit{The
Economic Journal}, \textbf{39}, 41--57.


\bibitem{} Ichimura, H. (1993) ``Semiparametric least squares (SLS) and
weighted SLS estimation of single index models,'' \textit{Journal of
Econometrics}, \textbf{58}, 71--120.

\bibitem{} Imbens, G. and Y. Xu (2024) ``LaLonde (1986) after nearly four
decades: Lessons learned,'' arXiv 2406.00827 (econ.EM).

\bibitem{} Jun, B and H. Song (2019) ``Tests for detecting probability mass
points,'' \textit{Korean Economic Review}, \textbf{35}, 205--248.

\bibitem{} Kankanala, S. and V. Zinde-Walsh (2024) ``Kernel-weighted
specification testing under general distributions,'' \textit{Bernoulli}, 
\textbf{30}, 1921--1944.


\bibitem{} K\"{o}hler, M., A. Schindler, and S. Sperlich (2014) ``A review
and comparison of bandwidth selection methods for kernel regression,'' 
\textit{International Statistical Review / Revue Internationale de
Statistique}, \textbf{82}, 243--274.

\bibitem{} Kotlyarova, Y, M. Schafgans, and V. Zinde-Walsh (2016)
``Smoothness: Bias and efficiency of non-parametric kernel estimators,'' in 
\textit{Advances in Econometrics: Essays in Honor of Aman Ullah}, Vol. 36,
G. Gonzales-Rivera, R.C. Hill and T.-H. Lee, eds. 561--589.

\bibitem{} Kurisu, D., Otsu, T., and M. Xu (2025) ``Nonparametric Causal
Inference with Functional Covariates,'' \textit{Journal of Business and
Economic Statistics}, 1--14. https://doi.org/10.1080/07350015.2025.2501563

\bibitem{} LaLonde, R. (1986) ``Evaluation the Econometric Evaluations of
Training Programs with Experimental Data,'' \textit{American Economic Review}%
, \textbf{76}, 604--620.


\bibitem{} Li, C., D. Ouyang, and J.S. Racine (2009) ``Nonparametric
regression with weakly dependent data: the discrete and continuous regressor
case,'' \textit{Journal of Nonparametric Statistics}, \textbf{21}, 697--711.

\bibitem{} Li, Q. and J.S. Racine (2007) \textit{Nonparametric econometrics:
Theory and practice}, Princeton University Press.

%

\bibitem{} Mandelbrot, B. (1997) \textquotedblleft Fractals and scaling in
finance, discontinuity, concentration, risk,\textquotedblright\ \textit{%
Selecta}, Volume E, Springer.

\bibitem{} Masry, E. (2005) ``Nonparametric regression estimation for
dependent functional data: asymptotic normality,'' \textit{Stochastic
Processes and their Applications}, \textbf{115}, 155--177.

\bibitem{} Nadaraya, E. (1965) ``On non-parametric estimates of density
functions and regression curves,'' \textit{Theory of Probability and its
Applications}, \textbf{10}, 186--190.

\bibitem{} Olson, C.A. (1998) ``A comparison of parametric and
semiparametric estimates of the effect of spousal health insurance coverage
on weekly hours worked by wives,'' \textit{Journal of Applied Econometrics}, 
\textbf{13}, 543--565.

\bibitem{} Pollard, D. (2001) ``A User's Guide to Measure Theoretic
Probability,'' Cambridge series in Statistical and Probabilistic Mathematics.

\bibitem{} Racine, J. and Q. Li (2004) ``Nonparametric estimation of
regression function with both categorical and continuous data,'' \textit{%
Journal of Econometrics}, \textbf{119}, 99--130.

\bibitem{} Ramsay, J. O. and B.W. Silverman (2005) ``Functional Data
Analysis,'' Springer, New York.


\bibitem{} Rosenbaum, P.R. and D.B. Rubin (1983) ``The central role of the
propensity score in observational studies for causal effects,'' \textit{%
Biometrika}, \textbf{70}, 41--55.

\bibitem{} Sain, S.R. (1994) ``Adaptive kernel density estimation,'' \textit{%
Computational Statistics and Data Analysis}, \textbf{39}, 165--186.


\bibitem{} Shen, G. (2002) ``Fractal dimension and fractal growth of
urbanized areas,'' \textit{International Journal of Geographical Information
Science}, \textbf{16}, 419---437.


\bibitem{} Takayasu, M. and H. Takaysu (2009) ``Fractals and Economics'', in
Encyclopedia of Complex Systems in Finance and Econometrics, Meyers, R.A.,
Eds., 444-463, Springer.


\bibitem{} Tibshirani, J. and S.Athey (2024) ``Package `grf' '', %
\url{https://cran.r-project.org/web/packages/grf/grf.pdf}.

\bibitem{} Vol'berg, A.L. and S.V. Konyagin (1988) ``On measures with the
doubling condition,'' \textit{Mathematics of the USSR-Izvestiya}, \textbf{30}%
, 629--638.

\bibitem{} Wager, S. and S. Athey (2018) ``Estimation and inference of
heterogeneous treatment effects using random forests,'' \textit{Journal of
the American Statistical Association}, \textbf{113}, 1228--1242.

\bibitem{} Wang, M.-C. and J. van Ryzin (1981) ``A class of smooth
estimators for discrete distributions,'' \textit{Biometrika}, \textbf{68},
301--309.

\bibitem{} Watson, G. S. (1964) ``Smooth regression analysis,'' \textit{%
Sankhya: The Indian Journal of Statistics}, Series (1961-2002), \textbf{26},
359--372.\vspace{.4in}
\end{thebibliography}

\begin{thebibliography}{99}

\bibitem{} Ahlfors, Lars (1966) \textit{Lectures on quasiconformal mappings}%
,\ Princeton University Press.

\bibitem{} Davydov, Y.A. (1968) ``Convergence of distributions generated by
stationary stochastic processes,'' \textit{Theory of Probability and its
Applications}, \textbf{13}, 691--696.

\bibitem{} Ferraty F., A. Mas, and P. Vieu (2007) ``Nonparametric regression
on functional data: Inference and practical aspects,'' \textit{Australian
and New Zealand Journal of Statistics}, \textbf{49}, 267--286.

\bibitem{} Ferraty F. and P. Vieu (2006) \textit{Nonparametric functional
data analysis: Theory and Practice}, Springer, New York.

\bibitem{} Hall, P., Q. Li, and J.S. Racine (2007) ``Nonparametric estimation of regression functions in the presence of irrelevant regressors,'' \textit{The Review of Economics and Statistics}, \textbf{89}, 784--789.

\bibitem{} Hotelling, H. (1929) ``Stability in competition,'' \textit{The
Economic Journal}, \textbf{39}, 41--57.

\bibitem{} Li, C., D. Ouyang, and J.S. Racine (2009) ``Nonparametric
regression with weakly dependent data: the discrete and continuous regressor
case,'' \textit{Journal of Nonparametric Statistics}, \textbf{21}, 697--711.

\bibitem{} Masry, E. (2005) ``Nonparametric regression estimation for
dependent functional data: asymptotic normality,'' \textit{Stochastic
Processes and their Applications}, \textbf{115}, 155--177.

\end{thebibliography}

\begin{thebibliography}{99}

\bibitem{} Hayfield, T. and J.S. Racine (2008) ``Nonparametric econometrics:
The np package,'' \textit{Journal of Statistical Software}, \textbf{27},
1--32.

\bibitem{} Wang, M.-C. and J. van Ryzin (1981) ``A class of smooth
estimators for discrete distributions,'' \textit{Biometrika}, \textbf{68},
301--309.

\end{thebibliography}

\begin{thebibliography}{99}
\bibitem{} Hayfield, T. and J.S. Racine (2008) ``Nonparametric econometrics:
The np package,'' \textit{Journal of Statistical Software}, \textbf{27},
1--32.

\bibitem{} Imbens, G. and Y. Xu (2024) ``LaLonde (1986) after nearly four
decades: Lessons learned,'' arXiv 2406.00827 (econ.EM).

\bibitem{} LaLonde, R. (1986) ``Evaluation the Econometric Evaluations of
Training Programs with Experimental Data,'' \textit{American Economic Review}%
, \textbf{76}, 604--620.

\bibitem{} Tibshirani, J. and S.Athey (2024) ``Package `grf' '', %
\url{https://cran.r-project.org/web/packages/grf/grf.pdf}.

\end{thebibliography}
\end{document}